\immediate\write18{makeindex \jobname.nlo -s nomencl.ist -o \jobname.nls}
\documentclass[english,preprint,superscriptaddress,showkeys,showpacs,nofootinbib]{revtex4-1}
\usepackage{lmodern}
\usepackage[T1]{fontenc}
\usepackage[utf8]{inputenc}
\usepackage{setspace}
\usepackage{textcomp}
\usepackage{indentfirst}
\usepackage{placeins}
\usepackage{babel}
\usepackage{booktabs}
\usepackage{graphicx}
\usepackage{fullpage}
\usepackage{supertabular}
\usepackage{comment}
\usepackage{amsmath}
\usepackage{amssymb}
\usepackage{eurosym}
\usepackage{float}
\usepackage[caption=false]{subfig}
\usepackage{textgreek}
\usepackage{epstopdf}
\usepackage{bigints}
\usepackage{array}
\usepackage{url}
\usepackage{nomencl}
\makenomenclature

\renewcommand{\nompreamble}{In brackets the number of the equation where the symbol is first used}

\begin{document}
\title{Hybrid mean field and real space model
for vacancy diffusion-mediated annealing of radiation defects}
\date{\today}
\author{I. ~Rovelli}
\email[Corresponding author: ]{ir1014@ic.ac.uk}
\affiliation{Department of Physics, Imperial College London, Exhibition Road, London SW7 2AZ, United Kingdom}
\affiliation{Culham Centre for Fusion Energy, UK Atomic Energy Authority, Abingdon, Oxfordshire OX14 3DB, United Kingdom}
\author{S.L. ~Dudarev}
\affiliation{Culham Centre for Fusion Energy, UK Atomic Energy Authority, Abingdon, Oxfordshire OX14 3DB, United Kingdom}
\author{A.P. ~Sutton}
\affiliation{Department of Physics, Imperial College London, Exhibition Road, London SW7 2AZ, United Kingdom}
\begin{abstract}
In a fusion or advanced fission reactor, high energy neutrons induce the formation of extended defect clusters in structural component materials, degrading their properties over time. Such damage can be partially recovered via a thermal annealing treatment. 
Therefore, for the design and operation of fusion and advanced fission nuclear energy systems it is critical to estimate and predict the annealing timescales for arbitrary configurations of defect clusters.\\
In our earlier paper  [I. Rovelli, S. L. Dudarev, and A. P. Sutton, J. Mech. Phys. Solids {\bf103}, 121 (2017)] we extended the Green function formulation by Gu, Xiang {\it et al.} [Y. Gu, Y. Xiang, S. S. Quek, and D. J. Srolovitz, J. Mech. Phys. Solids {\bf83}, 319 (2015)] for the climb of curved dislocations, to include the evaporation and growth of cavities and vacancy clusters, and take into account the effect of free surfaces.\\
In this work, we further develop this model to include the effect of radiation defects that are below the experimental detection limit, via a mean field approach coupled with an explicit treatment of the evolution of discrete defect clusters distributed in real space.\\
We show that randomly distributed small defects screen diffusive interactions between larger discrete clusters. The evolution of the coupled system is modelled self-consistently.
We also simulate the evolution of defects in an infinite laterally extended thin film, using the Ewald summation of screened Yukawa-type diffusive propagators.
\end{abstract}

\maketitle

\nomenclature[1a]{$R$}{Radius of dislocation loop or cavity (\ref{eq:sym1})}
\nomenclature[1b]{$t$}{Time (\ref{eq:sym2})}
\nomenclature[1c]{$D_v$}{Vacancy diffusion coefficient (\ref{eq:sym1})}
\nomenclature[1d]{$D_v^0$}{Vacancy diffusion coefficient pre-exponential (\ref{eq:tau})}
\nomenclature[1db]{$T$}{Absolute temperature (\ref{eq:sym3})}
\nomenclature[1dc]{$T_m$}{Melting temperature (in text, page \pageref{sym4})}
\nomenclature[1e]{$c$}{Vacancy concentration per atomic site (\ref{eq:Gu1})}
\nomenclature[1f]{$c_0$}{Equilibrium vacancy concentration (\ref{eq:sym1})}
\nomenclature[1ga]{$c_\infty$}{Vacancy concentration at infinity (\ref{eq:sym3})}
\nomenclature[1gb]{$c_S$}{Vacancy concentration at thin film surface (\ref{eq:BCTF})}
\nomenclature[1h]{$c_\Sigma$}{Vacancy concentration at cavity surface (\ref{eq:sym3})}
\nomenclature[1i]{$c_\Delta$}{Vacancy concentration at dislocation loop line (\ref{eq:BCloop})}
\nomenclature[1j]{$c_b$}{Background vacancy concentration due to macroscopic defect clusters (\ref{eq:micro})}
\nomenclature[1k]{$\bar{c}$}{Configuration-averaged vacancy concentration with respect to the small dislocation loops population (\ref{eq:sym6})}
\nomenclature[1l]{$c_{[i]}$}{Vacancy concentration omitting the contribution from the $i$th dislocation loop (\ref{eq:micro})}
\nomenclature[1m]{$\bar{c}_{[i]}^\text{eff}$}{Effective averaged vacancy concentration experienced by the $i$th dislocation loop (\ref{eq:ceff})}
\nomenclature[1n]{$c_\text{avg}$}{Spatial average of the vacancy concentration in the medium (\ref{eq:cavgg})}
\nomenclature[1o]{$\gamma$}{Surface energy per unit area (\ref{eq:sym3})}
\nomenclature[1p]{$\Omega$}{Atomic volume (\ref{eq:sym3})}
\nomenclature[1q]{$k_B$}{Boltzmann's constant (\ref{eq:sym3})}
\nomenclature[1s]{$\mu$}{Shear modulus (\ref{eq:sym4})}
\nomenclature[1t]{$\nu$}{Poisson's ratio (\ref{eq:sym4})}
\nomenclature[1u]{$r_d$}{Dislocation core radius (\ref{eq:vloop})}
\nomenclature[1v]{$E_\text{v}$}{Vacancy formation energy (\ref{eq:tau})}
\nomenclature[1w]{$E_\text{m}$}{Vacancy migration barrier (\ref{eq:tau})}
\nomenclature[1x]{$b_e$}{Edge component of dislocation Burgers vector (\ref{eq:vloop})}
\nomenclature[1y]{$\tau_e$}{Characteristic timescale of defect cluster evaporation (\ref{eq:tau})}
\nomenclature[1z]{$v_\text{cl}$}{Dislocation climb velocity (\ref{eq:Gu1})}
\nomenclature[2a]{$f_\text{cl}$}{Dislocation climb force (\ref{eq:BCloop})}
\nomenclature[2b]{$\mathcal{V}$}{Defect cluster volume (\ref{eq:sym2})}
\nomenclature[2c]{$G$}{Free space steady-state diffusion Green's function (\ref{eq:sym2})}
\nomenclature[2d]{$G_Y$}{Yukawa (screened) Green's function (\ref{eq:sym8})}
\nomenclature[2e]{$J_n$}{Vacancy current normal to cavity surface (\ref{eq:simple})}
\nomenclature[2f]{$\Sigma$}{Locus of points belonging to a cavity surface (\ref{eq:simple})}
\nomenclature[2g]{${\xi_\Delta}$}{Prefactor to the vacancy concentration generated by a circular dislocation loop (\ref{eq:micro})}
\nomenclature[2ha]{${\overline{\xi_\Delta}}$}{Size average of the prefactor to the vacancy concentration generated by a circular dislocation loop (\ref{eq:AandB})}
\nomenclature[2hb]{$\overline{\xi_\Delta c}$}{Size average of the product of the prefactor to the vacancy concentration generated by a circular dislocation loop and the boundary condition at the dislocation loop line (\ref{eq:AandB})}
\nomenclature[2i]{$S_\text{l}$}{Perturbation to the vacancy field in the neighborhood of the $i$th loop arising from the condition of local thermodynamic equilibrium. (\ref{eq:sym6})}
\nomenclature[2j]{$S^i_\text{nl}$}{Non-local contribution to the vacancy field near the $i$th loop from all the other loops in the system. (\ref{eq:sym6})}
\nomenclature[2k]{$p$}{Probability density function of the positions and radii of all the dislocation loops in the system (\ref{eq:sym5})}
\nomenclature[2l]{$p_1$}{Probability density function of the position and radius of a single loop (\ref{eq:sym7})}
\nomenclature[2m]{$f$}{Size distribution function for a single loop (\ref{eq:sym9})}
\nomenclature[2n]{$F$}{Time-dependent size distribution function for a single loop (\ref{eq:continuity})}
\nomenclature[2o]{$\phi$}{Scaled non-dimensional size distribution function for a single loop (\ref{eq:sym10})}
\nomenclature[2p]{$v_\text{il}$}{Drift velocity in size space of interstitial dislocation loops (\ref{eq:cavgg})}
\nomenclature[2q]{$v_\text{vl}$}{Drift velocity in size space of vacancy dislocation loops (\ref{eq:sym11})}
\nomenclature[2r]{$v_\text{c}$}{Drift velocity in size space of cavities (\ref{eq:sym11})}
\nomenclature[2s]{$\rho$}{Number density of dislocation loops in the medium (\ref{eq:sym12})}
\nomenclature[2t]{$\rho_c$}{Number density of cavities in the medium (\ref{eq:coupling})}
\nomenclature[2u]{$\overline{\dot{V}}_c$}{Average rate of change of the volume of cavities (\ref{eq:coupling})}
 \nomenclature[2v]{$K$}{Effective diffusional interaction between cavities in a thin film (\ref{eq:YukawaKernel})}
 \nomenclature[2x]{$\mathcal{T}$}{Scattering operator of random scatterers (\ref{eq:Tseries})}
   \nomenclature[2y]{$\mathcal{S}$}{Self energy reflecting the averaged effect of random scatterers (\ref{eq:Sigmaseries})}

\section{Introduction}
High energy neutrons are the most prominent source of radiation damage to metallic plasma-facing components (PFCs) of fusion reactors, or structural components of advanced fission power plants. Atoms displaced by the neutrons generate vacancies and interstitials that aggregate to form prismatic vacancy and interstitial dislocation loops and voids \cite{VanRenterghem2016,Ferroni2015,Fukumoto2013,Nagasaka2005,Byun2014}. To design annealing protocols for the recovery of neutron irradiation damage \cite{Ferroni2015}, we have to be able to predict the evolution of arbitrary populations of such defects as a function of temperature and time.\\

In an earlier paper \cite{Rovelli2017} we presented a formalism that extends the non-local dislocation climb model of Gu, Xiang {\it et al.} \cite{Gu2015} to simulate the high temperature evolution of discrete dislocation loops and vacancy clusters inside a finite medium. One of the motivations for developing a real-space model, in contrast to a rate-theory approach, is that the sizes of clusters of vacancies or interstitials produced by neutron irradiation appear to obey a power-law probability distribution \cite{Sand2013,Yi2015} of the form $f(n)\sim n^{-s}$, with $n$ the number of interstitials or vacancies in the cluster. For $s<2$ the average of the size distribution is not defined, and so a representative cluster needed in a rate theory model does not exist. A real-space model can also treat local variations in the number density of clusters, for instance those arising from \emph{depleted zones}\cite{Beavan1971} at grain boundaries and free surfaces.\\

\begin{center}
\begin{figure}[h!]
\centering
\includegraphics[width=0.6\columnwidth]{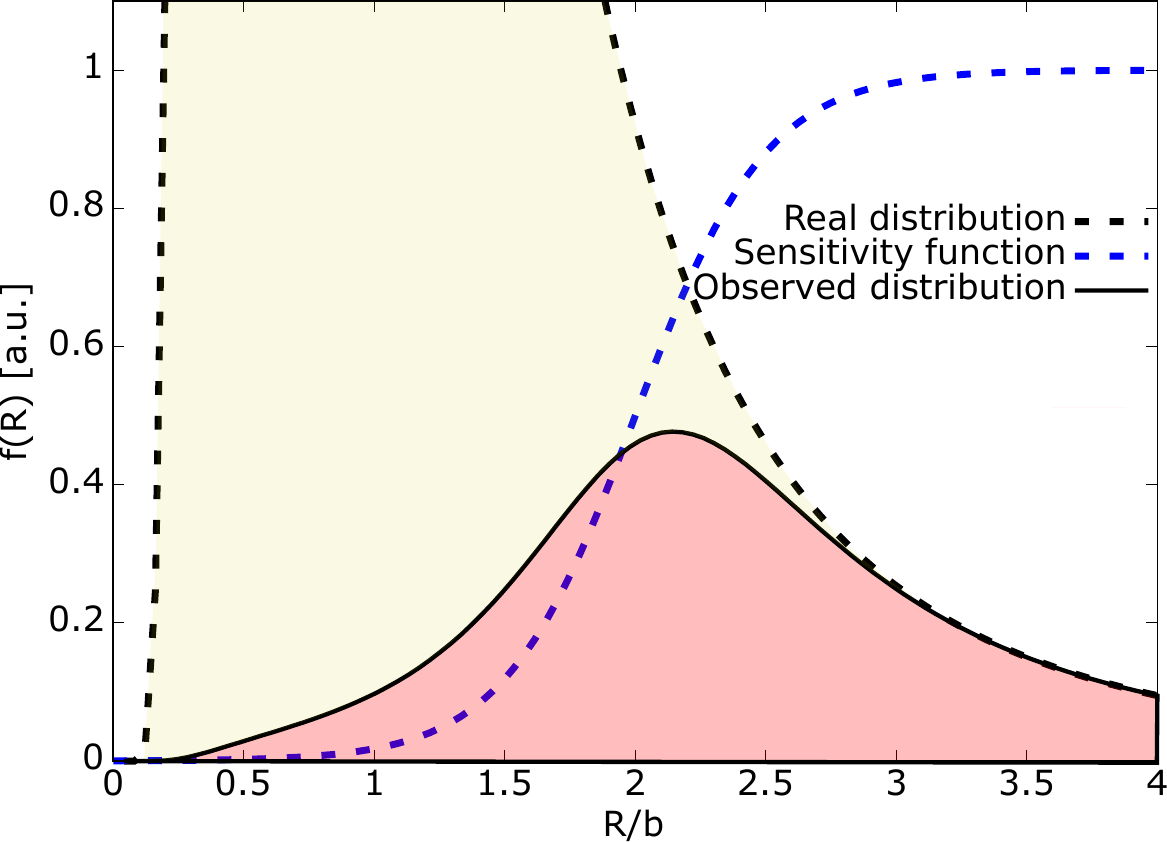}
\caption{Qualitative explanation of the observed Gaussian-like size distribution of defect clusters in neutron irradiated tungsten. The dashed black line represents the true, power-law like, size distribution. The dashed blue line represents the experimental sensitivity function plotted as a function of cluster size. The solid black line is the observed distribution, given by the product of the true distribution and the sensitivity function. The red shaded area is proportional to the number density of the observed clusters, while the yellow shaded area is proportional to the number density of mostly ``invisible'' clusters.}
\label{fig:RealDistro}
\end{figure}
\end{center}

It is likely that the experimentally determined number of the smallest defect clusters is always underestimated owing to detection limits of the instrumentation. 
The sizes of experimentally observed defect clusters is usually assumed to obey a Gaussian-like distribution. 
Such a distribution can be understood as a product of the true distribution with a sigmoid function representing the instrumental sensitivity, as sketched in Fig. \ref{fig:RealDistro}. \\
Therefore, experimental data on radiation induced defect clusters is generally incomplete and the number density of the smallest ``invisible'' defect clusters can be much greater than the observed population, as recently noted by Liu {\it et al.} \cite{Liu2017}.
It would be useful, therefore, to describe the unknown small-size cluster population as an effective mean field, which evolves self-consistently with the observable cluster population. 
This approach would provide an efficient way to investigate the effect of invisible clusters on the distribution of sizes of visible clusters. 
It turns out that the mean field is governed  by only a few low-order moments of the size distribution of visible clusters, which can be determined by the experimentally measured evolution of the observable population. \\
In this paper we present a hybrid model that couples the evolution of an experimentally visible, discrete population of defect clusters in real space with a mean field, representing the experimentally invisible clusters.\\

In sec.\ref{sec:timescales} we estimate the expected timescales for the evaporation of nanometric cavities as a function of temperature in W, Fe and Be.
In sec. \ref{sec:MFdef} we introduce the mean field formulation by averaging the positions and sizes of the small clusters, using a technique developed in scattering theory by Edwards \cite{Edwards1958}. 
In this way we obtain a general expression, eq.~(\ref{eq:MF2}), for the vacancy concentration in the presence of an arbitrary distribution of small clusters.\\
We then present in appendix \ref{app:pert} a perturbative approach that expands eq.~(\ref{eq:MF2}) in terms of an increasing number of scattering events, deriving an analogue of the self-energy expansion for condensed matter systems.
In the case of homogeneous and uncorrelated distributions, we provide closed form expressions for the self-energy up to the $3^{\text{rd}}$ order terms in eq.~(\ref{eq:selfexplicit}).\\

By considering the first order approximation to the effective self-energy, we show that the effective Green's function governing the interaction between the observable clusters takes  the form of a Yukawa propagator,  eq.~(\ref{eq:YukawaGreen}). 
Thus, we show that diffusive interactions mediated by vacancies between larger clusters are screened by the mean-field of small clusters.
This enables us in eq.~(\ref{eq:MFsinglefinal}) to provide a closed form expression for the vacancy field where only observable clusters are treated explicitly.\\
The rates of growth of cavities can be calculated self-consistently in the same manner as in our previous model \cite{Rovelli2017} by evaluating the vacancy field at the cluster positions with defined  boundary conditions. We couple the evolution of the mean field with large clusters via eq.~(\ref{eq:continuity}) and eq.~(\ref{eq:coupling}).\\

The theory is developed for an infinite medium. However, the majority of experimental data on radiation-induced defect clusters has so far been obtained using thin film samples.\\
In sec.\ref{sec:TF} we present an extension of the theory to a thin film infinitely extended in the lateral direction, making use of a variant of Ewald summation developed for Yukawa potentials \cite{Salin2000}. In sec.\ref{sec:AllMF} we briefly present a mean-field treatment of all defect clusters.\\
Finally, in sec.\ref{sec:Numerics} we present numerical simulations to illustrate applications of the theory to the evolution of distributions of cluster sizes in irradiated thin films of W, Be and Fe.\\
A list of mathematical symbols and notations used in the present work is provided after the appendices.

\section{Preliminary estimate of governing timescales}
\label{sec:timescales}
An estimate of the timescale required for the complete evaporation of a nanometric defect cluster can be obtained by considering an isolated spherical cavity in a vacancy atmosphere in local thermodynamic equilibrium. Assuming a steady state solution\footnote{The applicability of this assumption is discussed quantitatively in \cite{Rovelli2017}} for the vacancy concentration throughout space, the rate of change of the cavity radius $R$ is:
\begin{equation}
\frac{dR}{dt}=\frac{D_v}{R}\left[c_0-c_\Sigma(R)\right],
\label{eq:sym1}
\end{equation}
where $D_v$ is the vacancy diffusion coefficient, $c_\Sigma$ is the local vacancy concentration at the cavity surface, and $c_0$ is the equilibrium vacancy concentration per atomic site, given by $c_0=\exp(-E_v/k_B T)$ where $E_v$ is the vacancy formation energy. $c_\Sigma(R)$ is determined by the condition of local equilibrium, i.e. there is no change in free energy if vacancies attach to or detach from the cavity, which leads to the Gibbs-Thomson expression:
\begin{equation}
c_\Sigma(R)=c_0 \exp \left(\frac{2\gamma\Omega}{R k_B T}\right),
\label{eq:sym3}
\end{equation}
where $\gamma$ is the cavity surface energy per unit area and $\Omega$ is the atomic volume. Therefore, if we consider a system of cavities of average radius $\bar{R}$ evolving adiabatically with a vacancy field in local thermodynamic equilibrium we can estimate the characteristic timescale for cavity evaporation as follows:
\begin{equation}
\tau_\text{e}=\frac{\bar{R}^2}{D_v [c_0-c_\Sigma(\bar{R})]}=\frac{\bar{R}^2}{D_v^0} \exp\left(\frac{E_\text{v}+E_\text{m}}{k_B T}\right) \left[\exp \left(\frac{2\gamma\Omega}{\bar{R} k_B T}\right)-1\right]^{-1},
\label{eq:tau}
\end{equation}
\begin{center}
\begin{figure}[h!]
\centering
\includegraphics[width=0.7\columnwidth]{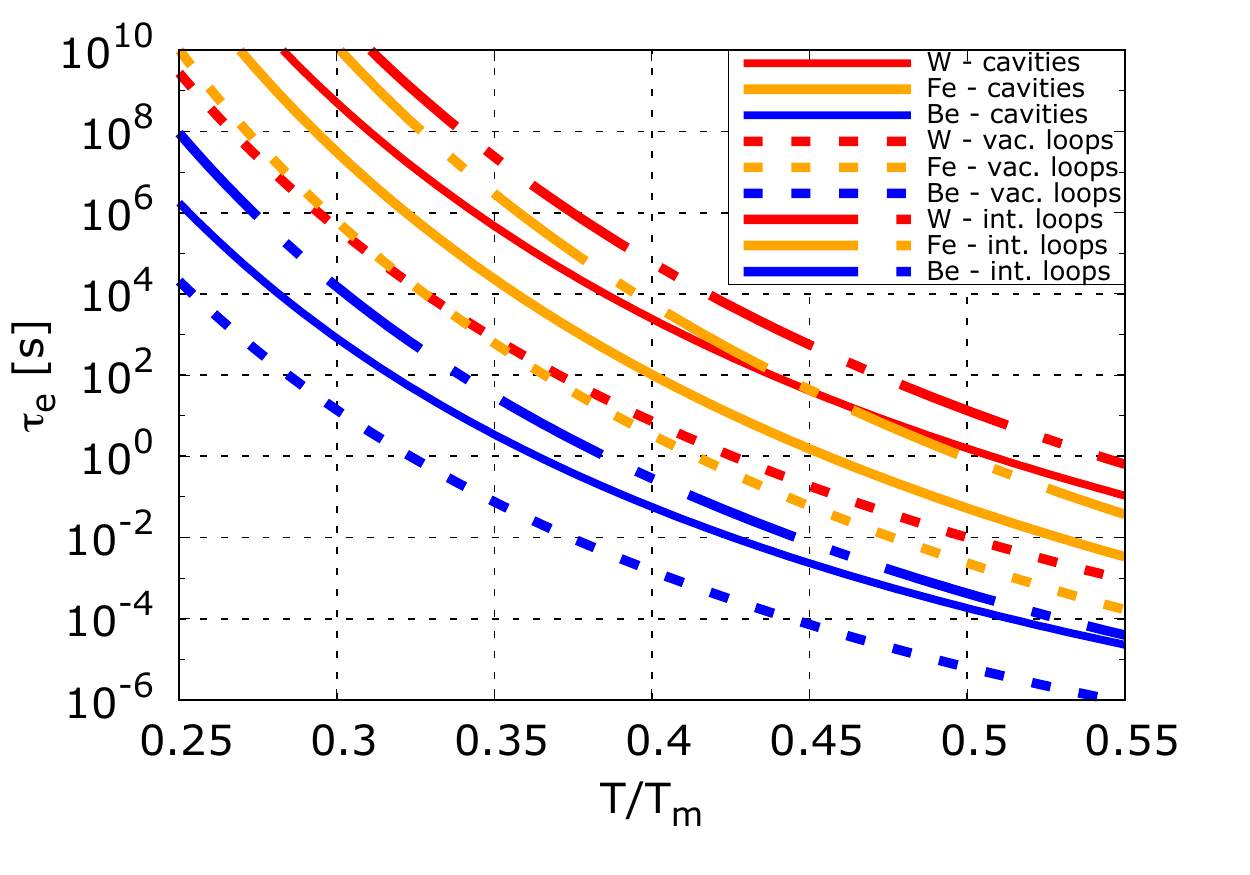}
\caption{Estimated timescales $\tau_\text{e}$ for the annihilation of $\sim1$ nm cavities (solid lines), vacancy dislocation loops (dashed lines) and interstitial dislocation loops (solid-dashed lines) in W (red), Fe (yellow) and Be (blue), as functions of homologous temperature $T/T_\text{m}$.}
\label{fig:Timescales}
\end{figure}
\end{center}
where $E_v$ and $E_m$ are respectively the vacancy formation and migration energies, and $D_v^0$ is the pre-exponential factor of the diffusion coefficient.\\
In a similar way, we can estimate the characteristic timescale of dislocation loop evaporation by considering the rate of change of the radius of an isolated circular prismatic loop \cite{Gu2015}:
\begin{equation}
\frac{dR}{dt}=\pm \frac{2 \pi D_v}{b_e \ln(8R/r_d)}\left[c_0-c_\Delta(R)\right],
\label{eq:vloop}
\end{equation}
where the plus/minus sign denote respectively the case of a vacancy or interstitial loop, $b_e$ is the edge (out of plane) component of the dislocation Burgers vector, $r_d$ is the radius of the dislocation core and $c_\Delta$ is the vacancy concentration infinitesimally close to the dislocation line. The assumption of local equilibrium between the dislocation loop and the vacancy field surrounding it leads to the following relation:
\begin{equation}
c_\Delta(R)=c_0\exp\left[-\frac{f_\text{cl}(R) \Omega}{b_e k_B T}\right],
\label{eq:BCloop}
\end{equation}
where, for a circular prismatic loop
\begin{equation}
f_\text{cl}(R)=\mp \frac{\mu b_e^2}{4 \pi (1-\nu)R} \left[\ln\left(\frac{8R}{r_d} \right)-1 \right]
\label{eq:sym4}
\end{equation}
is the climb force per unit length of the dislocation\footnote{In this treatment we neglect the additional contributions to the climb force due to stresses imposed on the system as a whole or arising from other loops. This is a good approximation because we are considering very small loops (a few nanometers wide) for which the self-interaction stress dominates. We have defined the climb force as the projection of the Peach-Koehler force in the direction of the cross product between the dislocation line direction and the Burgers vector. Therefore, the climb force acting on a vacancy loop is the negative of that acting on an interstitial loop.},  $\nu$ is Poisson's ratio and $\mu$ is the shear modulus and the minus/plus signs distinguish between vacancy type an interstitial type, respectively. We can therefore estimate the characteristic timescale for the evaporation of a dislocation loop of radius $\bar{R}$ as:
\begin{equation}
\tau_\text{e}=\pm\frac{b_e\bar{R}\ln(8\bar{R}/r_d)}{2 \pi D_v^0} \exp\left(\frac{E_\text{v}+E_\text{m}}{k_B T}\right) \left\lbrace \exp \left[\pm\frac{\mu \Omega b_e}{4 \pi (1-\nu) k_B T}\left(\ln\left(\frac{8\bar{R}}{r_d}\right)-1 \right)\right]-1\right\rbrace ^{-1}.
\label{eq:tauloop}
\end{equation}
The parameters used for the investigated materials throughout the present work are given in table \ref{tab:parameters}.\\

\begin{table}[h]
\begin{center}
\begin{ruledtabular}
    \begin{tabular}{ c  c c c c c c c c }
    Material & $\gamma$ [eV/nm$^2$] & $\mu$ [GPa] & $\nu$ & $b$ [nm] & $\Omega$ [nm$^3$] &  $E_\text{v}$ [eV] & $E_\text{m}$ [eV] &  $D_\text{v}^0$ [nm$^2$/s] \\ \hline
    W  & 20.4 $^\text{a}$ & 161 $^\text{b}$ & 0.28 $^\text{b}$ & 0.27 $^\text{b}$ & 0.016 $^\text{b}$ & 3.56 $^\text{c}$ & 1.78 $^\text{c}$ & 4.0$\times10^{12}$ $^\text{d}$ \\
    Fe & 15.3 $^\text{a}$ & 82 $^\text{b}$ & 0.29 $^\text{b}$ & 0.29 $^\text{b}$ & 0.012 $^\text{b}$ & 2.07 $^\text{c}$ & 0.65 $^\text{c}$ & 2.8$\times10^{14}$ $^\text{e}$\\
    Be & 12.5 $^\text{a}$ & 132 $^\text{b}$ & 0.03 $^\text{b}$ & 0.18 $^\text{b}$ & 0.008 $^\text{b}$ & 0.95 $^\text{c}$ & 0.81 $^\text{c}$ & 5.7$\times10^{13}$ $^\text{f}$\\ 
    \end{tabular}
    \end{ruledtabular}
\end{center}
\caption{Material parameters used in the present work. Parameters for Be are averages of basal and non-basal values. References: $^\text{a}$\cite{Vitos1998}, $^\text{b}$\cite{KayeLaby}, $^\text{c}$\cite{DudarevDFT2013}, $^\text{d}$\cite{Mundy1978}, $^\text{e}$\cite{Iijima1988}, $^\text{f}$\cite{Dupouy1966}. 1 eV/nm$^2 \sim 0.16$ J/m$^2$.}
\label{tab:parameters}
\end{table}
We applied eq.~(\ref{eq:tau}) to bcc iron, tungsten and beryllium, which are candidate materials for nuclear fusion engineering applications, with $\bar{R}=1$ nm, which is within the range of experimentally observed sizes of radiation-induced cavities and dislocation loops. The computed $\tau_\text{e}$ are plotted in Fig. \ref{fig:Timescales} as a function of the homologous temperature $T/T_\text{m}$, where $T_\text{m}$ \label{sym4} is the melting point. It is evident that $\tau_\text{e}$ can vary over many orders of magnitude: from milliseconds to years, depending on temperature and material properties.\\
We note that although W shares the same crystal structure as Fe in Fig. \ref{fig:Timescales}, there is a systematic difference of at least one order of magnitude in the timescale $\tau_e$ for cavities and dislocation loops, even after scaling the temperature to the relevant melting point. \\

\begin{center}
\begin{figure}[h!]
\centering
\subfloat[]{\includegraphics[width=0.515\textwidth]{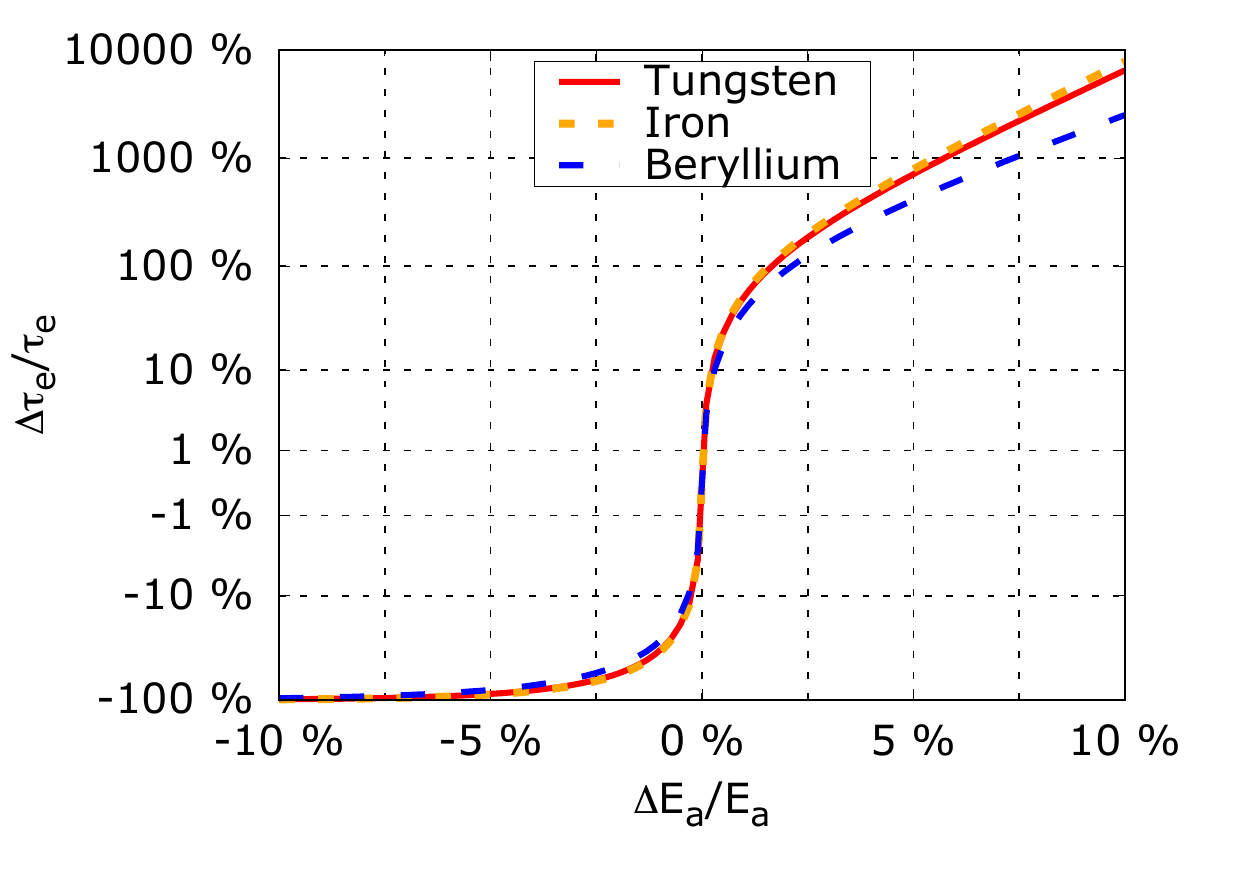}}
\hspace{-20pt}
\subfloat[]{\includegraphics[width=0.51\textwidth]{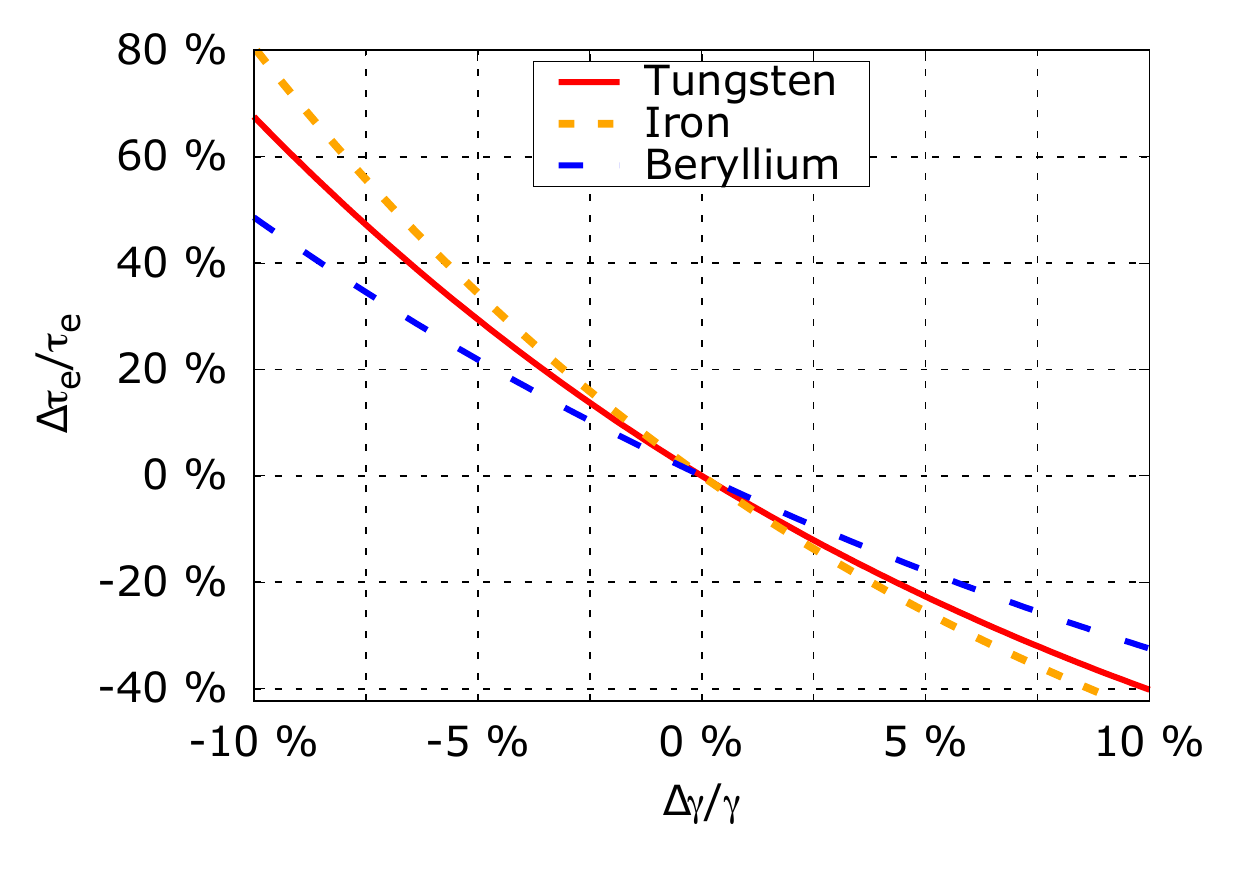}}
\caption{Relative change of the estimated timescale $\tau_e$ for cavity evaporation at $T/T_\text{m}=0.4$ with respect to small deviations of activation energy $E_a=E_\text{v}+E_\text{m}$ (a) and surface energy  $\gamma$ (b). In plot (a) the scale of  the $y$ axis is linear between -$1 \%$ and $1 \%$ and logarithmic elsewhere.}
\label{fig:Sensitivity}
\end{figure}
\end{center}

In Fig. \ref{fig:Sensitivity} we plot the dependence of the relative change of predicted timescales $\Delta \tau_e/\tau_e$ for cavity evaporation on relative changes in the activation energy for diffusion $\Delta E_a/E_a$, where $E_a=E_\text{v}+E_\text{m}$, and relative changes in the surface energy $\Delta \gamma/\gamma$, of up to $\pm10\%$, at $T=0.4 \: T_\text{m}$. In Fig. \ref{fig:Sensitivity}(a) we see that an overestimation of $E_a$ by just $\sim 2 \%$ leads to a relative error of at least $100\%$ for $\tau_e$ in all three materials.
This extreme sensitivity should be kept in mind when attempting to compare experimentally observed timescales with theoretical estimates. Activation energies for diffusion can be affected by a number of factors which may not be under experimental control, such as the presence of bound impurity-vacancy complexes, or  a dependence of the entropy of activation on temperature, which may become significant at higher temperatures \cite{Glensk2014}. The sensitivity to errors in the assumed surface energy is less extreme, but also a source of uncertainty because the surface energy per unit area of clusters with radius as small as 1~nm or less may differ significantly from the surface energies per unit area of larger clusters.

\section{Definition of the mean field formalism}
\label{sec:MFdef}
\subsection{Adiabatic vacancy field in the presence of a small defect cluster}
Consider an infinite crystal containing a circular prismatic dislocation loop of radius $R$, with its center at the origin and lying in the $x-y$ plane. The loop may be of vacancy or interstitial character. Since the loop lies in the $x-y$ plane the component, $b_e$, of the Burgers vector normal to the loop (i.e. along the $z$-axis) is constant at all points around the loop. Let $c(\mathbf{x})$ be the vacancy concentration per atomic site, i.e. $c(\mathbf{x})$ is the (dimensionless) probability of finding a vacancy at a lattice site at $\mathbf{x}$, and let us assume that vacancies are the only mobile point defects. If we further assume that the evolution of the vacancy field is adiabatic with respect to changes in the loop configuration\footnote{this assumption implies that local equilibrium exists between each dislocation segment and the surrounding vacancy field. For quantitative bounds on this assumption see \cite{Rovelli2017}}, we can write \cite{Gu2015} the following equation:
\begin{equation}
c(\mathbf{x})=-\frac{b_e}{4 \pi D_v}  \oint_{\Gamma} \frac{v_\text{cl}(\mathbf{x}')}{|\mathbf{x}-\mathbf{x}' |} dl'+c_\infty,
\label{eq:Gu1}
\end{equation}
where $\Gamma$ denotes the dislocation line, $v_\text{cl}(\mathbf{x})$ is the dislocation climb velocity, and $c_\infty$ is the vacancy concentration at infinity. We define $v_\text{cl}(\mathbf{x})$ as the projection of the velocity of a point on $\Gamma$ along the vector defined by the cross product between the tangent vector to the dislocation line at $\mathbf{x}$ and the Burgers vector.\\ Throughout this paper we adhere to the definition of the Burgers vector used by Hirth and Lothe \cite{Hirth1982}, i.e. given a circular path $C$, drawn around a dislocation line according to the right hand rule with respect to the dislocation line direction, the Burgers vector $\mathbf{b}$ is defined as:
\begin{equation}
\mathbf{b}=\oint_C \frac{\partial \mathbf{u}}{\partial l} dl
\end{equation}
where $\mathbf{u}$ is the displacement field arising from the dislocation. This implies that the cross product between the tangent vector to the dislocation line and the Burgers vector points inward toward the center of a prismatic interstitial loop, and outward away from the center of a prismatic vacancy loop.
We warn the reader that the opposite convention is used by Landau \cite{Landau1970}, Trinkaus \cite{Trinkaus1972} and others, where the direction of the Burgers vector is reversed with respect to the dislocation line. We also define the normal direction $\hat{\mathbf{n}}$ to the surface enclosed by a dislocation loop according to the right hand rule with respect to the dislocation line direction, so that $\mathbf{b}$ and $\hat{\mathbf{n}}$ are parallel for a prismatic vacancy loop and anti-parallel for a prismatic interstitial loop.\\
We point out that eq.~(\ref{eq:Gu1}) is the scalar form of a general vector equation (see appendix \ref{app:loopvolume} for details):
\begin{equation}
c(\mathbf{x})=-\frac{1}{4 \pi D_v}  \oint_{\Gamma} \frac{\mathbf{v}(\mathbf{x}')}{|\mathbf{x}-\mathbf{x}' |}\cdot\left(d\mathbf{l}'\times\mathbf{b}\right)+c_\infty=\oint_{\Gamma} G(|{\bf x}-{\bf x}'|)\mathbf{v}(\mathbf{x}')\cdot\left(d\mathbf{l}'\times\mathbf{b}\right)+c_\infty,
\end{equation}
where $G(|{\bf x}|)=-1/4\pi D_v|{\bf x}|$ is the free space Green's function of the steady-state diffusion equation, $D_v \nabla^2 G(|{\bf x}|)=\delta({\bf x})$, $\mathbf{v}(\mathbf{x}')$ is the vector velocity of a point $\mathbf{x}'\in\Gamma$, $\mathbf{b}$ is the Burgers vector of the dislocation loop, and the differential $d\mathbf{l}'$ is tangential to the dislocation line at $\mathbf{x}'$.\\
By evaluating eq.~(\ref{eq:Gu1}) on $\Gamma$ and using eq.~(\ref{eq:BCloop}) we can self-consistently find an analytical solution for $v_\text{cl}(\mathbf{x})$, leading to a closed-form expression for the vacancy concentration in the medium:
\begin{equation}
c(\mathbf{x})=c_\infty-\frac{1}{2 \ln(8R/r_d)}  \oint_{\Gamma} \frac{1}{|\mathbf{x}-\mathbf{x}' |}   \left[c_\infty-c_\Delta(R) \right] dl'.
\end{equation}
Let $r$ denote the distance between $\mathbf{x}$ and the center of the dislocation loop and let $\phi=\cos^{-1}(z/r)$, where $z$ is the component of $\mathbf{x}$ out of the $x-y$ plane. In the limit of $r \gg R$ we can expand the above integrand in powers of $R/r$ to obtain:
\begin{equation}
c(r,\phi)=c_\infty-\left[c_\infty-c_\Delta(R)\right]\frac{\pi}{\ln \left(8 R/r_d\right)} \left(\frac{R}{r}\right) \left[ 1+\frac{1}{2}\left(\frac{R}{r}\right)^2 \left(\frac{3\sin^2\phi}{2} -1\right)+ O\left( \frac{R}{r} \right)^3 \right],
\label{eq:cloopseries}
\end{equation}
where $\phi=\cos^{-1}(z/r)$, and $z$ is the component of $\mathbf{x}$ out of the $x-y$ plane.
The above expression shows that to second order in $R/r$, a dislocation loop, or in fact any defect cluster, can be treated as a spherically symmetric vacancy source or sink.\\
\\
On the other hand, the vacancy field sufficiently far from an isolated spherical cavity at the origin can be \emph{exactly} interpreted as originating from a point-like source. Indeed, at all $r>R$, Newton's shell theorem shows that the vacancy concentration is a function of the distance to the center of the cavity $r$, and is given by \cite{Rovelli2017,Dudarev2000}:
\begin{equation}
c(r) = c_\infty-\left[c_\infty-c_\Sigma(R)\right]\left(\frac{R}{r}\right).
\label{eq:ccavNewt}
\end{equation}
We note that the above expressions are examples of a more general equation that admits a clear physical interpretation. The vacancy field at large distances from \emph{any} point defect cluster situated at the origin, which can be approximated as an isotropic point-like source, is characterized by the rate of change of the volume of the cluster\footnote{with the caveat that $\mathcal{V}$ is negative for vacancy clusters or voids because their growth reduces the overall vacancy concentration in solution, and positive for interstitial clusters because their growth increases the overall vacancy concentration in solution assuming there are no free interstitial atoms.} $\mathcal{V}$:
\begin{equation}
c(\mathbf{x})=c_\infty-\frac{\text{d}\mathcal{V}}{\text{d}t} G(|\mathbf{x}|).
\label{eq:sym2}
\end{equation}
This statement can be proven straightforwardly for a spherical cavity, where:
\begin{equation}
\frac{\text{d}\mathcal{V}_\text{cav}}{\text{d}t}={d \over dt}\left(-{4\pi\over 3}R^3 \right)=-4\pi R^2 \frac{dR}{dt}=-4\pi D_v R \left[c_\infty-c_\Sigma(R)\right],
\end{equation}
which, upon substitution in eq.~(\ref{eq:sym2}), leads to eq.~(\ref{eq:ccavNewt}).\\ For a prismatic dislocation loop we prove in appendix \ref{app:loopvolume} that:
\begin{equation}
\frac{d\mathcal{V}_\text{loop}}{dt}=-\oint_{\Gamma}\mathbf{v}(\mathbf{x}')\cdot\left(d\mathbf{l}'\times\mathbf{b}\right)=-
\oint_{\Gamma}\mathbf{b}\cdot\left(\mathbf{v}(\mathbf{x}') \times d\mathbf{l}'\right)=-\oint_\Gamma v_\text{cl}(\mathbf{x}') b_e(\mathbf{x}') dl',
\end{equation}
which, assuming constant $v_\text{cl}(\mathbf{x}')$ and $b_e({\mathbf{x}'})$ over $\Gamma$, becomes:
\begin{equation}
\frac{d\mathcal{V}_\text{loop}}{dt}=-2 \pi R v_\text{cl} b_e=-\frac{4 \pi^2 R D_v}{\ln(8R/r_d)}\left[c_\infty-c_\Delta(R)\right],
\end{equation}
giving the leading order term of eq.~(\ref{eq:cloopseries}) upon substitution in eq.~(\ref{eq:sym2}).
\\

To simplify the presentation, we will consider only interstitial dislocation loops forming the invisible cluster population. This assumption is justified by the fact that the energy gain associated with the formation of a vacancy cluster is smaller than that of an interstitial cluster \cite{Gilbert2008,Alexander2016}, and in general a vacancy cluster has to be of appreciable size to remain stable at a finite temperature \cite{Dudarev2004}.\\
Also, in general $c_\infty$ should depend on time, following adiabatically the evolution of all the clusters in the system.
However, in a real finite system $c_\infty$ is governed at equilibrium by the surface energy and geometry of its external boundaries (free surfaces, grain boundaries), and can thus be considered constant in time for practical applications, provided that the morphology of external boundaries does not change appreciably during the evolution.

\subsection{Representing the invisible dislocation loops by a mean field fully coupled to the visible clusters}
Consider $N$ cavities and $n$ interstitial prismatic loops in an infinite medium. 
The cavities are sufficiently large and they are visible experimentally, but the prismatic loops are too small to be detected. In the following the evolution of the cavities will be considered explicitly as discrete objects, but the evolution of the loops will be treated through a mean-field. The evolution of the cavities and the mean field will be fully coupled. \\

Let $c_b(\mathbf{x})$ be the vacancy concentration field obtained by considering only the cavities. It is given by the equation:
\begin{equation}
D_v \nabla^2 c_b(\mathbf{x})=\Omega J_n (\mathbf{x}) \sum_{i=1}^N \delta \left[\Sigma_i (t) \right],
\label{eq:simple}
\end{equation}
where $\Sigma_i (t)$ denotes the surface of the $i$-th cavity at time $t$, $J_n$ is the vacancy current normal to a cavity surface, where the normal direction is considered pointing from the bulk towards the center of the cavity,  i.e. $J_n(\mathbf{x}) > 0$ if vacancies are entering the cavity at $\mathbf{x}$. $\delta(\Sigma)$ represents a delta function evaluated on a surface, defined as:
\begin{equation}
\int_V \varphi(\mathbf{x}) \delta(\Sigma) dV=\int_\Sigma \varphi(\mathbf{x}) dS,
\end{equation}
where $\varphi(\mathbf{x})$ is an arbitrary trial function and $V$ is an arbitrary volume containing the surface $\Sigma$.\\

Consider the invisible interstitial loops. Let $\mathbf{x}_i$ and $R_i$ denote respectively the center and the radius of the $i$th loop. To simplify the notation and make clear the parametric dependencies of the various functions, we define the following sets:

\begin{equation}
\begin{split}
\mathcal{X}&=\left\lbrace\mathbf{x}_i : i=1,...,n \right\rbrace,\\
\mathcal{R}&=\left\lbrace R_i : i=1,...,n \right\rbrace,\\
\mathcal{X}_{[i]}&=\left\lbrace \mathbf{x}_j : j=1,...,n ; \: j \neq i \right\rbrace,\\
\mathcal{R}_{[i]}&=\left\lbrace R_j :  j=1,...,n ; \: j \neq i \right\rbrace
\end{split}
\end{equation}
and we introduce the shorthand notations:
\begin{equation}
d\mathcal{X}=\prod_{i=1}^{n} d \mathbf{x}_i, \quad
d\mathcal{R}=\prod_{i=1}^{n} d R_i, \quad
d\mathcal{X}_{[i]}=\prod_{j \neq i}^{n} d \mathbf{x}_j, \quad
d\mathcal{R}_{[i]}=\prod_{j \neq i}^{n} d R_i.
\end{equation}
The total vacancy field obtained by also considering the effect of the $n$ dislocation loops can then be expressed as \cite{Edwards1958}:
\begin{equation}
c(\mathbf{x}; \mathcal{X},\mathcal{R})=c_b(\mathbf{x})-\sum_{i=1}^n \frac{d \mathcal{V}_i}{dt} G(|\mathbf{x}-\mathbf{x}_i|)=c_b(\mathbf{x})+\sum_{i=1}^n \xi_\Delta(R_i) G(|\mathbf{x}-\mathbf{x}_i|) \left[c_{[i]}(\mathbf{x}_i; \mathcal{X},\mathcal{R})-c_\Delta(R_i) \right],
\label{eq:micro}
\end{equation}
where $\xi_{\Delta}(R)=4\pi^2 R D_v/\ln(8R/r_d)$. The vacancy field obtained by omitting the contribution arising from the $i$th loop is $c_{[i]}(\mathbf{x};\mathcal{X},\mathcal{R})$, and it is defined by the self-consistent condition:
\begin{equation}
c_{[i]}(\mathbf{x};\mathcal{X},\mathcal{R})=c_b(\mathbf{x})+\sum_{j \neq i}^n \xi_\Delta(R_j) G(|\mathbf{x}-\mathbf{x}_j|) \left[c_{[j]}(\mathbf{x}_j; \mathcal{X},\mathcal{R})-c_\Delta(R_j) \right]
\label{eq:selfcons}.
\end{equation}
We define the probability of finding the $n$ dislocation loops in the $n$ volume elements  $(\mathbf{x}_1+d\mathbf{x}_1,...,\mathbf{x}_n+d\mathbf{x}_n)$, and with loop radii in the ranges $(R_1+dR_1,...,R_n+dR_n)$, as:
\begin{equation}
p(\mathcal{X},\mathcal{R}) d\mathcal{X} d\mathcal{R}.
\label{eq:sym5}
\end{equation}
The average of eq.~(\ref{eq:micro}) with respect to the positions and radii of each loop can then be expressed as:
\begin{equation}
\begin{split}
\bar{c}(\mathbf{x})&=\int c(\mathbf{x}; \mathcal{X},\mathcal{R}) p(\mathcal{X},\mathcal{R}) d\mathcal{X}d\mathcal{R}\\
&=c_b(\mathbf{x})+\sum_{i=1}^n \int \left[S^i_\text{nl}(\mathbf{x}; \mathcal{X},\mathcal{R} )-S_\text{l}(\mathbf{x}; \mathbf{x}_i,R_i) \right] p(\mathcal{X},\mathcal{R}) d\mathcal{X}d\mathcal{R},
\end{split}
\label{eq:sym6}
\end{equation}
where $S^i_\text{nl}(\mathbf{x}; \mathcal{X},\mathcal{R} )$ and $S_\text{l}(\mathbf{x}; \mathbf{x}_i,R_i) $ are respectively the non-local and local contributions of the $i$th loop to the total concentration field, defined as:
\begin{equation}
\begin{split}
S^i_\text{nl}(\mathbf{x}; \mathcal{X},\mathcal{R} )&=\xi_\Delta(R_i)  G(|\mathbf{x}-\mathbf{x}_i|) c_{[i]}(\mathbf{x}_i; \mathcal{X},\mathcal{R}),\\
S_\text{l}(\mathbf{x}; \mathbf{x}_i,R_i)&=\xi_\Delta(R_i)  G(|\mathbf{x}-\mathbf{x}_i|) c_\Delta(R_i).
\end{split}
\end{equation}
$S_\text{l}$ is the perturbation to the vacancy field in the neighborhood of the $i$'th loop arising from the condition of local thermodynamic equilibrium. The non-local term $S^i_\text{nl}$ describes the contribution to the  vacancy field near loop $i$ from all the other loops in the system.\\

By introducing the one-loop probability density function:
\begin{equation}
p_1(\mathbf{x}_i,R_i)=\int p(\mathcal{X},\mathcal{R}) d\mathcal{X}_{[i]}d\mathcal{R}_{[i]},
\label{eq:sym7}
\end{equation}
we can express the average of the local contribution as:
\begin{equation}
\bar{S}_\text{l}(\mathbf{x})=\int  S_\text{l}(\mathbf{x}; \mathbf{x}_i,R_i)  p_1(\mathbf{x}_i,R_i)  d\mathbf{x}_i dR_i,
\end{equation}
which can be readily calculated without any knowledge of correlations in the positions or radii of different loops, requiring only the single loop spatial and size distribution $p_1(\mathbf{x},R)$.\\

We now average the non-local contribution. Let us define the conditional probability density $p(\mathbf{x}_i,R_i | \mathcal{X}_{[i]},\mathcal{R}_{[i]})$ by the relation:
\begin{equation}
p_1(\mathbf{x}_i,R_i)\, p(\mathbf{x}_i,R_i | \mathcal{X}_{[i]},\mathcal{R}_{[i]})=p( \mathcal{X},\mathcal{R})
\end{equation}
and the effective averaged vacancy field experienced by the $i$th loop as:
\begin{equation}
 \bar{c}^\text{\,eff}_{[i]}(\mathbf{x}; \mathbf{x}_i,R_i)=\int c_{[i]}(\mathbf{x}; \mathcal{X},\mathcal{R})  p(\mathbf{x}_i,R_i | \mathcal{X}_{[i]},\mathcal{R}_{[i]})  d\mathcal{X}_{[i]}d\mathcal{R}_{[i]}
\label{eq:ceff}.
\end{equation}
By employing these definitions, we may write the average of the non-local contribution as:
\begin{equation}
\bar{S}_\text{nl}(\mathbf{x})=\int \xi_\Delta(R_i) G(|\mathbf{x}-\mathbf{x}_i|) \bar{c}^\text{\,eff}_{[i]}(\mathbf{x}; \mathbf{x}_i,R_i) \: p_1(\mathbf{x}_i,R_i)  d\mathbf{x}_i dR_i.
\end{equation}
In summary, the governing equation for the vacancy concentration, averaged over all possible positions and radii of the invisible interstitial loops, is given by:
\begin{equation}
\bar{c}(\mathbf{x})=c_b(\mathbf{x})+n \int \xi_\Delta(R_i) G(|\mathbf{x}-\mathbf{x}_i|)  \: p(\mathbf{x}_i;R_i) \left[ \bar{c}^\text{\,eff}_{[i]}(\mathbf{x}; \mathbf{x}_i,R_i)-c_\Delta(R_i) \right]  d\mathbf{x}_i dR_i.
\label{eq:MF2}
\end{equation}
The integral in this equation is the mean field of the invisible loops in which the visible clusters sit.  Information about correlations between the positions and sizes of the invisible loops is contained in the effective field $\bar{c}^\text{eff}_{[i]}$.

\subsection{The simplest approximation of the field $\bar{c}^\text{eff}_{[i]}$ of eq.~(\ref{eq:ceff}) and screening}

A formal expansion of the effective field defined in eq.~(\ref{eq:ceff}) is derived in appendix \ref{app:pert}, which provides the theoretical foundation to treat correlations between the positions of the invisible loops to arbitrary degrees of accuracy. In this section we consider the simplest approximation, which is to neglect all these correlations.  We may then characterize the loops entirely through the size distribution function $f(R)$, which is related to the single-loop probability density function as follows:
\begin{equation}
 p_1(\mathbf{x},R)=n^{-1} f(R).
 \label{eq:sym9}
\end{equation}
$f(R)dR$ is the number of dislocation loops per unit volume with radii between $R$ and $R+dR$.\\

Provided the concentration of invisible loops is sufficiently large, it is reasonable to assume that the average vacancy field seen by dislocation loops is equal to the configuration averaged field, i.e. $\bar{c}(\mathbf{x}) \approx \bar{c}^\text{\,eff}_{[i]}$. In these approximations eq.~(\ref{eq:MF2}) for the configuration-averaged concentration takes on the form:
\begin{equation}
\bar{c}(\mathbf{x})=c_b(\mathbf{x})+ \int dV' G(|\mathbf{x}-\mathbf{x}'|) \int_b^\infty \xi_\Delta(R) f(R)   \left[ \bar{c}(\mathbf{x}')-c_\Delta(R) \right]  dR,
\label{eq:MF3}
\end{equation}
where the Burgers vector $b$ acts as a lower bound on the possible dislocation loop size.
We now define the averages:
\begin{equation}
\begin{split}
&\overline{\xi_\Delta}=\int_b^\infty dR f(R) \xi_\Delta(R),\\
&\overline{\xi_\Delta c_\Delta}=\int_b^\infty dR f(R) \xi_\Delta(R) c_\Delta(R),
\end{split}
\label{eq:AandB}
\end{equation}
where both have the dimensions of inverse time.
Eq.~(\ref{eq:MF3}) then becomes:
\begin{equation}
\bar{c}(\mathbf{x})=c_b(\mathbf{x})+  \int dV' G(|\mathbf{x}-\mathbf{x}'|) \left[ \overline{\xi_\Delta}\bar{c}(\mathbf{x}')-\overline{\xi_\Delta c_\Delta} \right].
\label{eq:MF4}
\end{equation}
We apply the operator $D_v \nabla^2$ to both sides of eq.~(\ref{eq:MF4}), obtaining:
\begin{equation}
D_v \nabla^2 \bar{c}(\mathbf{x})=D_v \nabla^2 c_b (\mathbf{x})+\int dV' \delta(\mathbf{x}-\mathbf{x'}) \left[\overline{\xi_\Delta} \bar{c}(\mathbf{x}')-\overline{\xi_\Delta c_\Delta} \right],
\label{eq:screened}
\end{equation}
which, along with eq.~(\ref{eq:simple}), can be reorganised as:
\begin{equation}
(D_v \nabla^2-\overline{\xi_\Delta}) \bar{c}(\mathbf{x})=\Omega J_n (\mathbf{x}) \sum_{i=1}^N \delta \left[\Sigma_i (t) \right]-\overline{\xi_\Delta c_\Delta}.
\label{eq:Debye}
\end{equation}
Eq.~(\ref{eq:Debye}) has  the mathematical structure of an inhomogeneous Debye-H{\"u}ckel equation.\\
The Yukawa Green's function, $G_Y(|\mathbf{x}-\mathbf{x}'|;\overline{\xi_\Delta})$, is defined by the equation:
\begin{equation}
(D_v \nabla^2-\overline{\xi_\Delta}) G_Y(|\mathbf{x}-\mathbf{x}'|;\overline{\xi_\Delta})=\delta(\mathbf{x}-\mathbf{x}'),
\label{eq:sym8}
\end{equation}
for which the solution is:
\begin{equation}
G_Y(|\mathbf{x}-\mathbf{x}'|;\overline{\xi_\Delta})=-\,\frac{\exp\left\{-\sqrt{\overline{\xi_\Delta}/D_v}\,\left|\mathbf{x}-\mathbf{x'}\right|\right\}}{4 \pi D_v \left|\mathbf{x}-\mathbf{x'}\right|}.
\label{eq:YukawaGreen}
\end{equation}
Thus, the diffusive interaction between cavities in eq.~(\ref{eq:screened}) is \emph{screened} by the mean field of the dislocation loops.  The screening length, $\sqrt{D_v/\xi_\Delta}$, limits the range of direct diffusional interaction between cavities. In this continuum treatment, once the effect of the loop population is replaced by a mean field, the medium between the cavities becomes a net adsorber for vacancies. This reflects the reality at the atomic scale where the mean free path of propagating vacancies is reduced by the presence of distributed sinks and sources. An analogous picture was obtained in theories of Ostwald ripening involving \emph{finite} volume fractions of second phase particles \cite{Marqusee1984,Fradkov1996,Wang2004,Wang2008}, where similar screened diffusion-mediated interactions were found. \\

In a dilute configuration of cavities each of them can be approximated as a point source with an effective sink strength given by $Q_i= 4 \pi R^2_i \dot{R}_i$.  A formal solution of eq.~(\ref{eq:Debye}) is then:
\begin{equation}
\bar{c}(\mathbf{x})=\sum_{i=1}^N Q_i
G_Y(|\mathbf{x}-\mathbf{x}_i|;\overline{\xi_\Delta})-\overline{\xi_\Delta c_\Delta} \int dV'
G_Y(|\mathbf{x}-\mathbf{x}_i|;\overline{\xi_\Delta})+c_\infty,
\end{equation}
where $\mathbf{x}_i$ denotes the center of the $i-$th cavity. We note that:
\begin{equation}
- \overline{\xi_\Delta c_\Delta} \int dV'
G_Y(|\mathbf{x}-\mathbf{x}'|;\overline{\xi_\Delta})=\frac{\overline{\xi_\Delta c_\Delta}}{\overline{\xi_\Delta}}=\frac{\int_{b}^\infty dR \, \xi_\Delta(R) f(R) c_\Delta(R)}{\int_{b}^\infty dR \, \xi_\Delta(R) f(R) }=\left\langle c_\Delta \right\rangle,
\end{equation}
where the average $\langle...\rangle$ is defined with respect to the weighted distribution $\tilde{f}(R)=\xi_\Delta(R) f(R)$. \\
Thus, we arrive at the following self-consistent relation between the rates of change of the cavity radii $\dot{R}_i$, the vacancy field and the distribution of the interstitial dislocation loops $f(R)$:
\begin{equation}
\bar{c}(\mathbf{x})=-\sum_{i=1}^N R^2_i \dot{R}_i  \frac{\exp\left[-\sqrt{\overline{\xi_\Delta}/D_v}|\mathbf{x}-\mathbf{x}_i|\right]}{{ D_v}  |\mathbf{x}-\mathbf{x}_i|}+c_\infty+\left\langle c_\Delta \right\rangle
\label{eq:MFsinglefinal}.
\end{equation}
To solve this equation the vacancy concentration has to be evaluated at the position of each cavity, satisfying the boundary condition of local thermodynamic equilibrium, i.e.:
\begin{equation}
c_\Sigma(R_i)=-\sum_{\substack{j=1 \\ j\neq i}}^N R^2_j \dot{R}_j  \frac{\exp\left[-\sqrt{\overline{\xi_\Delta}/D_v}|\mathbf{x}_i-\mathbf{x}_j|\right]}{{ D_v}  |\mathbf{x}_i-\mathbf{x}_j|}-\frac{R_i \dot{R}_i}{D_v}+c_\infty+\left\langle c_\Delta \right\rangle, \qquad i=1,..,N,
\label{eq:system}
\end{equation}
where $-\frac{R_i \dot{R}_i}{D_v}$ is a finite size correction accounting for the self-diffusional interaction of a cavity with itself, representing vacancies propagating between different points on the surface of the same cavity. As a comparison, we recall the analogous of the above system of equations for a system where the $n$ dislocation loops are explicitly considered as discrete objects, given by the set of $N+n$ equations \cite{Rovelli2017}:
\begin{equation}
\begin{split}
c_{\Sigma}(R_i)&=-\sum_{\substack{j=1 \\ j\neq i}}^{N}\frac{R_j^2 \dot{R}_j}{D_v|\mathbf{x}_i-\mathbf{x}_j|}-\frac{R_i \dot{R}_i}{D_v}+\sum_{k=N+1}^{N+n}\frac{b_eR_k \dot{R}_k}{2D_v|\mathbf{x}_i-\mathbf{x}_k|}+c_\infty, \qquad i=1,..,N\\
c_{\Delta}(R_i)&=-\sum_{j=1}^{N}\frac{R_j^2 \dot{R}_j}{D_v|\mathbf{x}_i-\mathbf{x}_j|}+\sum_{\substack{k=N+1 \\ k\neq i}}^{N+n}\frac{b_eR_k \dot{R}_k}{2 D_v|\mathbf{x}_i-\mathbf{x}_k|}-\frac{b_e\ln(\frac{8R_i}{r_d}) \dot{R}_i}{2\pi D_v }+c_\infty, \qquad i=N+1,..,N+n,
\end{split}
\end{equation}
where the indices from 1 to $N$ denote cavities and from $N+1$ to $N+n$ denote dislocation loops. 
By inverting the linear system defined by eq.~(\ref{eq:system}), the set of $\dot{R}_i$ can be then calculated provided $f(R)$ is known.\\

The screening length changes with time because it depends on the evolution of the distribution of the interstitial loop sizes. Let the time-dependent size distribution function be $F(R,t)$. A continuity equation in particle-size space for the distribution $F(R,t)$ can be written as:
\begin{equation}
\frac{\partial F(R,t)}{\partial t}+\frac{\partial \left[ F(R,t) v_\text{il}(R,t)\right]}{\partial R}=\dot{n}(R,t),
\label{eq:continuity}
\end{equation}
where $\dot{n}(R,t)$ is the net rate at which loops of radius $R$ are created, as a result of coalescence of existing loops and nucleation of new loops. The growth rate $v_\text{il}(R,t)$ of a loop of radius $R$ at time $t$ is given by:
\begin{equation}
v_\text{il}(R,t)=-\left[c_\text{avg}(t)-c_\Delta(R)\right] \frac{2 \pi D_v}{b_e \ln (8 R/r_d)},
\label{eq:cavgg}
\end{equation}
where $c_\text{avg}(t)$ is the spatially averaged vacancy concentration. Differentiating with respect to $R$:
\begin{equation}
\frac{\partial v_\text{il}(R,t)}{\partial R}=\frac{2 \pi D_v}{b_e \ln(8R/r_d)} \left\lbrace \frac{\left[c_\text{avg}(t)-c_\Delta(R)\right] }{R \ln(8 R/r_d)}+\frac{ \mu b_e \Omega c_\Delta(R)}{4 \pi (1-\nu) R^2 k_B T} \left[\ln\left(\frac{8R}{r_d}\right)-2 \right]\right\rbrace.
\end{equation}
In the homogeneous mean field treatment, the term $c_\text{avg}(t)$ can be obtained by averaging eq.~(\ref{eq:MFsinglefinal}) with respect to $\mathbf{x}$:
\begin{equation}
\begin{split}
c_\text{avg}(t)&=- \lim_{V \rightarrow \infty} \left\lbrace \sum_{i=1}^N \frac{R^2_i(t) \dot{R}_i(t)}{V D_v}  \int_V d^3 x \frac{\exp\left[-\sqrt{\overline{\xi_\Delta}/D_v}|\mathbf{x}-\mathbf{x}_i|\right]}{ |\mathbf{x}-\mathbf{x}_i|} \right\rbrace +c_\infty+\left\langle c_\Delta \right\rangle\\
&=\left\langle c_\Delta \right\rangle+c_\infty- \frac{4 \pi \rho_c}{\overline{\xi_\Delta}}\left( \frac{1}{N} \sum_{i=1}^N  R^2_i(t) \dot{R}_i(t) \right)=\left\langle c_\Delta \right\rangle+c_\infty- \frac{\overline{\dot{V}}_c(t)  \rho_c}{\overline{\xi_\Delta}},
\label{eq:coupling}
\end{split}
\end{equation}
where $\rho_c$ denotes the number density of cavities and $\overline{\dot{V}}_c(t)$ denotes the average rate of change of the volume of cavities, at time $t$. It is clear that in the limiting case of $\overline{\xi_\Delta}=0$ the above expression diverges, and the integral of the Green's function has to be performed up to a suitable cutoff distance.\\

\begin{center}
\begin{figure}[h!]
\centering
\includegraphics[width=0.8\columnwidth]{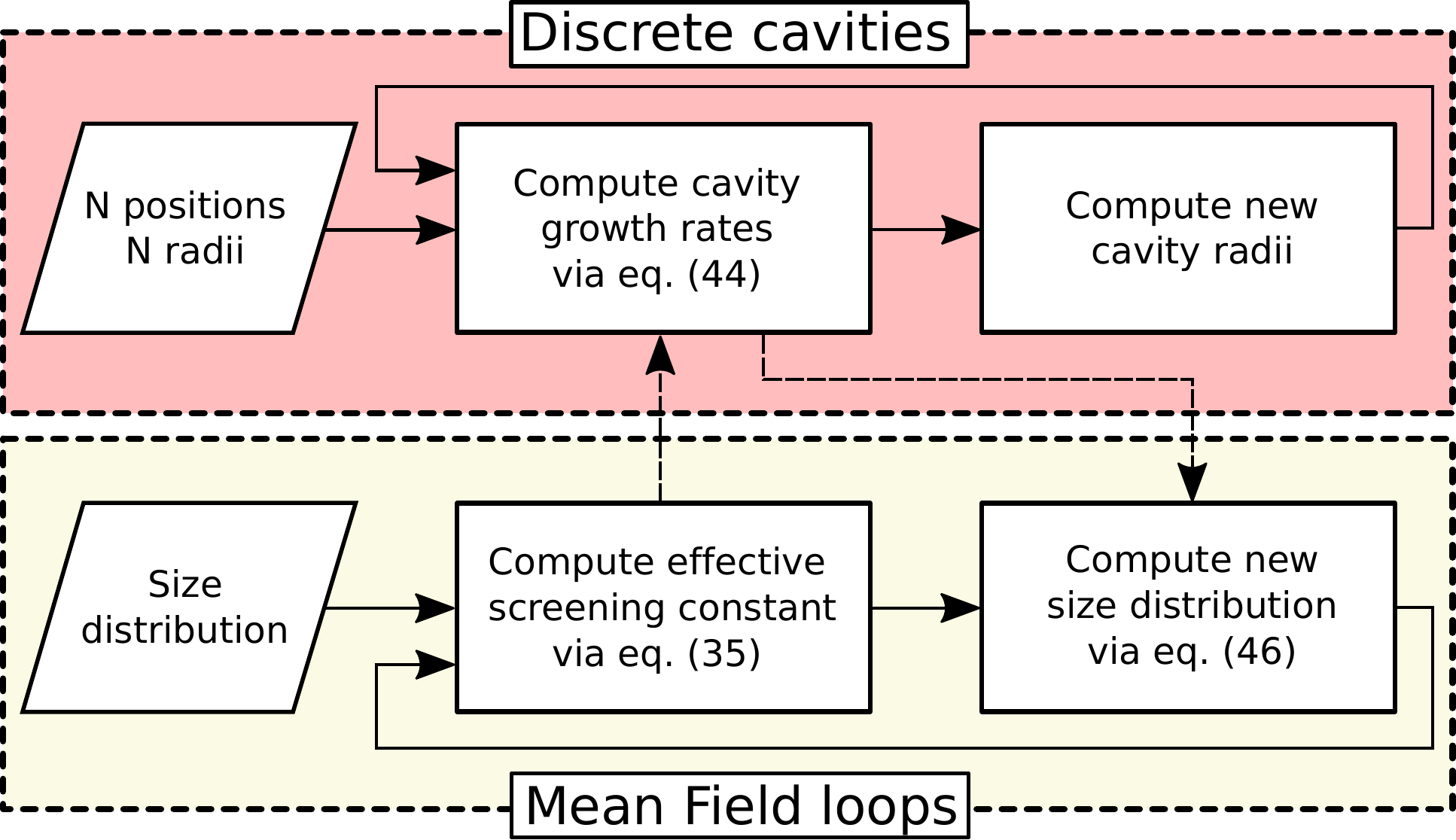}
\caption{Flowchart summarising a scheme to compute the evolution of the coupled mean field of interstitial loops and discrete cavities.}
\label{fig:flowchart}
\end{figure}
\end{center}

In summary, while the mean field of interstitial loops screens the diffusive interaction between cavities, the rate of growth of cavities also determines the evolution of the mean field: the cavities and the mean field are coupled. In Fig. \ref{fig:flowchart} we show a flowchart that summarises a  scheme to compute the system evolution, highlighting the couplings between the cavities and the mean field.
\subsection{Numerical estimate of the screening length}
In this section we calculate the initial value, at $t=0$, of the screening length of the effective interaction between cavities in the presence of a mean field of interstitial dislocation loops.
It is helpful to introduce the scaled distribution function:
\begin{equation}
A \phi\left(\frac{R}{b},t\right)=F(R,t),
\label{eq:sym10}
\end{equation}
where $A$ is a normalisation factor with the dimensions of length$^{-4}$, defined with respect to the size distribution at $t=0$ as:
\begin{equation}
A=\rho(0)  \left(b \int_1^{\infty} du \:\phi(u,0) \right)^{-1},
\label{eq:sym12}
\end{equation}
where $\rho(t)$ is the number density of the mean field clusters at time t.
The screening length $\Delta l = \sqrt{D_v/\overline{\xi_\Delta}}$, at time $t=0$, is given by:
\begin{equation}
\begin{split}
&\Delta l(0) =  \left( \int_{b}^\infty dR \: \xi_\Delta(R) F(R,t)/D_v  \right)^{-\frac{1}{2}}=\frac{1}{2 \pi b\sqrt{A}} \left( \int_{1}^\infty du \: u \: \phi(u,0)/\ln(8 bu/r_d) \right)^{-\frac{1}{2}}.
\end{split}
\end{equation}
\begin{center}
\begin{figure}[h!]
\centering
\includegraphics[width=0.7\columnwidth]{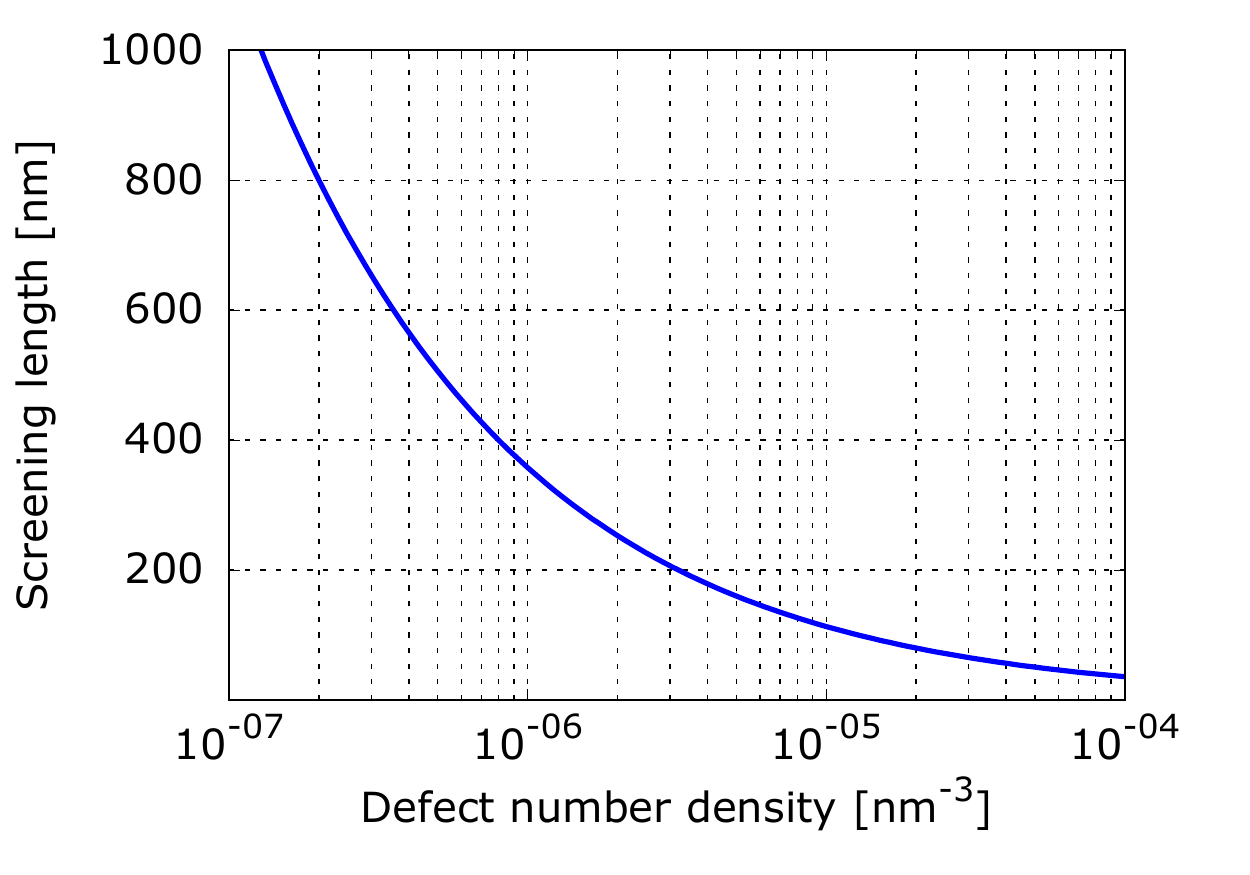}
\caption{Plot of the prefactor $(2\pi)^{-1} (\rho(0) b)^{-1/2}$ of eq.~(\ref{eq:scrlength}) as a function of loop number density $\rho$. The value of the Burgers vector $b$ is $0.27$ nm, which is representative of tungsten.}
\label{fig:attenuation}
\end{figure}
\end{center}
For the initial loop size distribution we assume a power law with an exponentially decaying cutoff at small loop radii, i.e.:
\begin{equation}
\phi(u,0)=e^{-c_1/u} u^{-c_2},
\end{equation}
where $c_1$ and $c_2$ are constants.
To ensure the existence of the integrals defining $\overline{\xi_\Delta}$ and $\overline{\xi_\Delta c}$ we require $c_2 > 2$. \\
The screening length can then be recast as:
\begin{equation}
\Delta l(0) \approx \frac{1}{ 2 \pi \sqrt{\rho(0) b}} \sqrt{\frac{ \int_1^{\infty} du \: \exp(-c_1/u)u^{-c_2}}{\int_1^{\infty} du \: \exp(-c_1/u) u^{-c_2+1}/\ln(8ub/r_d)}} \equiv \frac{\Pi(c_1,c_2)}{ 2 \pi \sqrt{\rho(0) b}},
\label{eq:scrlength}
\end{equation}
where we have separated the dependence on the loop number density from the shape of the size distribution contained in $\Pi(c_1,c_2)$.  For realistic loop densities of $10^{-6}-10^{-4}$ nm$^{-3}$ the factor $\left(2 \pi \sqrt{\rho(0) b}\right)  ^{-1}$ is of the order of $10^2-10^3$ nm (see Fig. \ref{fig:attenuation}), while $\Pi(c_1,c_2)$ is of the order of unity for reasonable values of $c_1$ and $c_2$. In particular, with $0.1<c1<1$ and $2<c2<4$, we have $0.68 \lesssim \Pi \lesssim 1.3$.
At these loop densities the screening length is comparable to the separation of the loops, which may be substantially smaller than the separation of cavities. Therefore the screening of the diffusive interaction between cavities by small dislocation loops plays a significant role in the evolution of the distribution of cavity sizes during an anneal.

\section{Extension to thin film geometry}
\label{sec:TF}
Trasmission electron microscopy is providing accurate data on the distributions of radiation defects \cite{Ferroni2015,Yi2013}. Since TEM samples are always thin films we provide in this section an extension of the theory to infinitely extended thin films.\\
\\
Consider a region $V \in \mathbb{R}^3$ infinitely extended in the $\hat{x}$ and $\hat{y}$ directions and of thickness $H$ in the $\hat{z}$ direction, i.e. $V=\left\lbrace (x,y,z) \in \mathbb{R}^3 : x \in \mathbb{R},y\in \mathbb{R},-H/2<z<H/2   \right\rbrace$. We define a $L \times L \times H$  cell $\Gamma(0)=\left\lbrace \mathbf{x}=x_1\mathbf{a}_1+x_2\mathbf{a}_2+x_3\mathbf{a}_3:-1/2<x_\alpha<1/2, \alpha=1,2,3 \right\rbrace$, where $\mathbf{a}_1=(L,0,0)$, $\mathbf{a}_2=(0,L,0)$, and $\mathbf{a}_3=(0,0,H)$ are the three basis vectors defining the cell. The cell $\Gamma(0)$ contains $N$ spherical cavities and a much larger number of smaller circular prismatic dislocation loops, with size distribution function $f(R)$, normalised to the loop number density. The primitive cell is infinitely replicated in the $\hat{x}$ and $\hat{y}$ directions.\\

Let $\Sigma_i$ denote the surface of the $i-$th cavity. We impose the following boundary conditions on the vacancy concentration field $c(\mathbf{x})$:
\begin{equation}
\begin{split}
c(x,y,-H/2)&=c(x,y,H/2)=c_S,\\
c(\mathbf{x}\in \Sigma_i)&=c_{\Sigma_i} \qquad i=1,...,N.
\end{split}
\label{eq:BCTF}
\end{equation}
\begin{center}
\begin{figure}[h!]
\centering
\includegraphics[width=0.8\columnwidth]{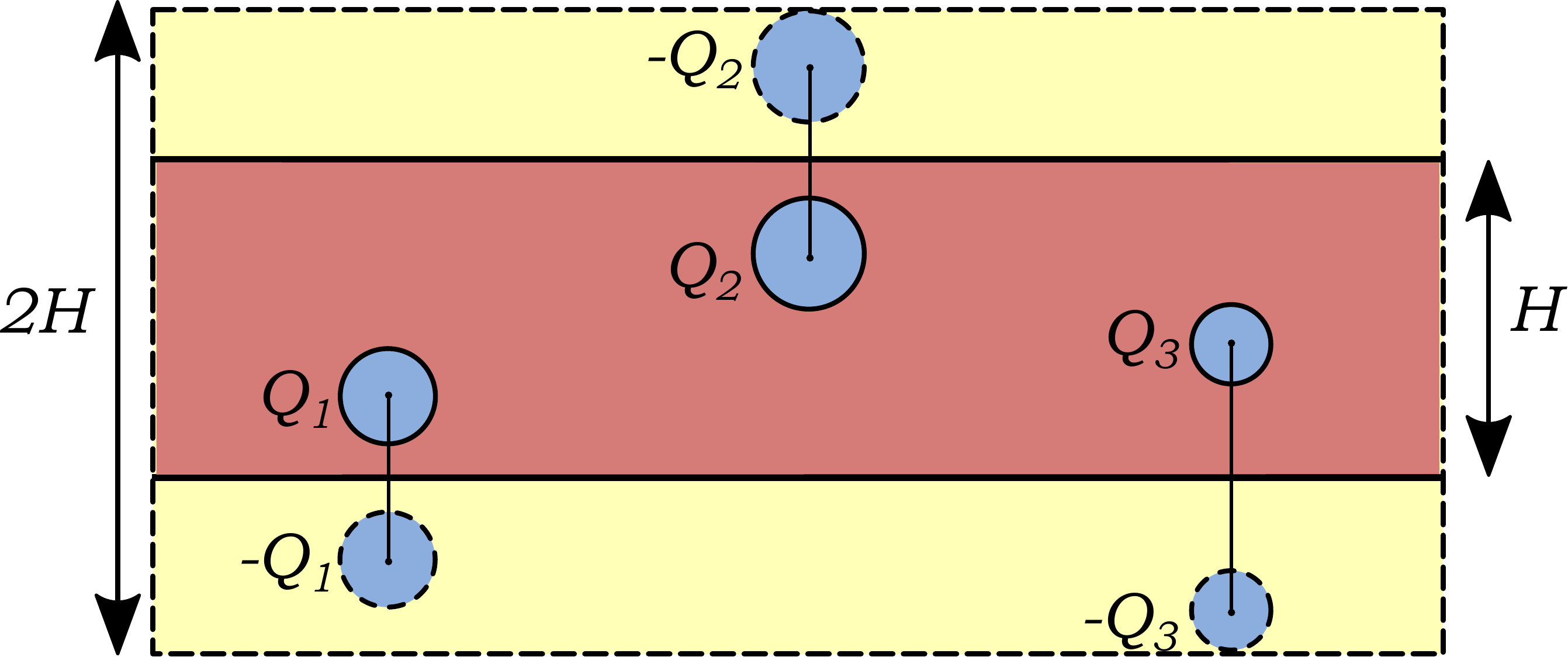}
\caption{2D sketch of the primitive cell structure for the thin film configuration. The red shaded region is  a repeat cell of the thin film. Image clusters (portrayed with dashed boundaries) of opposite effective charge are constructed in the yellow shaded regions and the original primitive cell is extended accordingly. Periodic boundary conditions are imposed on the dashed boundaries and in the directions perpendicular to the drawing. The size of cavities with respect to the simulation cell is exaggerated for clarity.}
\label{fig:Ewald}
\end{figure}
\end{center}
Using the terminology of electrostatics, each cavity can be associated with an effective charge $Q$, related to  rate of growth of the cavity volume $\mathcal{V}$:
\begin{equation}
Q_i=-\frac{d\mathcal{V}_i}{dt}=4 \pi R_i^2 \frac{dR_i}{dt},
\end{equation}
where $R_i$ is the radius of the $i$th cavity. As noted in the previous section, to first order the effect of the mean field of dislocation loops is to introduce a screened interaction with screening length $\sqrt{D_v/\overline{\xi_\Delta}(t)}$, and to displace the resultant concentration field by an amount

\[
\left\langle c_\Delta \right\rangle(t)=\frac{\int dR \xi_\Delta(R) F(R,t) c_\Delta(R)}{\int dR  \xi_\Delta(R) F(R,t)}.\\
\]

The boundary conditions at $c(z=\pm H/2)=c_S$ can be satisfied using the method of images, by introducing an infinite number of periodic images of the thin film in the $\hat{z}$ directions. The images are constructed as follows: we extend the primitive cell by $H/2$ in the positive and negative $\hat{z}$ directions; for each cavity with  charge $Q$ in the original primitive cell located at $(x,y,z)$ we add a virtual cavity of charge $-Q$ at $(x,y,\text{sgn}(z)H-z)$. A new primitive cell may then be defined containing $2N$ interacting cavities: $\Gamma(0)=\big\lbrace \mathbf{x}=x_1\mathbf{a}_1+x_2\mathbf{a}_2+x_3\mathbf{a}_3:-1/2<x_\alpha<1/2, \alpha=1,2,3 \big\rbrace$, with primitive vectors $\mathbf{a}_1=(L,0,0)$, $\mathbf{a}_2=(0,L,0)$, and $\mathbf{a}_3=(0,0,2H)$. A 2D sketch of such a construction is presented in Fig. \ref{fig:Ewald}.
This new primitive cell is charge neutral by construction, and it is repeated infinitely many times along $z$.\\

The self consistent system of equations that determines the set of $\left\lbrace Q_i \right\rbrace_{i=1,...,N}$ at time $t$ is therefore given by:
\begin{equation}
\begin{split}
c_{\Sigma_i}=\sum_{\substack{j=1\\j\neq i}}^{N} Q_j \left[ \sideset{}{'}\sum_{\mathbf{m}}  G_Y(|\mathbf{x}_i-\mathbf{m}+\mathbf{x}_j|;\overline{\xi_\Delta})-\sum_{\mathbf{m}} G_Y(|\mathbf{x}_i-\mathbf{m}+{\mathbf{\bar{x}}}_j|;\overline{\xi_\Delta}) \right]-&\frac{Q_i}{4 \pi D_v R_i}+c_S+\left\langle c_\Delta \right\rangle,\\
&i=1,..,N,
\end{split}
\label{eq:NoEwald}
\end{equation}
where $\mathbf{m}=m_1\mathbf{a}_1+m_2\mathbf{a}_2+m_3\mathbf{a}_3$, $m_\alpha \in \mathbb{Z}$, $\alpha=1,2,3$, $\mathbf{\bar{x}}_j=(x_j,y_j,\text{sgn}(z_j)H-z_j)$ and primed sums denote that the term $\mathbf{m}=\mathbf{0}$ is not included when $\mathbf{x}_i=\mathbf{x}_j$.\\

The sums over $\mathbf{m}$  are convergent for finite values of the screening length.\\
However, it is possible that the mean field loops disappear during an anneal, leading to an infinite screening length and therefore to a $1/r$ diffusive interaction between the cavities. 
In that case, the sums over $\mathbf{m}$ are only \emph{conditionally} convergent.\\
Since the new primitive cell is charge neutral, we may use a variant of the Ewald summation technique to derive an absolutely convergent series for all screening lengths. 
This can be achieved by introducing the following kernel:
\begin{equation}
\begin{split}
&K(\mathbf{x}_i,\mathbf{x}_j;\epsilon)=-\sum_\mathbf{k} \frac{\exp\left[-(k^2+\epsilon^2)/4\beta^2 \right]}{ VD_v( k^2+\epsilon^2)}e^{i\mathbf{k}\cdot\mathbf{x}_i}\left[e^{-i \mathbf{k}\cdot \mathbf{x}_j}-e^{-i \mathbf{k}\cdot \mathbf{\bar{x}}_j} \right]+\\&-\frac{1}{8 \pi D_v} \Bigg\lbrace \sideset{}{'}\sum_\mathbf{m} \frac{1}{|\mathbf{x}_i-\mathbf{x}_j+\mathbf{m}|} \Bigg[ \text{erfc}\left(\beta |\mathbf{x}_i-\mathbf{x}_j+\mathbf{m}|+\frac{\epsilon}{2\beta} \right)e^{\epsilon|\mathbf{x}_i-\mathbf{x}_j+\mathbf{m}|}\\
&+\text{erfc}\left(\beta |\mathbf{x}_i-\mathbf{x}_j+\mathbf{m}|-\frac{\epsilon}{2\beta} \right)e^{-\epsilon|\mathbf{x}_i-\mathbf{x}_j+\mathbf{m}|}    \Bigg]\\
&-\sum_\mathbf{m} \frac{1}{|\mathbf{x}_i-\mathbf{\bar{x}}_j+\mathbf{m}|} \Bigg[ \text{erfc}\left(\beta |\mathbf{x}_i-\mathbf{\bar{x}}_j+\mathbf{m}|+\frac{\epsilon}{2\beta} \right)e^{\epsilon|\mathbf{x}_i-\mathbf{\bar{x}}_j+\mathbf{m}|}\\
&+\text{erfc}\left(\beta |\mathbf{x}_i-\mathbf{\bar{x}}_j+\mathbf{m}|-\frac{\epsilon}{2\beta} \right)e^{-\epsilon|\mathbf{x}_i-\mathbf{\bar{x}}_j+\mathbf{m}|}    \Bigg] \Bigg \rbrace+\frac{\delta_{ij}}{4 \pi D_v} \left[\frac{2 \beta e^{-\frac{\epsilon^2}{4 \beta^2}}}{\sqrt{\pi}} - \epsilon \: \text{erfc} \left(\frac{\epsilon}{2 \beta}\right)\right].
\end{split}
\label{eq:YukawaKernel}
\end{equation}
The set of self-consistent equations to be solved become:
\begin{equation}
\begin{split}
c_{\Sigma_i}=\sum_{\substack{j=1\\j\neq i}}^{N} Q_j K\left(\mathbf{x}_i,\mathbf{x}_j;\sqrt{\overline{\xi_\Delta}/D_v}\right)-\frac{Q_i}{4 \pi D_v R_i}+c_S+\left\langle c_\Delta \right\rangle, \qquad i=1,..,N,
\end{split}
\end{equation}
which is the analogue of eq.~(\ref{eq:system}) for the thin film configuration. A detailed derivation of  $K(\mathbf{x}_i,\mathbf{x}_j;\epsilon)$ is given in appendix \ref{app:TF}.\\ We note that the numerical scheme to compute the temporal evolution of cavities remains the same as that presented at the end of sec.3.\\
By taking the limit in the kernel of an infinite screening length we obtain:
\begin{equation}
\begin{split}
&\lim_{\epsilon \rightarrow 0} K(\mathbf{x}_i,\mathbf{x}_j,\epsilon)=-\sum_\mathbf{k} \frac{\exp\left[-k^2/4\beta^2 \right]}{ VD_v k^2}e^{i\mathbf{k}\cdot\mathbf{x}_i}\left[e^{-i \mathbf{k}\cdot \mathbf{x}_j}-e^{-i \mathbf{k}\cdot \mathbf{\bar{x}}_j} \right]\\
&-\frac{1}{4 \pi D_v} \left[ \sideset{}{'}\sum_\mathbf{m} \frac{\text{erfc}\left(\beta |\mathbf{x}_i-\mathbf{x}_j+\mathbf{m}| \right)}{|\mathbf{x}_i-\mathbf{x}_j+\mathbf{m}|} -\sum_\mathbf{m} \frac{\text{erfc}\left(\beta |\mathbf{x}_i-\mathbf{\bar{x}}_j+\mathbf{m}| \right)}{|\mathbf{x}_i-\mathbf{\bar{x}}_j+\mathbf{m}|}    \right]+\delta_{ij} \frac{2 \beta}{4 \pi^{3/2} D_v},
\end{split}
\end{equation}
which is the usual expression for the Ewald sum for the unscreened Green's function $G(|\mathbf{x}-\mathbf{x}'|)=-\left(4 \pi D_v |\mathbf{x}-\mathbf{x}'|\right)^{-1}$. Thus the kernel $K(\mathbf{x}_i,\mathbf{x}_j;\epsilon)$ enables us to treat within the same set of equations the evolution of cavities, whatever the screening length.
\section{Full mean field description}
\label{sec:AllMF}
In this section we consider both dislocation loops and cavities as contributors to a mean field.  This may be useful whenever information on cluster distributions at all length scales is incomplete. Consider the vacancy field generated by a spherical cavity in an infinite homogeneous medium in which the vacancy concentration far from the cavity is $c_\infty$:
\begin{equation}
c(r)=c_\infty-\left[c_\infty-c_\Sigma(R)\right]\frac{R}{r}=c_\infty+\xi_\Sigma(R)G(r) \left[c_\infty-c_\Sigma(R)\right],
\end{equation}
where $\xi_\Sigma(R)=4 \pi D_v R $ and:
\begin{equation}
c_\Sigma(R)=c_0 \exp\left[\frac{2 \gamma \Omega}{R k_B T} \right].
\end{equation}
Consider the case of a spatially homogeneous, uncorrelated distribution of loops and cavities. We define the number density respectively of vacancy loops, interstitial loops and cavities, with radii in the range $(R+dR)$, as $f_\text{v}(R)dR$, $f_\text{i}(R)dR$ and $f_c(R)dR$.\\ Following a similar treatment to that used in sec.3, we can write a mean field equation analogous to that of eq.~(\ref{eq:MF3}):
\begin{equation}
\begin{split}
\bar{c}(\mathbf{x})&=c_\infty+ \int dV' G(|\mathbf{x}-\mathbf{x}'|)  \int_{b}^\infty dR \: \bigg\lbrace \: \xi_{\Delta}(R) f_\text{v}(R)  \left[\bar{c}(\mathbf{x}')-c^\text{v}_\Delta(R) \right]\\
&+\xi_{\Delta}(R) f_\text{i}(R) \left[\bar{c}(\mathbf{x}')-c^\text{i}_\Delta(R) \right]+\xi_{\Sigma}(R) f_c(R)   \left[\bar{c}(\mathbf{x}')-c_\Sigma(R) \right] \bigg\rbrace,
\label{eq:everycluster}
\end{split}
\end{equation}
where the boundary condition $c^\text{v}_\Delta(R)$ for vacancy loops differs from that for interstitial loops, $c^\text{i}_\Delta(R)$, by the sign in the exponent, i.e. $c^\text{v}_\Delta(R)\cdot c^\text{i}_\Delta(R)=c_0^2$.\\

We define the quantities: $c_1=c_\Delta^v$, $c_2=c_\Delta^i$, $c_3=c_\Sigma$, $f_1=f_v$, $f_2=f_i$, $f_3=f_c$, $\xi_1=\xi_2=\xi_\Delta$ and $\xi_3=\xi_\Sigma$. Eq.~(\ref{eq:everycluster}) then assumes the compact form:
\begin{equation}
\bar{c}(\mathbf{x})=c_\infty+ \sum_{\alpha=1}^3 \int dV' G(|\mathbf{x}-\mathbf{x}'|)  \int_{b}^\infty dR \: \xi_\alpha(R) f_\alpha(R)  \left[\bar{c}(\mathbf{x}')-c_\alpha(R) \right].
\end{equation}
We define:
\begin{equation}
\begin{split}
\overline{\xi} &=\sum_{\alpha=1}^{3} \overline{\xi_\alpha}=\sum_{\alpha=1}^{3} \int_{b}^\infty dR \: \xi_{\alpha}(R) f_{\alpha}(R), \\
\overline{\xi c} &=\sum_{\alpha=1}^{3} \overline{\xi_\alpha c_\alpha}=\sum_{\alpha=1}^{3} \int_{b}^\infty dR  \xi_\alpha(R) f_\alpha(R)c_\alpha(R),
\end{split}
\end{equation}
so that:
\begin{equation}
\bar{c}(\mathbf{x})=c_\infty+\overline{\xi} \int dV' G(|\mathbf{x}-\mathbf{x}'|)\bar{c}(\mathbf{x}')-\overline{\xi c} \int dV' G(|\mathbf{x}-\mathbf{x}'|).
\end{equation}
In momentum space, using the Fourier transform definition ${f}(\mathbf{x})=(2 \pi)^{-3/2} \int d\mathbf{q} e^{i \mathbf{q}\cdot\mathbf{x}} \tilde{f}(\mathbf{q}) $, this becomes:
\begin{equation}
\tilde{\bar{c}}(\mathbf{q})=\tilde{c}_\infty+\overline{\xi}  \tilde{G}(\mathbf{q}) \tilde{\bar{c}}(\mathbf{q})-(2\pi)^{3/2} \overline{\xi c}\tilde{G}(\mathbf{q}) \delta(\mathbf{q}),
\end{equation}
leading to the solution:
\begin{equation}
\tilde{\bar{c}}(\mathbf{q})=\frac{\tilde{c}_\infty-(2\pi)^{3/2}\overline{\xi c}\tilde{G}(\mathbf{q}) \delta(\mathbf{q})} {1-\overline{\xi} \tilde{G}(\mathbf{q})}.
\end{equation}
In a homogeneous infinite medium, we have $G(|\mathbf{x}-\mathbf{x}'|)=-1/(4 \pi D_v |\mathbf{x}-\mathbf{x}'|)$ leading to:
\begin{equation}
\tilde{\bar{c}}(\mathbf{q})=\delta(\mathbf{q}) \frac{D_v c_\infty+(2\pi)^{3/2}\overline{\xi c}/q^2 } {D_v+\overline{\xi} /q^2}=\frac{\delta(q)}{2 \pi} \frac{D_v c_\infty+(2\pi)^{3/2}\overline{\xi c}/q^2 } {D_v q^2+\overline{\xi}},
\end{equation}
and, transforming back in real space:
\begin{equation}
\begin{split}
&\int \frac{d\mathbf{q}}{(2 \pi)^{3/2}} e^{i \mathbf{q}\cdot\mathbf{x}}  \tilde{\bar{c}}(\mathbf{q})\\
&={\frac{1}{(2 \pi)^{5/2}}} \int_0^\infty dq \int_0^\pi \sin\phi \: d\phi \int_{-\pi}^\pi d\theta   \delta(q) \frac{ D_v q^2 c_\infty +(2\pi)^{3/2}\overline{\xi c}}{D_v q^2+\overline{\xi}} e^{i q r \cos\phi}=\frac{\overline{\xi c}}{\overline{\xi}},
\label{eq:concvst}
\end{split}
\end{equation}
which implicitly depends on time through the size distribution functions $f_\alpha(R)$.  We point out that for  $f_\alpha(R)=\delta_{\alpha2} f(R)$ we recover the result of the previous sections.\\
The evolution laws for the size distributions are analogous to eq.~(\ref{eq:continuity}), where the growth velocities are given by:
\begin{equation}
\begin{split}
v_\text{vl}(R,t)&=\left[\bar{c}(t)-c_\Delta^\text{v}(R)\right] \frac{2 \pi D_v}{b \ln (\frac{8 R}{r_d})},\\
v_\text{il}(R,t)&=-\left[\bar{c}(t)-c_\Delta^\text{i}(R)\right] \frac{2 \pi D_v}{b \ln (\frac{8 R}{r_d})},\\
v_\text{c}(R,t)&=-\left[\bar{c}(t)-c_\Sigma(R)\right] \frac{D_v}{R},
\label{eq:sym11}
\end{split}
\end{equation}
respectively for vacancy loops, interstitial loops and cavities.\\
As a consequence, the average vacancy concentration depends on time through the time-dependent size distribution functions $F_\alpha(R,t)$ as:
\begin{equation}
 \bar{c}(t)=\frac{\overline{\xi c}(t)}{\overline{\xi}(t)}=\frac{\sum_{\alpha=1}^{3} \int_{b}^\infty dR \: \xi_{\alpha}(R) F_{\alpha}(R,t)}{\sum_{\alpha=1}^{3} \int_{b}^\infty dR \: \xi_{\alpha}(R) F_{\alpha}(R,t) c_\alpha(R)}.
\end{equation}
A numerical scheme for the computation of the cluster size distribution as a function of time can then be summarised as follows:
\begin{enumerate}
\item Starting from the size distributions $f_\text{v}(R,t_0)$,  $f_\text{i}(R,t_0)$ and $g(R,t_0)$, compute $c(t_0)$ via eq.~(\ref{eq:concvst})
\item Compute the new size distributions at time $t_0+dt$ using eq.~(\ref{eq:continuity})
\item Reiterate from step 1.
\end{enumerate}

\section{Numerical simulations}
\label{sec:Numerics}
\begin{center}
\begin{figure}[h!]
\centering
\subfloat[]{\includegraphics[width=0.385\textwidth]{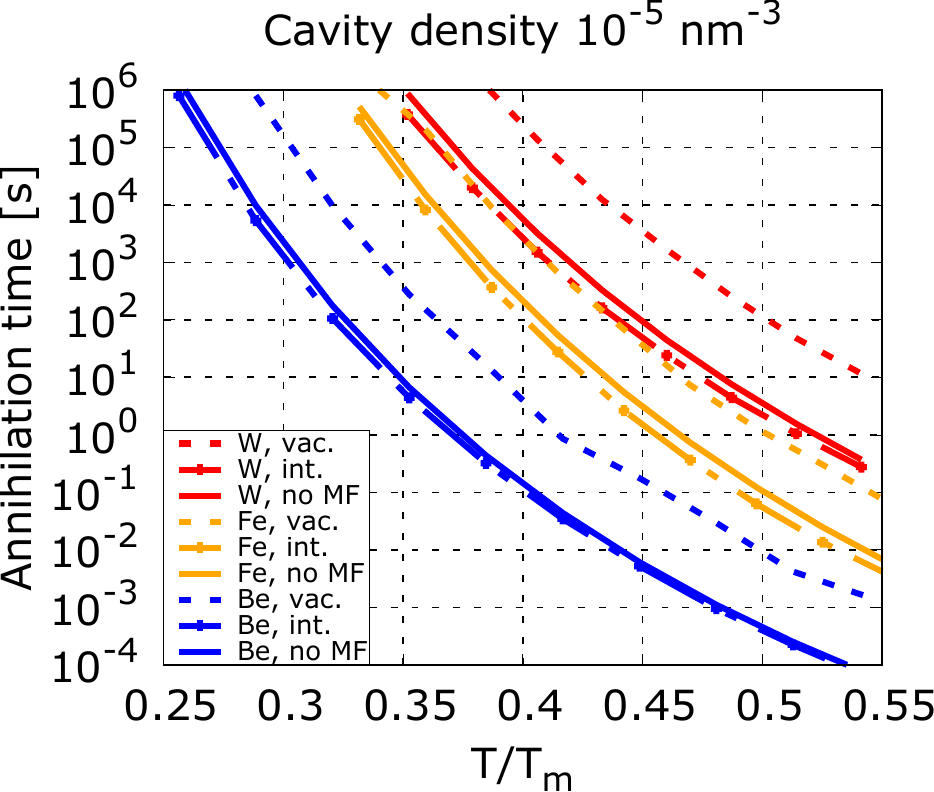}}
\hspace{-40pt}
\subfloat[]{\includegraphics[width=0.385\textwidth]{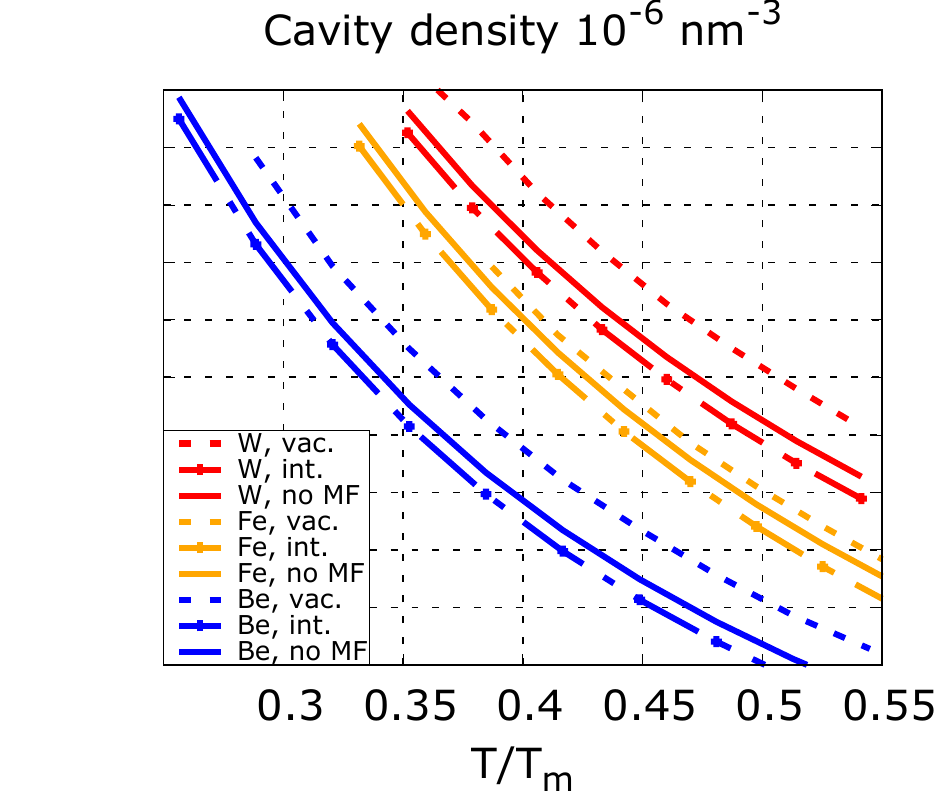}}
\hspace{-40pt}
\subfloat[]{\includegraphics[width=0.385\textwidth]{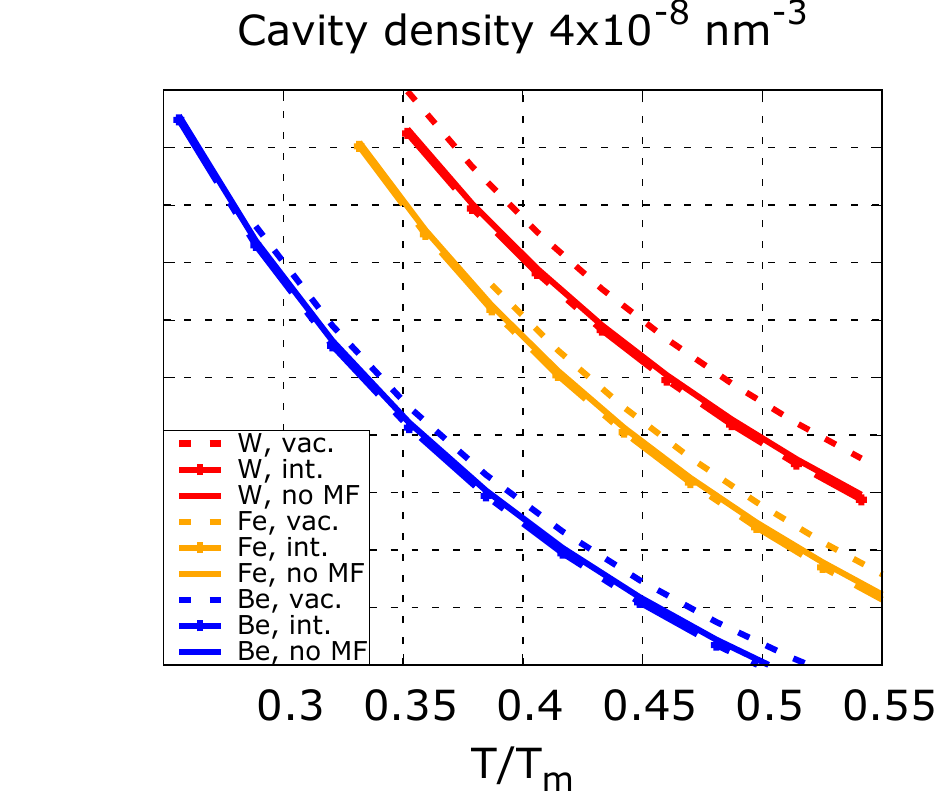}}
\caption{Time taken to remove all cavities in thin films of thickness $200$~nm in tungsten (red), bcc iron (orange) and beryllium (blue), as a function of annealing homologous temperature, for three number densities of cavities: (a)~$10^{-5}$ nm$^{-3}$, (b)~$10^{-6}$ nm$^{-3}$, and (c)~$4\cdot10^{-8}$ nm$^{-3}$. In each case there are either no dislocation loops present (solid lines), or only vacancy loops (dashed lines) or only interstitial loops (dotted solid-dashed lines) . Diffusion is assumed to occur by a vacancy mechanism only.}
\label{fig:Simulations}
\end{figure}
\end{center}
We present simulations of an anneal of an infinite thin film of thickness $H=200$ nm with 30 cavities per periodic cell, using the technique presented in sec.\ref{sec:TF}, for tungsten, iron, and beryllium. The sizes  $L$ of the primitive cell in both the $x$ and $y$ directions are respectively 133.64 nm, 422.58 nm and 2112.89 for the simulations with cavity densities of $10^{-5}$ nm$^{-3}$, $10^{-6}$ nm$^{-3}$ and $4\cdot10^{-8}$ nm$^{-3}$.\\ We compared the effect of three mean field conditions: containing only interstitial loops, or only vacancy loops, or no loops. The initial size distribution of the loops was $\phi(R)=(b/R)^{c_2} \exp(-c_1 b/R)$, $c_1=0.2$, $c_2=3$ and normalisation such that the loop number density at $t=0$ was $10^{-4}$ nm$^{-3}$, therefore the average separation between loops was $\sim22$ nm.\\
The cavities were assigned to random positions in the periodic cell, satisfying the number density and a minimum separation between cavities of $10$~nm, with sizes taken from a gaussian distribution of mean 1 nm and standard deviation 0.1~nm. 
No vacancy supersaturation was assumed in the film, so that its free surfaces were assigned a constant vacancy concentration equal to the equilibrium value $c_0$.\\

In Fig. \ref{fig:Simulations} we plot the time taken to remove all the cavities, as a function of homologous annealing temperature $T/T_\text{m}$.

We see that the timescale estimates given in sec.\ref{sec:timescales} are in broad agreement with numerical simulations, for all three investigated materials. As already noted in our previous work \cite{Rovelli2017}, we see that as the number density of cavities increases the time to annihilate all cavities also increases. 
This is due to the diffusive interaction between cavities: as the distances between cavities are reduced, the local vacancy concentration between cavities increases. 
The driving force for vacancy emission from each cavity is then reduced compared to the case of a dilute configuration of cavities.\\ 

The presence of a mean field of dislocation loops, as previously discussed, introduces a screening of the diffusive interaction between cavities \emph{and} increases of the overall background vacancy concentration. 
Screening accelerates the annihilation of cavities, while the increased vacancy concentration has the opposite effect, and can even induce a transient phase of cavity growth. Whether one effect or the other dominates depends on the type of dislocation loops that make up the mean field.\\

We have assumed that the only mobile point defects are vacancies. The concentrations of these point defects just outside interstitial and vacancy loops are highly asymmetrical due to the exponential dependence on the climb force and the change of sign of the Burgers vector between the loops. 
As a consequence, the vacancy concentration near interstitial loops is much less than for vacancy loops. Therefore the rate of evolution of vacancy loops is faster than that of interstitial loops.\\

The screening factor and the larger background vacancy concentration provided by vacancy loops suggest that they might affect the evolution of cavities to a higher degree than interstitial loops. 
However, it should also be kept in mind that the evolution by vacancy diffusion of vacancy loops is faster than that of interstitial loops with the same initial distributions of loop sizes, and by the end of the simulation a large fraction of the initial vacancy loop population has evaporated.\\
\begin{center}
\begin{figure}[h!]
\centering
\includegraphics[width=0.8\textwidth]{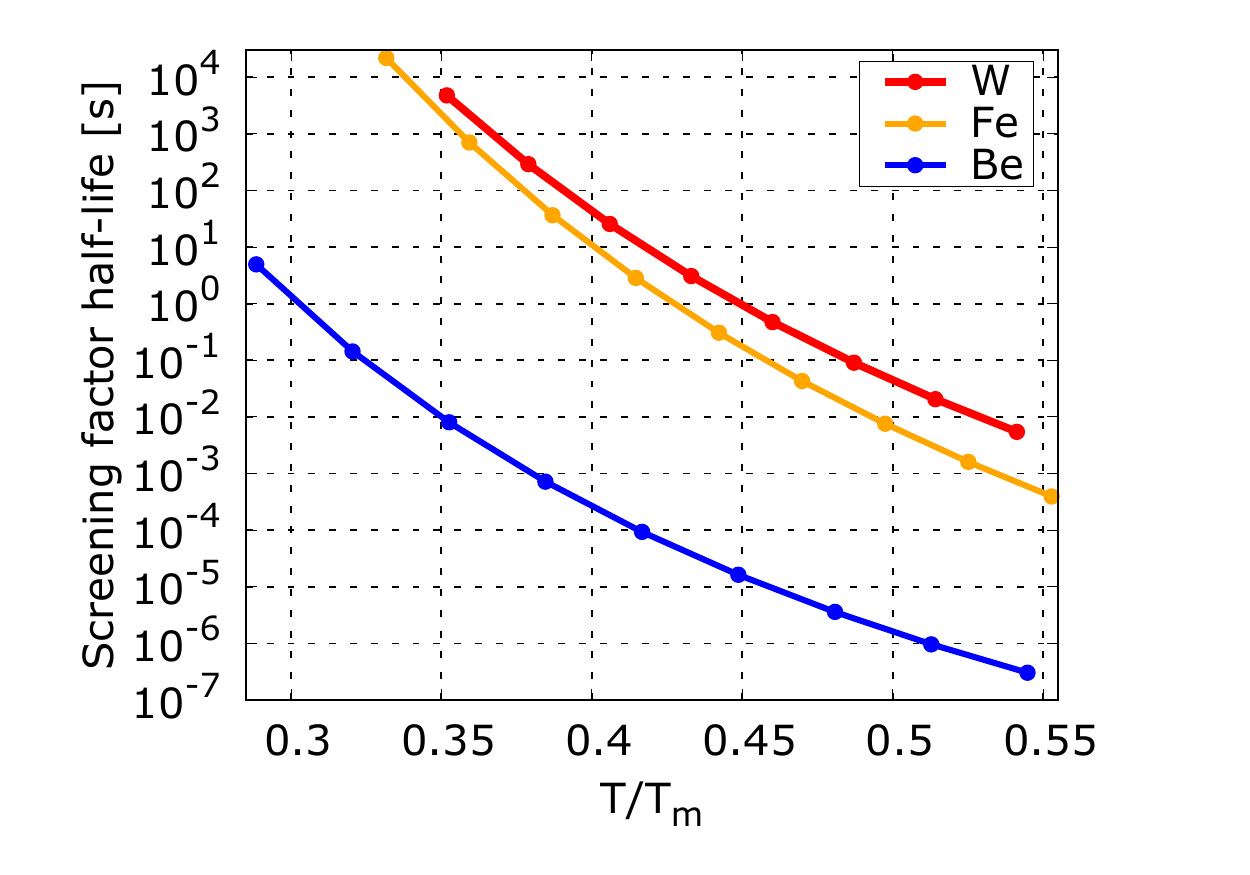}
\caption{Time taken to halve the screening factor $\sqrt{\xi/D_v}$ with respect to its initial value, in thin films of thickness $200$~nm in tungsten (red), bcc iron (orange) and beryllium (blue), with a mean field of vacancy loops, as a function of annealing homologous temperature.}
\label{fig:MFHL}
\end{figure}
\end{center}
To illustrate this behaviour, we plot in Fig. \ref{fig:MFHL} the time required for the vacancy loops screening coefficient $\sqrt{\xi/D_v}$  to halve with respect to its initial value, as a function of simulated temperature. 
We point out that the ``half-lives'' of the screening coefficient are always much shorter than the time required for all cavities to evaporate. The evolution of the vacancy loops mean field does not appear to be noticeably affected by the number density of cavities, at least in the investigated density range from $4\cdot10^{-8}$ to $10^{-5}$ nm$^{-3}$. \\

In simulations with only interstitial loops as part of the mean field, on the other hand, the initial size distribution function, the screening coefficient and the shift to the background vacancy concentration are practically constant with respect to the evolution of cavities.\\

According to our simulations, when the mean field is made of interstitial loops there is a systematic decrease of cavity evaporation timescales with respect to the reference values, for every investigated cavity number density. 
We must conclude that the screening effect, partially suppressing the diffusive interaction between cavities, is the dominant one for interstitial loops. The opposite holds true for the vacancy loops mean field simulations, where a systematic and very pronounced \emph{increase} in cavity evaporation timescales suggests that the extra background vacancy concentration plays the most important role.\\

\begin{center}
\begin{figure}[h!]
\centering
\subfloat[]{\includegraphics[width=0.52\textwidth]{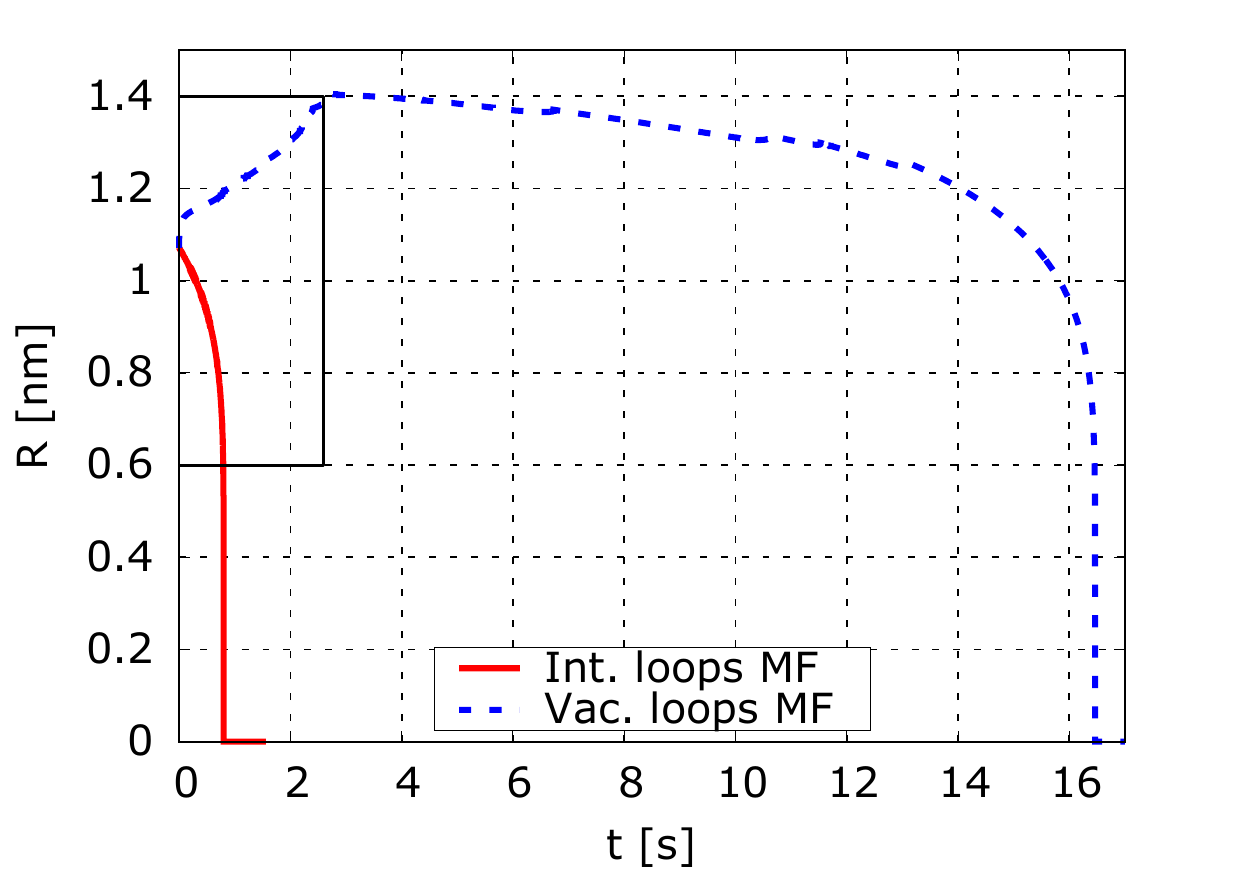}}
\hspace{-30pt}
\subfloat[]{\includegraphics[width=0.52\textwidth]{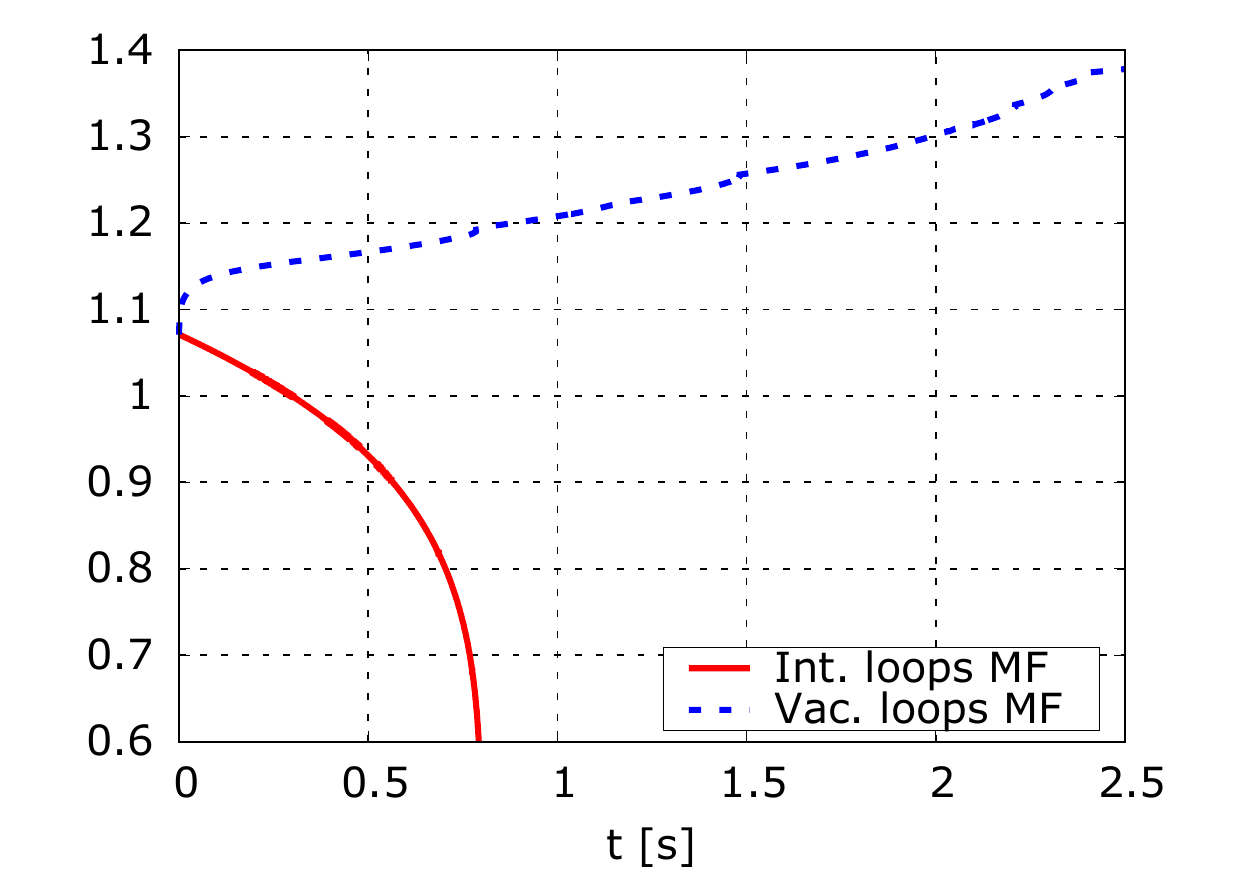}}
\caption{Evolution of the radius of a single cavity at $(0.37L,\,0.58L,\,0.39H)$ as a function of time for tungsten at 1800 K and for a cavity number density of $10^{-6}$ nm$^{-3}$. Results are compared for the case of an interstitial loop mean field (solid red lines), and a vacancy loop mean field (dashed blue lines). Plot (b) is a magnification of the boxed area of plot (a).}
\label{fig:Zoom1}
\end{figure}
\end{center}

In order to further elucidate this point, we show in Fig. \ref{fig:Zoom1} the evolution in time of the size of a single cavity in tungsten at 1800 K and with a cavity number density of $10^{-6}$ nm$^{-3}$. The coordinates of the cavity in the primitive cell are $(0.37L,\,0.58L,\,0.39H)$. We note that in the initial phases of the simulation the additional background vacancy concentration induced by vacancy loops results in the cavity rapidly \emph{adsorbing} vacancies. 
As the mean field evolves, vacancy loops decrease in number and the extra vacancy concentration is reduced until the surface energy of cavities becomes the dominant driving force, leading to their evaporation. 
Due to the initial phase of growth of the cavity, the total annihilation time with the vacancy loops mean field is therefore larger than in the case of the interstitial loops mean field.\\
\vspace{-40pt}
\begin{center}
\begin{figure}[h!]
\centering
\subfloat[]{\includegraphics[width=0.52\textwidth]{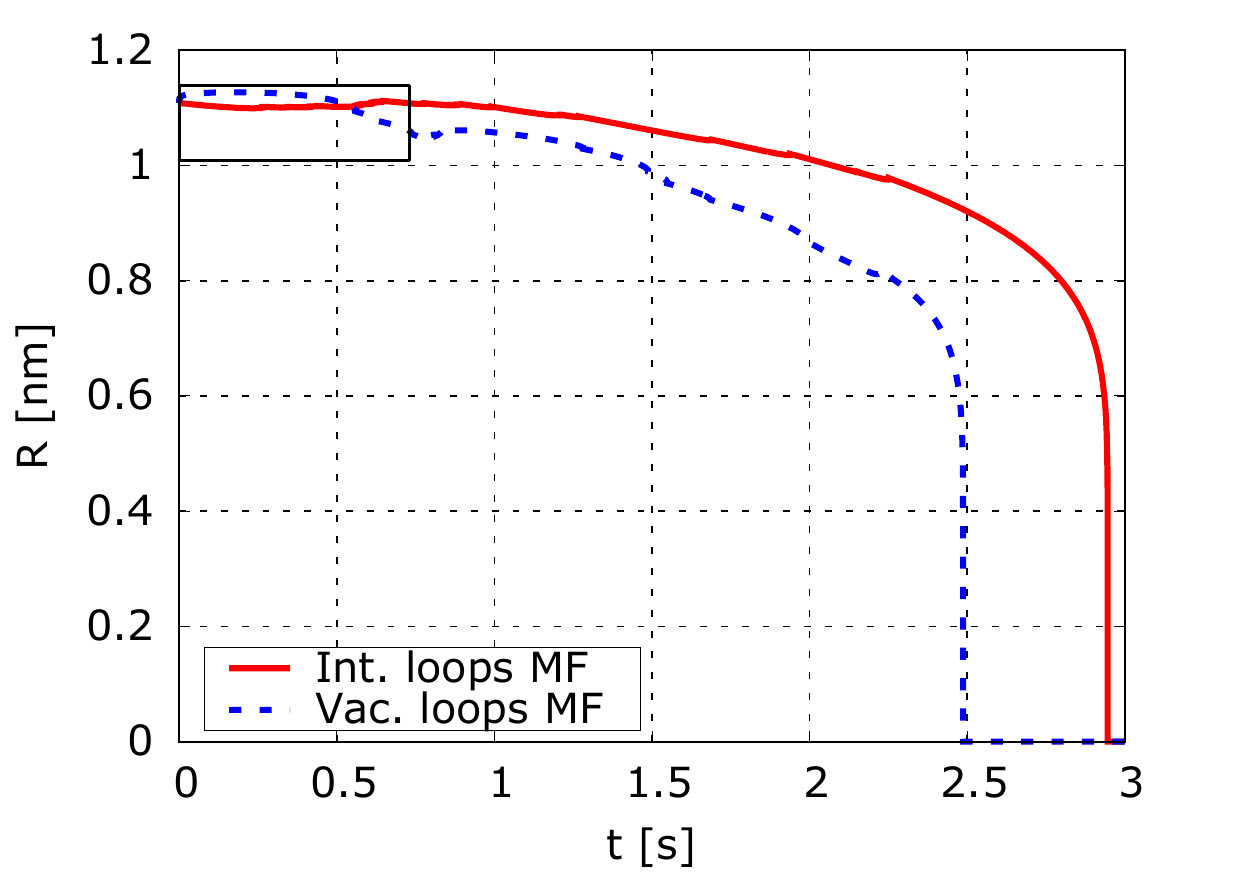}}
\hspace{-30pt}
\subfloat[]{\includegraphics[width=0.52\textwidth]{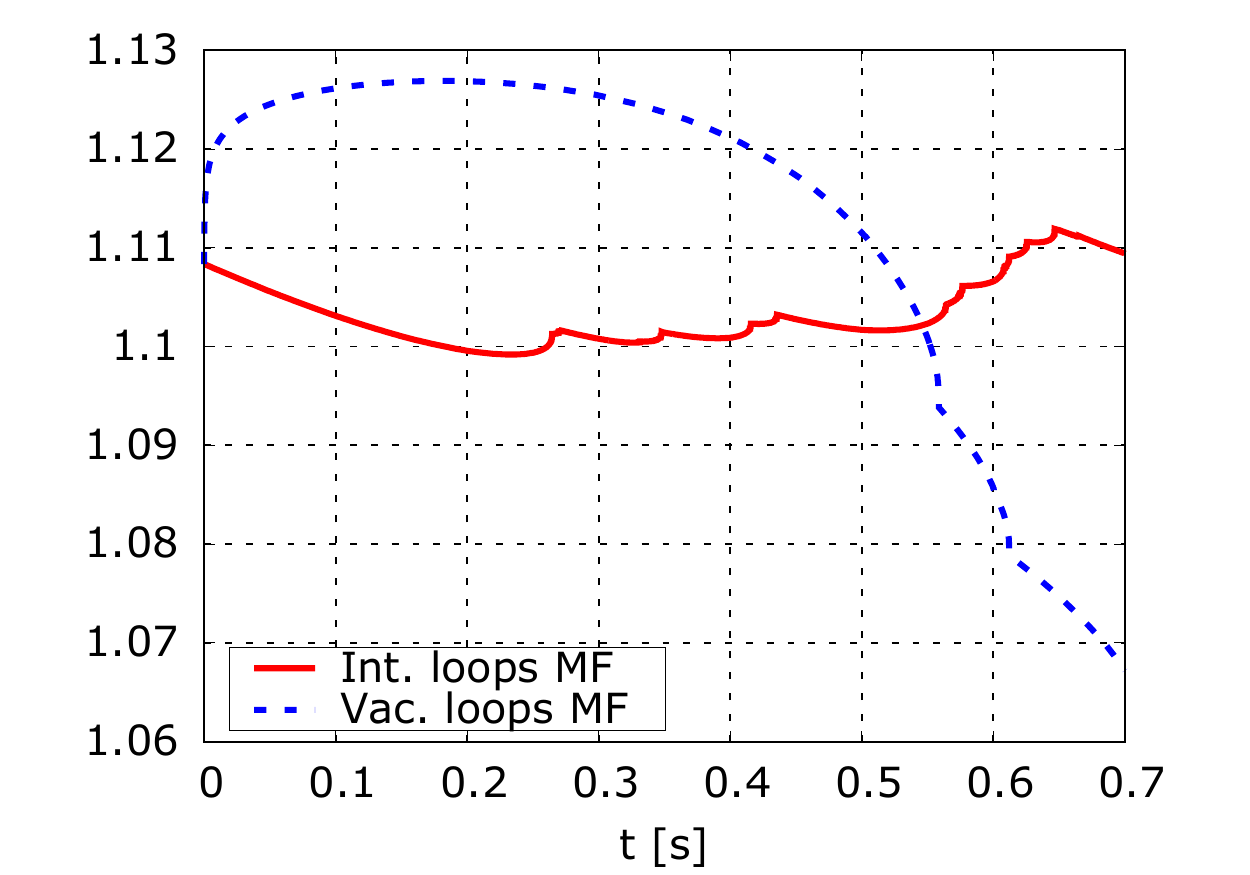}}
\caption{Same as Fig. \ref{fig:Zoom1} but for a cavity number density of $10^{-5}$ nm$^{-3}$. The cavity is located at coordinates $(0.36L,\,0.44L,\,0.65H)$ in the primitive cell.}
\label{fig:Zoom2}
\end{figure}
\end{center}
While we found this to hold true for all investigated cavity densities, the picture is slightly more complex when only considering the evolution of a single cavity of the ensemble. 
In particular, the \emph{same} cavity will not always evaporate faster with an interstitial loop mean field, especially in simulations with a high cavity density.\\ To illustrate such a case, we present In Fig. \ref{fig:Zoom2} the evolution of a single cavity with coordinates $(0.36L,\,0.44L,\,0.65H)$, with a cavity a number density $10^{-5}$ nm$^{-3}$ and at a temperature of 1800 K.
\begin{center}
\begin{figure}[h!]
\centering
\subfloat[]{\includegraphics[width=0.3\textwidth]{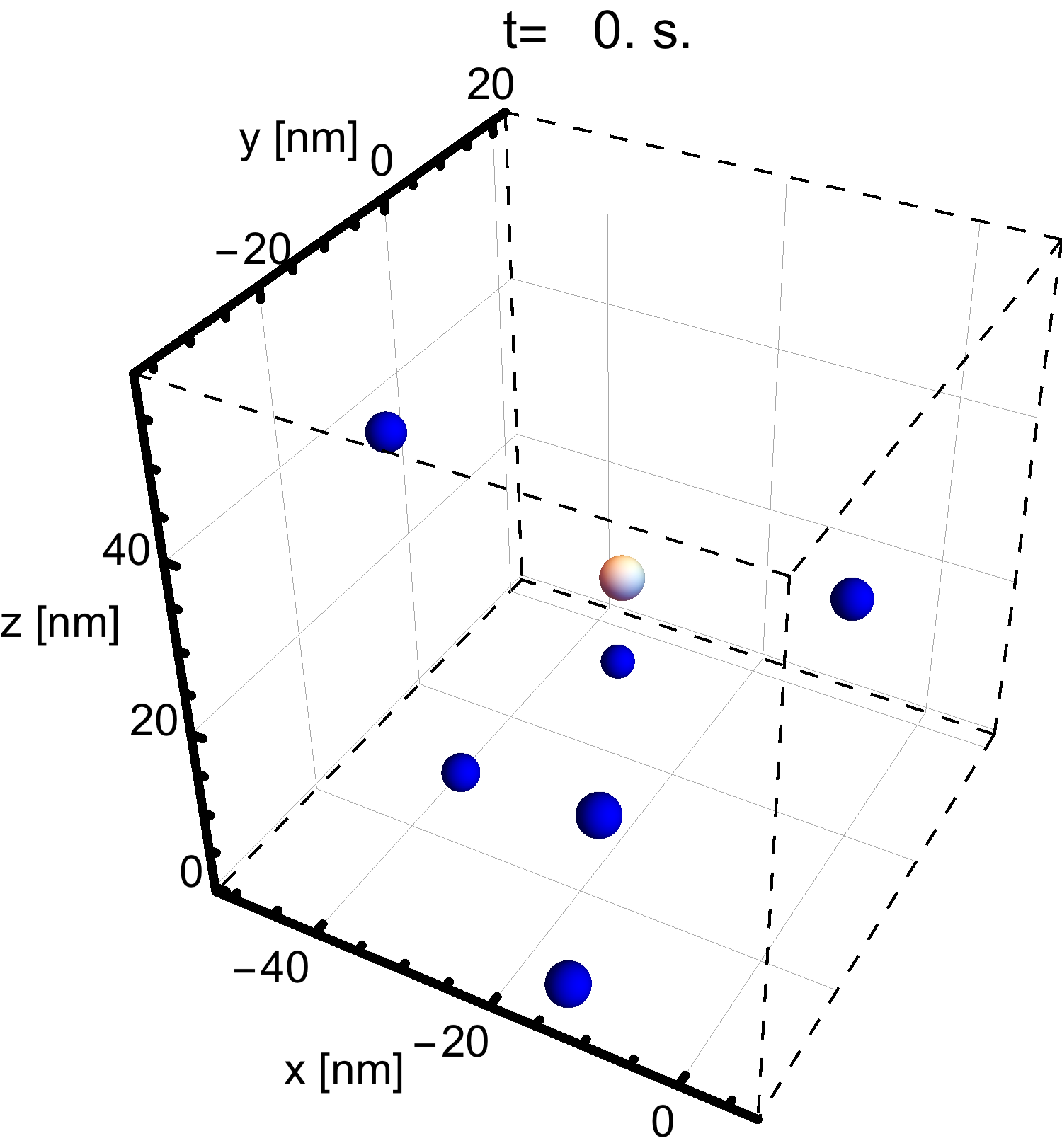}}
\subfloat[]{\includegraphics[width=0.3\textwidth]{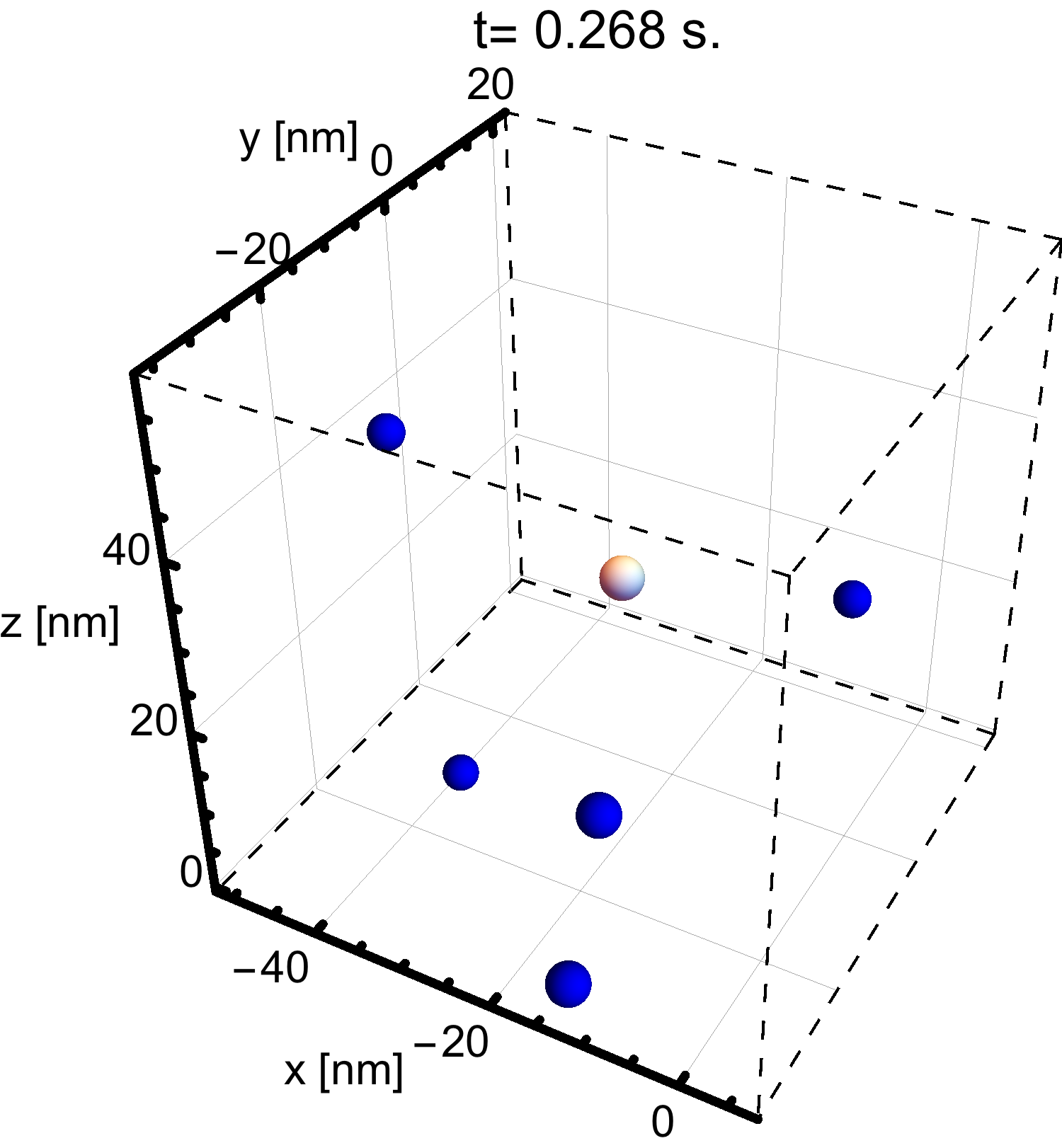}}
\subfloat[]{\includegraphics[width=0.3\textwidth]{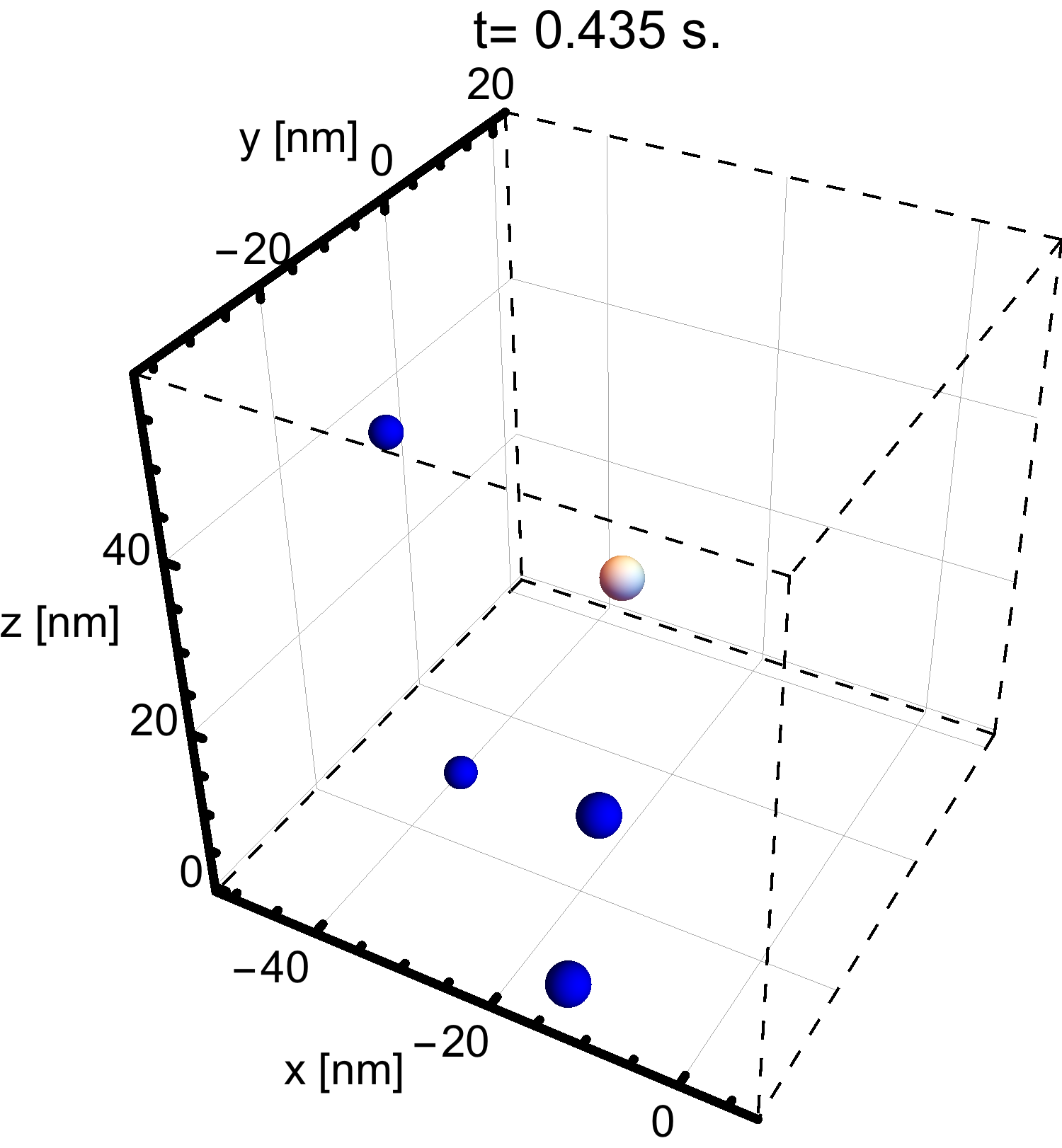}}\\
\subfloat[]{\includegraphics[width=0.3\textwidth]{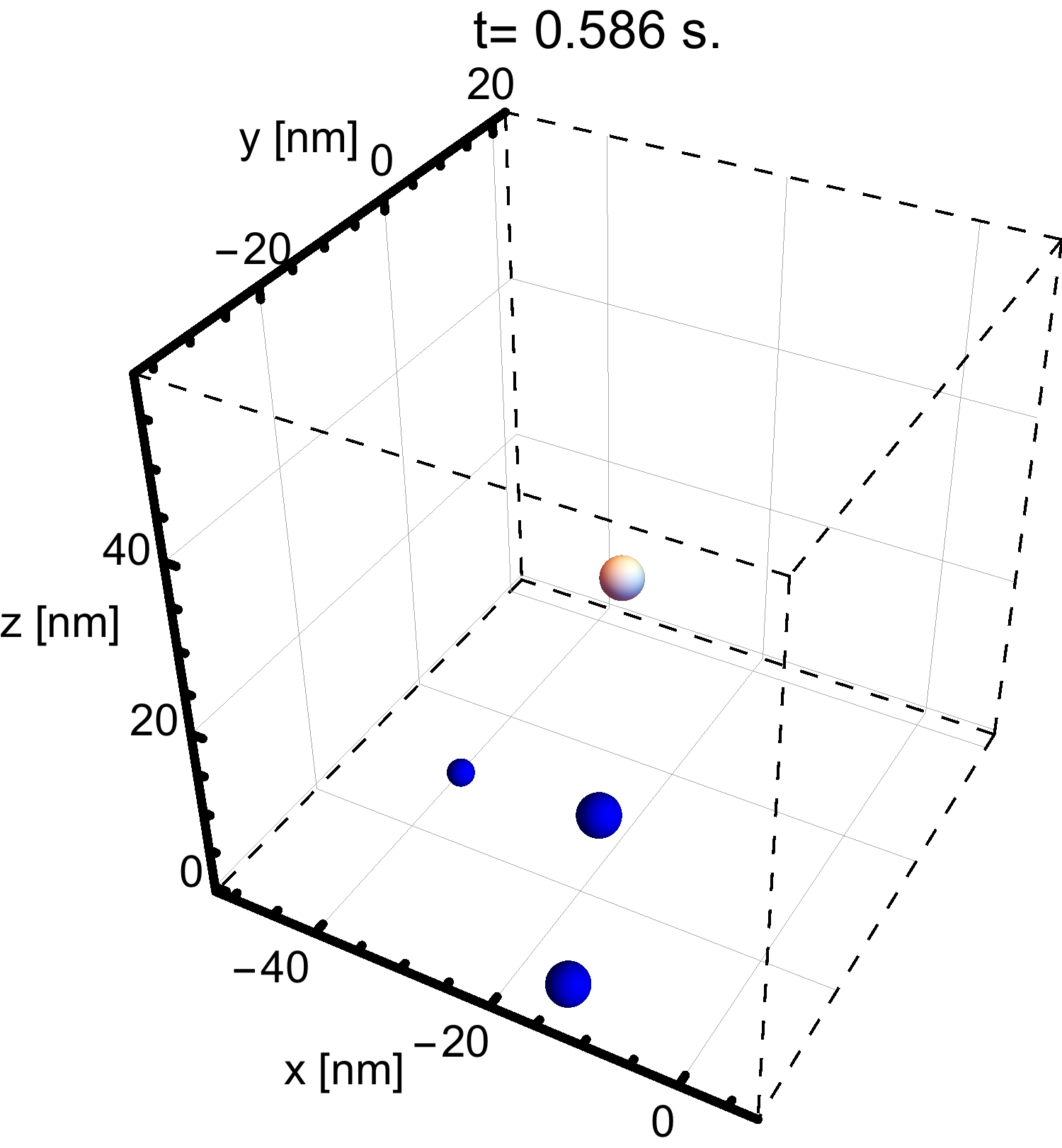}}
\subfloat[]{\includegraphics[width=0.3\textwidth]{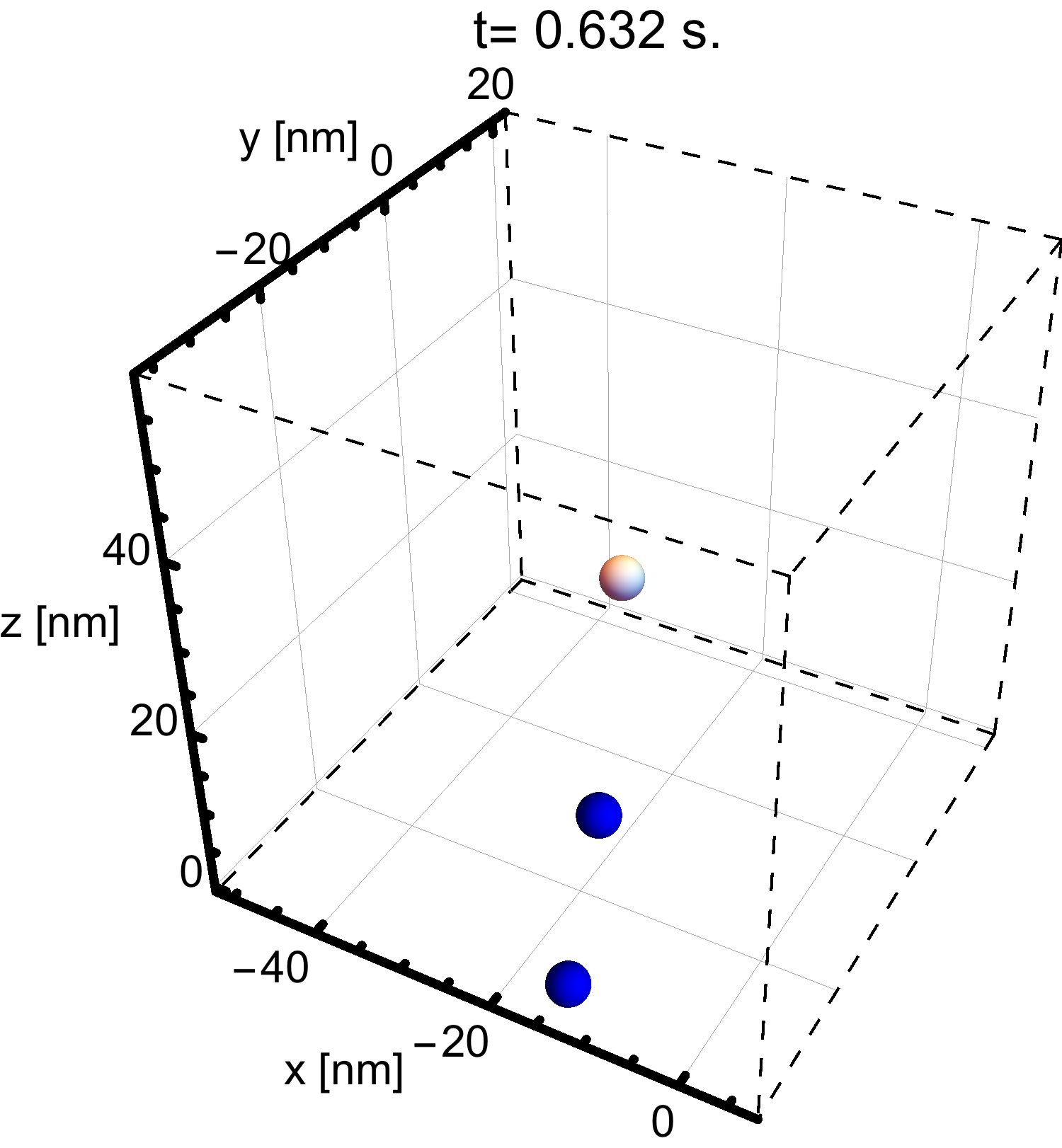}}
\subfloat[]{\includegraphics[width=0.3\textwidth]{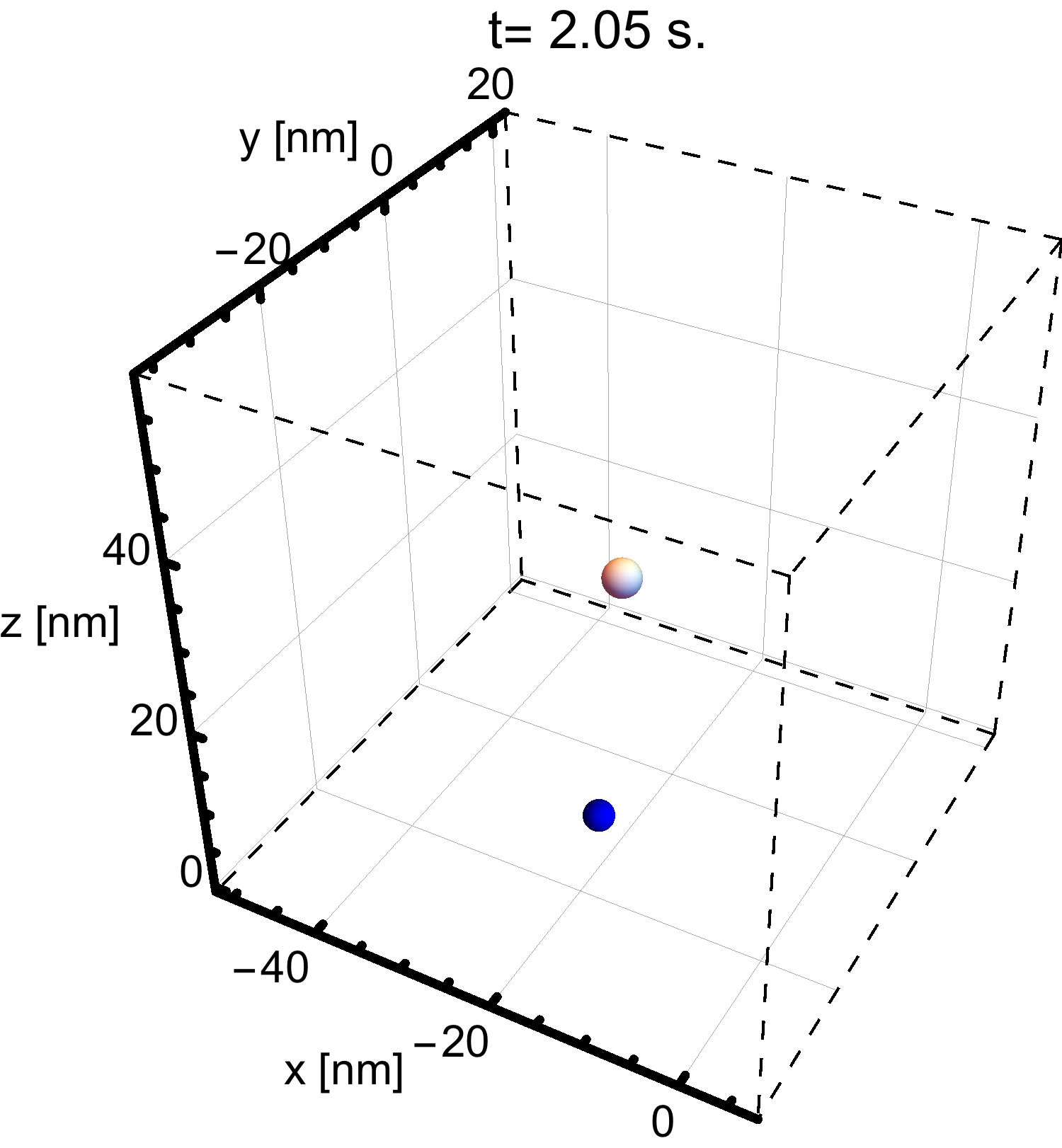}}
\caption{Snapshots of the evolution of the cavity of Fig. \ref{fig:Zoom2}, shaded in orange and white in the cell center, and neighboring cavities in blue at increasing times from (a) to (f). The plots show only a small portion of the  periodic cell. The sizes of the cavities have been increased by a factor of 2 for greater clarity.}
\label{fig:RealSpace}
\end{figure}
\end{center}
As we can see, the vacancy loops mean field still induces an initial phase of cavity growth (Fig. \ref{fig:Zoom2}), but the cavity evolves erratically in the interstitial loops mean field simulation. 
By comparing Fig. \ref{fig:Zoom2}b with a real space view of the simulation (Fig. \ref{fig:RealSpace}) we find this behaviour is most likely due to local interaction between cavities.\\
In particular, the erratic size increases and decreases of the cavity seem to be correlated  with the evaporation of cavities in its neighborhood. 
Such local effects are much less important with the vacancy loops mean field because, even though the shift in background vacancy concentration dominates, the screening coefficient is still substantially larger than for interstitial loops, as already mentioned. 
This suggests that at higher densities local interactions between cavities play a large role and might lead the dynamics of single cavities to deviate from the general behaviour with respect to mean field conditions.

\section{Conclusions}
\label{sec:Conclusions}
In this paper we derived a hybrid mean field and real space model that couples our earlier \cite{Rovelli2017}  non-local model of evolution of cavities produced by irradiation with a mean field representing dislocation loops smaller than the experimental detection limit. \\
The main result is that the mean field screens the diffusive interactions between cavities and adds to, or subtracts from, the vacancy concentration depending on whether the loops are vacancy or interstitial character respectively.\\ 
We presented a general scheme to implement higher order corrections to the model. The model was initially derived for an infinite medium, but it was then modified to treat the case of an infinitely extended thin film, which is more useful for comparing with available experimental data obtained by transmission electron microscopy. \\
Through preliminary numerical simulations we discussed some of the features of the model, highlighting in particular the interplay between the evolution of the mean field and the cavities, and the roles of the cavity number density and mean fields comprising vacancy vs. interstitial loops. \\

\section*{Acknowledgements}
IR and SLD were partly funded by the RCUK Energy Programme (grant No. EP/P012450/1). IR was also supported by the EPSRC Centre for Doctoral Training on Theory and Simulation of Materials at Imperial College London under grant number EP/L015579/1. This work was carried out within the framework of the EUROfusion Consortium and has received funding from the Euratom research and training programme 2014-2018 under grant agreement No. 633053. The views and opinions expressed herein do not necessarily reflect those of the European Commission. APS is grateful to the Blackett Laboratory for the provision of laboratory facilities.

\appendix
\section{Relaxation volume of a dislocation loop and its evolution in time}
\label{app:loopvolume}
\begin{center}
\begin{figure}[h!]
\centering
\includegraphics[width=0.6\textwidth]{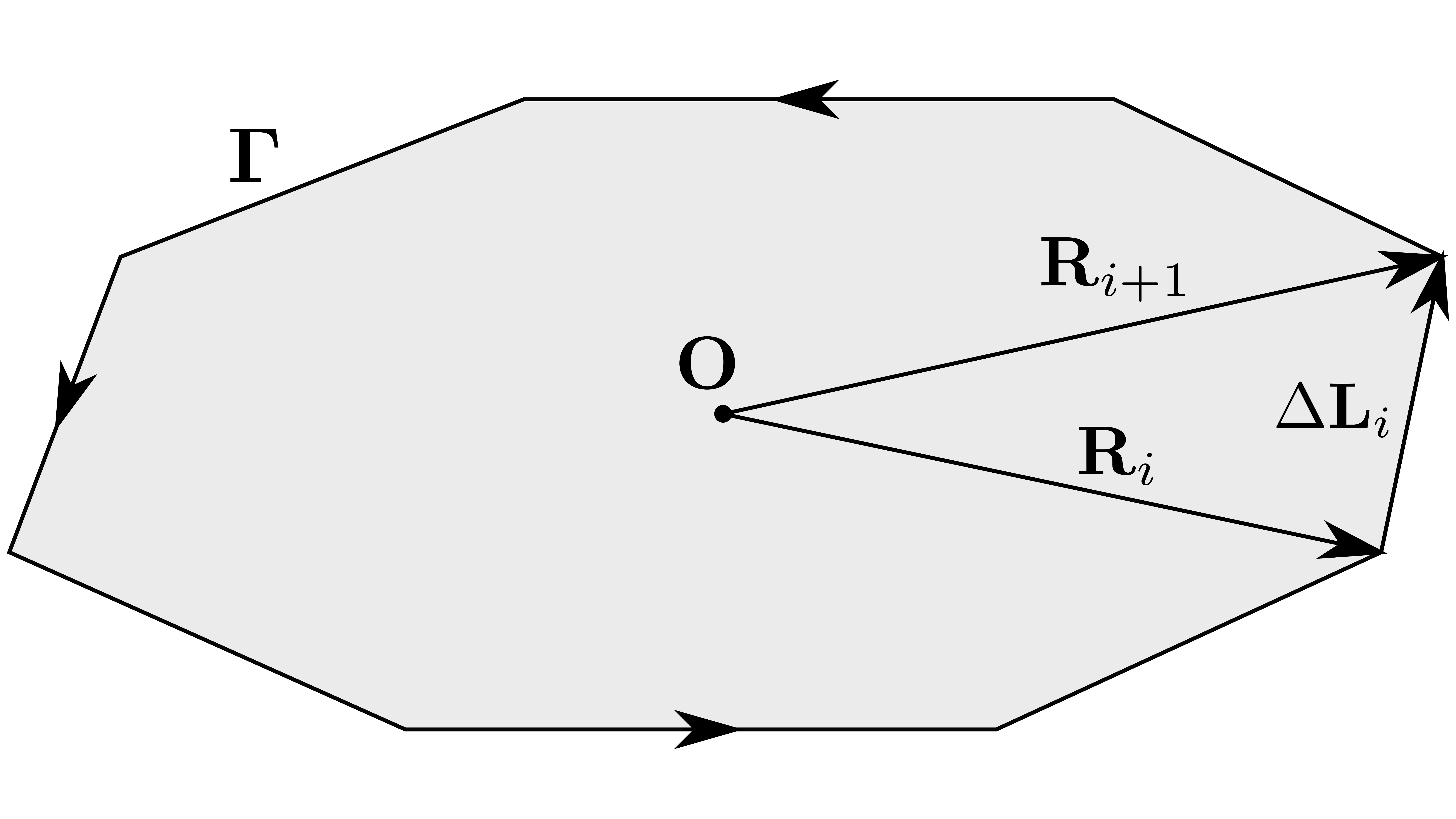}
\caption{Sketch of the vectors $\mathbf{R}_i$, $\mathbf{R}_{i+1}$ and $\Delta \mathbf{L}_i$ with respect to the boundary $\Gamma$ of the dislocation loop. The arrows on $\Gamma$ denote the direction of the dislocation line.}
\label{fig:looparea}
\end{figure}
\end{center}
Let us consider an arbitrarily shaped dislocation loop, not necessarily planar, with Burgers vector $\mathbf{b}$ and bounded by a piecewise linear dislocation line $\Gamma$ divided in $N$ segments. Given an arbitrary orthogonal coordinate system, let $\mathbf{R}_i$ and $\mathbf{R}_{i+1}=\mathbf{R}_i+\Delta\mathbf{L}_i$ be the vectors denoting the extremes of the $i$-th segment on $\Gamma$, with $\Delta \mathbf{ L}_i$ parallel to the dislocation line direction, and $\mathbf{R}_{N+1}=\mathbf{R}_1$. The vector area of the triangle identified by the origin $\mathbf{O}$, $\mathbf{R}_i$ and $\mathbf{R}_{i+1}$ is given by:
\begin{equation}
\mathbf{A}_i=\frac{1}{2} \left(\mathbf{R}_i \times \mathbf{R}_{i+1}\right).
\end{equation}
Summing the individual contributions given by the $N$ triangles, we obtain the total vector area of the loop:
\begin{equation}
\mathbf{A}=\sum_{i=1}^N \mathbf{A}_i=\frac{1}{2}\sum_{i=1}^N \left(\mathbf{R}_i \times \mathbf{R}_{i+1} \right),
\end{equation}
which, in order to be a well-defined quantity, has to be independent from the choice of the origin of coordinates. In order to prove this statement, let us shift the coordinate system by an arbitrary vector $\mathbf{R}_0$ and compute the new $\mathbf{A}'$:
\begin{equation}
\begin{split}
\mathbf{A}'&=\frac{1}{2}\sum_{i=1}^N \left[\left(\mathbf{R}_i + \mathbf{R}_0\right)\times \left(\mathbf{R}_{i+1} + \mathbf{R}_0 \right) \right]\\
&=\frac{1}{2}\sum_{i=1}^N \left(\mathbf{R}_i \times \mathbf{R}_{i+1} \right)+\frac{1}{2}\mathbf{R}_0 \times \underbrace{\sum_{i=1}^N \left(\mathbf{R}_{i+1} - \mathbf{R}_{i} \right)}_{0}=\mathbf{A}.
\end{split}
\end{equation}
Therefore $\mathbf{A}$ is indeed a well-defined measure for the vector area of a dislocation loop. 
A formula for the vector area of a dislocation loop can also be written in the form of a contour integral, see equation (27.11) of Ref. \cite{Landau1970}, as
\begin{equation}
\mathbf{A}={1\over 2}\oint (\mathbf{x}\times d\mathbf{l}),
\end{equation}
where $\mathbf{x} \in \Gamma$ and $d\mathbf{l}$ is everywhere tangential to the dislocation line. In the notations of Fig. \ref{fig:looparea} the above equation corresponds to the limit $N\rightarrow \infty$.\\
Using the convention for the Burgers vector and the dislocation line tangential vector adopted in \cite{Hirth1982}, we now define the relaxation volume of the loop as follows:
\begin{equation}
\mathcal{V}=-(\mathbf{b}\cdot \mathbf{A})=-\frac{1}{2} \mathbf{b} \cdot \sum_{i=1}^N \left(\mathbf{R}_i \times \mathbf{R}_{i+1} \right),
\end{equation}
which is correctly negative for vacancy prismatic loops, where $\mathbf{b}$ is parallel to $\mathbf{A}$, and positive for interstitial prismatic loops, where $\mathbf{b}$ is anti-parallel to $\mathbf{A}$.
We now consider the continuous limit for $|\Delta \mathbf{L}_i |\rightarrow 0$ and $N \rightarrow \infty$, leading to the line integral:
\begin{equation}
\mathcal{V} = -\frac{1}{2} \oint_\Gamma \mathbf{b} \cdot \left( \mathbf{x} \times d\mathbf{l} \right).
\end{equation}
Let us assume that the line $\Gamma$, defining the perimeter of the dislocation loop, is parameterized by variable $\varphi\in[0,1)$ and that it evolves with time, so that for $\mathbf{x}\in \Gamma$ we have $\mathbf{x}=\mathbf{x}(\varphi,t)$ and $d\mathbf{l}=\frac{\partial \mathbf{x}}{\partial \varphi}d\varphi$. We can then write $\mathcal{V}$(t) as:
\begin{equation}
\mathcal{V}(t)=-\frac{1}{2} \mathbf{b} \cdot \int_0^1  \left[\mathbf{x}(\varphi,t) \times \frac{\partial \mathbf{x}}{\partial \varphi}(\varphi,t) \right] d\varphi
\end{equation}
and its temporal rate of change as
\begin{equation}
\frac{d\mathcal{V}}{dt}=-\frac{1}{2} \mathbf{b} \cdot \left\lbrace \int_0^1 \left[\frac{d\mathbf{x}}{dt}\times \frac{\partial\mathbf{x}}{\partial\varphi} \right] d\varphi+\int_0^1 \left[\mathbf{x}\times \frac{\partial}{\partial\varphi} \left( \frac{d\mathbf{x}}{dt}\right) \right] d\varphi \right\rbrace.
\end{equation}
Using integration by parts, the second integral in the right-hand side can be rearranged as:
\begin{equation}
\int_0^1 \left[\mathbf{x}\times \frac{\partial}{\partial\varphi} \left( \frac{d\mathbf{x}}{dt}\right) \right] d\varphi=\underbrace{\left[\mathbf{x} \times \frac{d\mathbf{x}}{dt} \right] \bigg\rvert_{\varphi=0}^{\varphi=1}}_{0}-\int_0^1 \left[\frac{\partial\mathbf{x}}{\partial\varphi}\times  \frac{d\mathbf{x}}{dt} \right] d\varphi=\int_0^1 \left[ \frac{d\mathbf{x}}{dt} \times \frac{\partial\mathbf{x}}{\partial\varphi} \right] d\varphi
\end{equation}
so that
\begin{equation}
\frac{d\mathcal{V}}{dt}=-\mathbf{b} \cdot \int_0^1 \left[\frac{d\mathbf{x}}{dt}\times \frac{\partial\mathbf{x}}{\partial\varphi} \right] d\varphi=- \oint_\Gamma \mathbf{b} \cdot \left[\frac{d\mathbf{x}}{dt}\times d\mathbf{l} \right],
\label{eq:HirthVol}
\end{equation}
which is a slightly modified form of eq.(4-2) (where it was given without proof) found in \cite{Hirth1982}. We can recast eq.~(\ref{eq:HirthVol}) in a form more convenient for our applications:
\begin{equation}
\frac{d\mathcal{V}}{dt}=-\oint_\Gamma \mathbf{b} \cdot \left(\frac{d\mathbf{x}}{dt}\times d\mathbf{l} \right)=-\oint_\Gamma \mathbf{b}\cdot \left( \mathbf{v}\times d\mathbf{l}\right) =-\oint_\Gamma \mathbf{v}(\mathbf{x})\cdot\left(d\mathbf{l} \times \mathbf{b} \right)=-\oint_\Gamma v_\text{cl}(\mathbf{x})b_e(\mathbf{x})dl,
\end{equation}
where ${\mathbf{v}}(\mathbf{x})={d\mathbf{x}}/{dt}$ is the vector velocity of a point $\mathbf{x}\in\Gamma$,  $v_\text{cl}$ is the (scalar) dislocation climb velocity, $b_e$ is the edge component of the Burgers vector and $dl$ is the differential arc length on $\Gamma$.

\section{Perturbative expansion of the formal scattering series}
\label{app:pert}
Let us combine eqs.~(\ref{eq:micro}) and (\ref{eq:selfcons}) in order to obtain a formal scattering series for the concentration field:
\begin{equation}
\begin{split}
 &c(\mathbf{x}; \mathcal{X},\mathcal{R})=c_b(\mathbf{x})+\sum_{i=1}^n  \xi_\Delta(R_i) G(\mathbf{x},\mathbf{x}_i) \left[c_b(\mathbf{x}_i)-c_\Delta(R_i)\right] \\
 &+\sum_{i=1}^n \sum_{j \neq i} G(\mathbf{x},\mathbf{x}_i)  \xi_\Delta(R_i) G(\mathbf{x}_i,\mathbf{x}_j)  \xi_\Delta(R_j)  \left[c_b(\mathbf{x}_j)-c_\Delta(R_j)\right]\\
&+\sum_{i=1}^n \sum_{j \neq i}  \sum_{k \neq j}  G(\mathbf{x},\mathbf{x}_i) \xi_\Delta(R_i)    G(\mathbf{x}_i,\mathbf{x}_j) \xi_\Delta(R_j) G(\mathbf{x}_j,\mathbf{x}_k) \xi_\Delta(R_k) \left[c_b(\mathbf{x}_k)-c_\Delta(R_l)\right]+...
 \end{split}
\label{eq:cseries}
\end{equation}
We can now recast eq.~(\ref{eq:cseries}) in a manner  similar to classical scattering theory:
\begin{equation}
\begin{split}
 &c(\mathbf{x};\mathcal{X},\mathcal{R})=c_b(\mathbf{x})+\sum_{i =1}^n \int d\mathbf{x}' d\mathbf{x}'' dR \:   G(\mathbf{x},\mathbf{x}') \mathcal{T}_i(\mathbf{x}',\mathbf{x}'',R) \left[c_b(\mathbf{x}'')-c_\Delta(R)\right] \\
 &+\sum_{i =1}^n \sum_{j \neq i} \int d\mathbf{x}' d\mathbf{x}'' d\mathbf{x}''' d\mathbf{x}^{iv} dR dR' G(\mathbf{x},\mathbf{x}') \mathcal{T}_i(\mathbf{x}',\mathbf{x}'',R)  G(\mathbf{x}'',\mathbf{x}''') \mathcal{T}_j(\mathbf{x}''',\mathbf{x}^{iv},R') \left[c_b(\mathbf{x}^{iv})-c_\Delta(R')\right]+...,
 \end{split}
\label{eq:Tseries}
\end{equation}
where $\mathcal{T}$ is a scattering operator defined by:
\begin{equation}
\mathcal{T}_i(\mathbf{x},\mathbf{x'},R)=\xi_\Delta(R) \delta(\mathbf{x}-\mathbf{x}_i) \delta(\mathbf{x}'-\mathbf{x}_i) \delta(R-R_i).
\end{equation}
Equivalently, in momentum space, using the Fourier transform definition ${f}(\mathbf{x})=(2 \pi)^{-3/2} \int d\mathbf{q} e^{i \mathbf{q}\cdot\mathbf{x}} \tilde{f}(\mathbf{q}) $, we have:
\begin{equation}
\begin{split}
 &\tilde{c}(\mathbf{q};\mathcal{X},\mathcal{R})=\tilde{c}_b(\mathbf{q})+\sum_{i =1}^n \int \dj\mathbf{q}' dR \:   \tilde{G}(\mathbf{q}) \tilde{\mathcal{T}}_i(\mathbf{q}-\mathbf{q}',R) \left[\tilde{c}_b(\mathbf{q}')-(2 \pi)^{3/2}\delta(\mathbf{q}')c_\Delta(R)\right]\\
 &+\sum_{i =1}^n \sum_{j \neq i} \int \dj\mathbf{q}' \dj\mathbf{q}'' dR dR' \tilde{G}(\mathbf{q}) \tilde{\mathcal{T}}_i(\mathbf{q}-\mathbf{q}',R)  \tilde{G}(\mathbf{q}') \tilde{\mathcal{T}}_j(\mathbf{q}'-\mathbf{q}'',R') \left[[\tilde{c}_b(\mathbf{q}'')-(2 \pi)^{3/2}\delta(\mathbf{q}'')c_\Delta(R')\right]+...,
 \end{split}
\label{eq:TseriesqSpace}
\end{equation}
where:
\begin{equation}
\tilde{\mathcal{T}}_i(\mathbf{q},R)=\xi_\Delta(R) \delta(R-R_i) e^{-i \mathbf{q} \cdot \mathbf{x}_i}
\end{equation}
and $\dj\mathbf{q}=(2\pi)^{-3/2}d\mathbf{q}$.\\
We can similarly write a corresponding equation for the configuration-averaged concentration:
\begin{equation}
\begin{split}
&\bar{c}(\mathbf{x})=c_b(\mathbf{x})+\int d\mathbf{x}' d\mathbf{x}'' dR \:   G(\mathbf{x},\mathbf{x}') \mathcal{S}(\mathbf{x}',\mathbf{x}'',R) \left[\bar{c}(\mathbf{x}'')-c_\Delta(R)\right]\\
&=c_b(\mathbf{x})+\int d\mathbf{x}' d\mathbf{x}'' dR \:   G(\mathbf{x},\mathbf{x}') {\mathcal{S}}(\mathbf{x}',\mathbf{x}'',R) \left[{c}_b(\mathbf{x}'')-c_\Delta(R)\right]\\
&+\int d\mathbf{x}' d\mathbf{x}'' d\mathbf{x}''' d\mathbf{x}^{iv} dR dR' G(\mathbf{x},\mathbf{x}')  \mathcal{S}(\mathbf{x}',\mathbf{x}'',R) G(\mathbf{x}'',\mathbf{x}''')  \mathcal{S}(\mathbf{x}''',\mathbf{x}^{iv},R') \left[c_b(\mathbf{x}^{iv})-c_\Delta(R')\right]+...,
\end{split}
\label{eq:Sigmaseries}
\end{equation}
which also serves as a definition for the ``self-energy'' $\mathcal{S}$, physically representing the interaction of the mean field with itself. An analogous equation in momentum space takes the form:
\begin{equation}
\begin{split}
&\tilde{\bar{c}}(\mathbf{q})=\tilde{c}_b(\mathbf{q})+\int dR \:   \tilde{G}(\mathbf{q}) \tilde{\mathcal{S}}(\mathbf{q},R)\left[\tilde{\bar{c}}(\mathbf{q}')-(2 \pi)^{3/2}\delta(\mathbf{q}')c_\Delta(R)\right] \\
&=\tilde{c}_b(\mathbf{q})+\int  dR \:   \tilde{G}(\mathbf{q}) \tilde{\mathcal{S}}(\mathbf{q},R) \left[\tilde{c}_b(\mathbf{q}')-(2 \pi)^{3/2}\delta(\mathbf{q}')c_\Delta(R)\right]\\
&+\int dR dR' \tilde{G}(\mathbf{q})  \tilde{\mathcal{S}}(\mathbf{q},R) \tilde{G}(\mathbf{q})  \tilde{\mathcal{S}}(\mathbf{q},R') \left[\tilde{c}_b(\mathbf{q}')-(2 \pi)^{3/2}\delta(\mathbf{q}')c_\Delta(R)\right]+...
\end{split}
\label{eq:SigmaseriesqSpace}
\end{equation}
In order to determine the function $\tilde{\mathcal{S}}(\mathbf{q},R)$ we now have to  perform the configuration-average in terms of the scattering operators $\tilde{\mathcal{T}}_i(\mathbf{q},R)$, and  compare the resulting expression with eq.~(\ref{eq:SigmaseriesqSpace}).\\
By introducing the notation $\overline{f}=\int f(\mathcal{X},\mathcal{R}) p(\mathcal{X},\mathcal{R}) d\mathcal{X}d\mathcal{R}$, the configuration average of eq.~(\ref{eq:TseriesqSpace}) takes the form:
\begin{equation}
\begin{split}
&\tilde{\bar{c}}(\mathbf{q})=\tilde{c}_b(\mathbf{q})+\sum_{i =1}^n \int \dj\mathbf{q}' dR \:   \tilde{G}(\mathbf{q}) \overline{\tilde{\tilde{\mathcal{T}}}_i(\mathbf{q}-\mathbf{q}',R)} \left[\tilde{c}_b(\mathbf{q}')-(2 \pi)^{3/2}\delta(\mathbf{q}')c_\Delta(R)\right] \\
 &+\sum_{i =1}^n \sum_{j \neq i} \int \dj\mathbf{q}' \dj\mathbf{q}'' dR dR' \tilde{G}(\mathbf{q}) \overline{\tilde{\mathcal{T}}_i(\mathbf{q}-\mathbf{q}',R)  \tilde{G}(\mathbf{q}') \tilde{\mathcal{T}}_j(\mathbf{q}'-\mathbf{q}'',R')} \left[[\tilde{c}_b(\mathbf{q}'')-(2 \pi)^{3/2}\delta(\mathbf{q}'')c_\Delta(R')\right]+...
 \end{split}
\label{eq:Tseriesavg}
\end{equation}
Let us define $\tilde{\mathcal{S}}_j(\mathbf{q},R)$ as the contribution to $\tilde{\mathcal{S}}$ containing the product of exactly a number $j$ of $\tilde{\mathcal{T}}$ operators, i.e. $\tilde{\mathcal{S}}=\sum_j^\infty \tilde{\mathcal{S}}_j$.\\
By comparing eqs.~(\ref{eq:SigmaseriesqSpace}) and (\ref{eq:Tseriesavg}) we can determine each order of the self energy recursively:
\begin{equation}
\begin{split}
\tilde{\mathcal{S}}_1(\mathbf{q},R)=& \sum_{i=1}^n \int \dj\mathbf{q}' \overline{\tilde{\mathcal{T}}_i(\mathbf{q}-\mathbf{q}',R)}, \\
\tilde{\mathcal{S}}_2(\mathbf{q},R)=& \int \dj\mathbf{q}'' \dj\mathbf{q}'' dR'\sum_{i=1}^n \sum_{j\neq i} \overline{\tilde{\mathcal{T}}_i(\mathbf{q}-\mathbf{q}',R')  \tilde{G}(\mathbf{q}') \tilde{\mathcal{T}}_j(\mathbf{q}'-\mathbf{q}'',R)}\\
&- \int dR'\tilde{\mathcal{S}}_1(\mathbf{q},R') \tilde{G}(\mathbf{q})\tilde{\mathcal{S}}_1(\mathbf{q},R),\\
\tilde{\mathcal{S}}_3(\mathbf{q},R)=&\int \dj\mathbf{q}' \dj\mathbf{q}'' \dj\mathbf{q}''' dR' dR'' \times \\ &\sum_{i=1}^n \sum_{j\neq i}\sum_{k\neq j} \overline{\tilde{\mathcal{T}}_i(\mathbf{q}-\mathbf{q}',R'')  \tilde{G}(\mathbf{q}') \tilde{\mathcal{T}}_j(\mathbf{q}'-\mathbf{q}'',R'') \tilde{G}(\mathbf{q}'') \tilde{\mathcal{T}}_k(\mathbf{q}''-\mathbf{q}''',R)}\\
&-\int dR' dR''\tilde{\mathcal{S}}_1(\mathbf{q},R'') \tilde{G}(\mathbf{q}) \tilde{\mathcal{S}}_1(\mathbf{q},R') \tilde{G}(\mathbf{q}) \tilde{\mathcal{S}}_1(\mathbf{q},R)\\
&-\int dR' \Bigg[\tilde{\mathcal{S}}_1(\mathbf{q},R') \tilde{G}(\mathbf{q}) \tilde{\mathcal{S}}_2(\mathbf{q},R)+
\tilde{\mathcal{S}}_2(\mathbf{q},R') \tilde{G}(\mathbf{q}) \tilde{\mathcal{S}}_1(\mathbf{q},R)\Bigg]
\end{split}
\end{equation}
and so on. \\The general pattern is evident: $\tilde{\mathcal{S}}_n$ contains a term of the form $\overline{\tilde{\mathcal{T}}\tilde{G}\tilde{\mathcal{T}}\tilde{G}\tilde{\mathcal{T}}...}$, with respectively $n$ and $n-1$ $\tilde{\mathcal{T}}$ and $\tilde{G}$ operators, minus all the possible unique combinations of lower order $\tilde{\mathcal{S}}$ and $\tilde{G}$ terms containing exactly a number $n$ of $\tilde{\mathcal{T}}$ operators.
Let us assume that the loops are homogeneously distributed in space and that there are no correlations either in the positions or radii of different loops, allowing us to write:
\begin{equation}
 p_1(\mathbf{x},R)=n^{-1} f(R),
\end{equation}
where $f(R)dR$ is the number density of dislocation loops with radii in the range $(R,R+dR)$.  We also consider the thermodynamic limit of $n\rightarrow \infty$, with constant number density.  The first contributions to $\mathcal{S}$ can then be explicitly expressed as:
\begin{equation}
\begin{split}
\tilde{\mathcal{S}}_1(\mathbf{q},R)&=\xi_\Delta(R) f(R),\\
\tilde{\mathcal{S}}_2(\mathbf{q},R)&=-\overline{\xi_\Delta} \tilde{G}(\mathbf{q})\xi_\Delta(R) f(R), \\
\tilde{\mathcal{S}}_3(\mathbf{q},R)&=\overline{\xi_\Delta} \xi^2_\Delta(R) f(R) \int \dj\mathbf{q}'' \tilde{G}^2(\mathbf{q}).
\label{eq:selfexplicit}
\end{split}
\end{equation}

\section{Ewald summation of the diffusive interactions in a thin film geometry}
\label{app:TF}
Consider a periodic system in 3 dimensions with $N$ clusters with effective charges $\left\lbrace Q_i\right\rbrace_{i=1}^N$ and $N$ ``image'' clusters with effective charges $\left\lbrace-Q_i\right\rbrace_{i=1}^N$   per primitive cell.\\ 
The primitive cell is defined as:    $\Gamma(0)=\big\lbrace \mathbf{x}=x_1\mathbf{a}_1+x_2\mathbf{a}_2+x_3\mathbf{a}_3:-1/2<x_\alpha<1/2, \alpha=1,2,3 \big\rbrace$, with primitive vectors $\mathbf{a}_1=(L,0,0)$, $\mathbf{a}_2=(0,L,0)$, and $\mathbf{a}_3=(0,0,2H)$.\\

Let $\mathbf{m}=m_1\mathbf{a}_1+m_2\mathbf{a}_2+m_3\mathbf{a}_3$, $m_\alpha \in \mathbb{Z}$ and $\mathbf{\bar{x}}_j=(x_j,y_j,\text{sgn}(z_j)H-z_j)$ and $G_Y(\mathbf{x},\mathbf{x}';\overline{\xi_\Delta})=-e^{-\sqrt{\overline{\xi_\Delta}/D_v}}\left(4 \pi D_v |\mathbf{x}-\mathbf{x}'| \right)^{-1}$. We want to reformulate the expression:
\begin{equation}
K(\mathbf{x}_i,\mathbf{x}_j,\overline{\xi_\Delta})= \sideset{}{'}\sum_{\mathbf{m}}  G_Y(\mathbf{x}_i,\mathbf{m}+\mathbf{x}_j;\overline{\xi_\Delta})-\sum_{\mathbf{m}} G_Y(\mathbf{x}_i,\mathbf{m}+{\mathbf{\bar{x}}}_j;\overline{\xi_\Delta}) \qquad i=1,..,N
\end{equation}
as an absolutely convergent series in real and reciprocal space.\\
We first define a ``primitive cell effective Green's function'' as follows:
\begin{equation}
G^\Gamma_Y(\mathbf{x};\overline{\xi_\Delta})=\sum_{\mathbf{m}} G_Y(\mathbf{x}+\mathbf{m};\overline{\xi_\Delta}),
\end{equation}
which satisfies the equation:
\begin{equation}
(D_v \nabla^2-\overline{\xi_\Delta})G^\Gamma_Y(\mathbf{x};\overline{\xi_\Delta})= \delta^\Gamma(\mathbf{x})=\sum_\mathbf{m} \delta(\mathbf{x}+\mathbf{m}).
\end{equation}
We split the primitive cell Green's function: $G^\Gamma_Y=G^\Gamma_{Y,F}+G^\Gamma_{Y,D}$, where:
\begin{equation}
\begin{split}
&(D_v\nabla^2-\overline{\xi_\Delta})G^\Gamma_{Y,F}(\mathbf{x};\overline{\xi_\Delta})=\sum_\mathbf{m} \lambda \rho_\beta(\mathbf{x}+\mathbf{m}),\\
&(D_v\nabla^2-\overline{\xi_\Delta})G^\Gamma_{Y,D}(\mathbf{x};\overline{\xi_\Delta})=\sum_\mathbf{m} \left[ \delta(\mathbf{x}+\mathbf{m})- \lambda\rho_\beta(\mathbf{x}+\mathbf{m}) \right]
\end{split}
\end{equation}
and $\rho_\beta(\mathbf{x})$ is a normalised gaussian distribution:
\begin{equation}
\rho_\beta(\mathbf{x})=\left(\frac{\beta^2}{\pi} \right)^{\frac{3}{2}} e^{-\beta^2 |x|^2}.
\end{equation}
Let us introduce a reciprocal lattice with basis vectors: $\mathbf{b}_1=(1/L,0,0)$, $\mathbf{b}_2=(0,1/L,0)$ and $\mathbf{b}_3=(0,0,1/2H)$, and the general reciprocal space vector as: $\mathbf{k}=k_1\mathbf{b}_1+k_2\mathbf{b}_2+k_3\mathbf{b}_3$, $k_\alpha \in \mathbb{Z}$, $\alpha=1,2,3$.\\
In particular, $G^\Gamma_{Y,F}$ is more conveniently expressed in terms of reciprocal space functions. By Fourier transforming $\rho_\beta$ and using the relation $\sum_\mathbf{m} \delta(\mathbf{x}+\mathbf{m})=\sum_\mathbf{k} \exp(i \mathbf{k} \cdot \mathbf{x})$, we have:
\begin{equation}
G^\Gamma_{Y,F}(\mathbf{x},\overline{\xi_\Delta})=-\frac{\lambda}{V} \sum_\mathbf{k} \frac{\exp\left(i\mathbf{k}\cdot\mathbf{x}-k^2/4\beta^2 \right)}{D_v k^2+\overline{\xi_\Delta}},
\end{equation}
where $V$ is the volume of the primitive cell.
On the other hand, we have:
\begin{equation}
\begin{split}
G^\Gamma_{Y,D}(\mathbf{x};\overline{\xi_\Delta})=G^\Gamma_{Y}(\mathbf{x};\overline{\xi_\Delta})-G^\Gamma_{Y,F}(\mathbf{x};\overline{\xi_\Delta})=\sum_\mathbf{m} \left[G_Y(\mathbf{x}+\mathbf{m};\overline{\xi_\Delta})-\lambda \psi_\beta(\mathbf{x}+\mathbf{m}) \right],
\end{split}
\end{equation}
where the potential $\psi_\beta(\mathbf{x})$ is given by:
\begin{equation}
\begin{split}
&\psi_\beta(\mathbf{x})=\int d\mathbf{x}' G_Y(\mathbf{x},\mathbf{x}',\overline{\xi_\Delta}) \rho_\beta(\mathbf{x}')\\
=&-\frac{e^{\overline{\xi_\Delta}/4 D_v \beta^2}}{4 \pi D_v}\Bigg\lbrace \frac{e^{-\sqrt{\overline{\xi_\Delta}/D_v}|\mathbf{x}|}}{2|\mathbf{x}|}\left[\text{erf}\left(\beta |\mathbf{x}|+\frac{\sqrt{\overline{\xi_\Delta}/D_v}}{2\beta} \right)+\text{erf}\left(\beta |\mathbf{x}|-\frac{\sqrt{\overline{\xi_\Delta}/D_v}}{2\beta} \right) \right]\\
&-\sinh\left(\sqrt{\overline{\xi_\Delta}/D_v} |\mathbf{x}|\right) \text{erfc} \left( \beta |\mathbf{x}|+\frac{\sqrt{\overline{\xi_\Delta}/D_v}}{2\beta}\right) \Bigg\rbrace.
\end{split}
\end{equation}
The parameter $\lambda$ should be chosen in way that minimises the long tail of $G^\Gamma_{Y,D}$. In particular, as $|\mathbf{x}|\rightarrow\infty$, $\psi_\beta(\mathbf{x}) \simeq \frac{e^{\overline{\xi_\Delta}^2/4 \beta^2-\overline{\xi_\Delta}|\mathbf{x}|}}{|\mathbf{x}|}$, therefore the optimal choice is $\lambda=e^{-\overline{\xi_\Delta}^2/4 \beta^2}$ and $G^\Gamma_{Y}$ assumes the form:
\begin{equation}
\begin{split}
&G^\Gamma_{Y}(\mathbf{x};\overline{\xi_\Delta})=-\sum_\mathbf{k} \frac{\exp\left[-(k^2+\overline{\xi_\Delta}/D_v)/4\beta^2 \right]}{ V(D_v k^2+\overline{\xi_\Delta})}e^{i\mathbf{k}\cdot\mathbf{x}}\\
&-\frac{1}{8 \pi D_v} \sum_\mathbf{m} \frac{1}{|\mathbf{x}+\mathbf{m}|} \Bigg[ \text{erfc}\left(\beta |\mathbf{x}+\mathbf{m}|+\frac{\sqrt{\overline{\xi_\Delta}/D_v}}{2\beta} \right)e^{\sqrt{\overline{\xi_\Delta}/D_v}|\mathbf{x}+\mathbf{m}|}\\
&+\text{erfc}\left(\beta |\mathbf{x}+\mathbf{m}|-\frac{\sqrt{\overline{\xi_\Delta}/D_v}}{2\beta} \right)e^{-\sqrt{\overline{\xi_\Delta}/D_v}|\mathbf{x}+\mathbf{m}|}    \Bigg].
\end{split}
\end{equation}
We now have to remove the spurious self interactions:
\begin{equation}
\begin{split}
&\sideset{}{'}\sum_{\mathbf{m}}  G_Y(\mathbf{x}_i,\mathbf{m}+\mathbf{x}_i;\overline{\xi_\Delta})=
\lim_{\mathbf{x}\rightarrow \mathbf{x}_i}\left[ G^\Gamma_{Y}(\mathbf{x},\mathbf{x}_i;\overline{\xi_\Delta})-G_{Y}(\mathbf{x},\mathbf{x}_i;\overline{\xi_\Delta}) \right]\\
&=- \sum_\mathbf{k} \frac{\exp\left[-(k^2+\overline{\xi_\Delta}/D_v)/4\beta^2 \right]}{ V(D_v k^2+\overline{\xi_\Delta})}\\
&-\frac{1}{8 \pi D_v} \sum_\mathbf{m\neq 0} \frac{1}{|\mathbf{m}|} \Bigg[ \text{erfc}\left(\beta |\mathbf{m}|+\frac{\sqrt{\overline{\xi_\Delta}/D_v}}{2\beta} \right)e^{\sqrt{\overline{\xi_\Delta}/D_v}|\mathbf{m}|}+\text{erfc}\left(\beta |\mathbf{m}|-\frac{\sqrt{\overline{\xi_\Delta}/D_v}}{2\beta} \right)e^{-\sqrt{\overline{\xi_\Delta}/D_v}|\mathbf{m}|}    \Bigg]\\
&-\frac{1}{4 \pi D_v} \lim_{r \rightarrow 0} \frac{1}{r} \left[\text{erfc}\left(\beta r+\frac{\sqrt{\overline{\xi_\Delta}/D_v}}{2\beta} \right)\frac{e^{\sqrt{\overline{\xi_\Delta}/D_v}r}}{2}+\text{erfc}\left(\beta r-\frac{\sqrt{\overline{\xi_\Delta}/D_v}}{2\beta} \right)\frac{e^{-\sqrt{\overline{\xi_\Delta}/D_v}r}}{2}-e^{-\sqrt{\overline{\xi_\Delta}/D_v}r}\right].
\end{split}
\end{equation}
Expanding to first order with respect to $r$:
\begin{equation}
\begin{split}
&\text{erfc}\left(\beta r+\frac{\sqrt{\overline{\xi_\Delta}/D_v}}{2\beta} \right)\frac{e^{\sqrt{\overline{\xi_\Delta}/D_v}r}}{2}+\text{erfc}\left(\beta r-\frac{\sqrt{\overline{\xi_\Delta}/D_v}}{2\beta} \right)\frac{e^{-\sqrt{\overline{\xi_\Delta}/D_v}r}}{2}-e^{-\sqrt{\overline{\xi_\Delta}/D_v}r}\\
&\approx \left[ \sqrt{\frac{\overline{\xi_\Delta}}{D_v}} \text{erfc} \left(\frac{\sqrt{\overline{\xi_\Delta}}}{2 \sqrt{D_v} \beta}\right)-\frac{2 e^{-\frac{\overline{\xi_\Delta}}{4 D_v \beta^2}}}{\sqrt{\pi}} \right] r + O(r^2).
\end{split}
\end{equation}
We can now combine all the terms, yielding to an absolutely convergent series in real and reciprocal space:
\begin{equation}
\begin{split}
&K(\mathbf{x}_i,\mathbf{x}_j;\epsilon)=-\sum_\mathbf{k} \frac{\exp\left[-(k^2+\epsilon^2)/4\beta^2 \right]}{ VD_v( k^2+\epsilon^2)}e^{i\mathbf{k}\cdot\mathbf{x}_i}\left[e^{-i \mathbf{k}\cdot \mathbf{x}_j}-e^{-i \mathbf{k}\cdot \mathbf{\bar{x}}_j} \right]\\
&-\frac{1}{8 \pi D_v} \Bigg\lbrace \sideset{}{'}\sum_\mathbf{m} \frac{1}{|\mathbf{x}_i-\mathbf{x}_j+\mathbf{m}|} \Bigg[ \text{erfc}\left(\beta |\mathbf{x}_i-\mathbf{x}_j+\mathbf{m}|+\frac{\epsilon}{2\beta} \right)e^{\epsilon|\mathbf{x}_i-\mathbf{x}_j+\mathbf{m}|}\\
&+\text{erfc}\left(\beta |\mathbf{x}_i-\mathbf{x}_j+\mathbf{m}|-\frac{\epsilon}{2\beta} \right)e^{-\epsilon|\mathbf{x}_i-\mathbf{x}_j+\mathbf{m}|}    \Bigg]\\
&-\sum_\mathbf{m} \frac{1}{|\mathbf{x}_i-\mathbf{\bar{x}}_j+\mathbf{m}|} \Bigg[ \text{erfc}\left(\beta |\mathbf{x}_i-\mathbf{\bar{x}}_j+\mathbf{m}|+\frac{\epsilon}{2\beta} \right)e^{\epsilon|\mathbf{x}_i-\mathbf{\bar{x}}_j+\mathbf{m}|}\\
&+\text{erfc}\left(\beta |\mathbf{x}_i-\mathbf{\bar{x}}_j+\mathbf{m}|-\frac{\epsilon}{2\beta} \right)e^{-\epsilon|\mathbf{x}_i-\mathbf{\bar{x}}_j+\mathbf{m}|}    \Bigg] \Bigg \rbrace+\frac{\delta_{ij}}{4 \pi D_v} \left[\frac{2 \beta e^{-\frac{\epsilon^2}{4 \beta^2}}}{\sqrt{\pi}} - \epsilon \: \text{erfc} \left(\frac{\epsilon}{2 \beta}\right)\right].
\end{split}
\end{equation}

\printnomenclature

\newpage
\bibliography{Biblio}

\begin{thebibliography}{32}%
\makeatletter
\providecommand \@ifxundefined [1]{%
 \@ifx{#1\undefined}
}%
\providecommand \@ifnum [1]{%
 \ifnum #1\expandafter \@firstoftwo
 \else \expandafter \@secondoftwo
 \fi
}%
\providecommand \@ifx [1]{%
 \ifx #1\expandafter \@firstoftwo
 \else \expandafter \@secondoftwo
 \fi
}%
\providecommand \natexlab [1]{#1}%
\providecommand \enquote  [1]{``#1''}%
\providecommand \bibnamefont  [1]{#1}%
\providecommand \bibfnamefont [1]{#1}%
\providecommand \citenamefont [1]{#1}%
\providecommand \href@noop [0]{\@secondoftwo}%
\providecommand \href [0]{\begingroup \@sanitize@url \@href}%
\providecommand \@href[1]{\@@startlink{#1}\@@href}%
\providecommand \@@href[1]{\endgroup#1\@@endlink}%
\providecommand \@sanitize@url [0]{\catcode `\\12\catcode `\$12\catcode
  `\&12\catcode `\#12\catcode `\^12\catcode `\_12\catcode `\%12\relax}%
\providecommand \@@startlink[1]{}%
\providecommand \@@endlink[0]{}%
\providecommand \url  [0]{\begingroup\@sanitize@url \@url }%
\providecommand \@url [1]{\endgroup\@href {#1}{\urlprefix }}%
\providecommand \urlprefix  [0]{URL }%
\providecommand \Eprint [0]{\href }%
\providecommand \doibase [0]{http://dx.doi.org/}%
\providecommand \selectlanguage [0]{\@gobble}%
\providecommand \bibinfo  [0]{\@secondoftwo}%
\providecommand \bibfield  [0]{\@secondoftwo}%
\providecommand \translation [1]{[#1]}%
\providecommand \BibitemOpen [0]{}%
\providecommand \bibitemStop [0]{}%
\providecommand \bibitemNoStop [0]{.\EOS\space}%
\providecommand \EOS [0]{\spacefactor3000\relax}%
\providecommand \BibitemShut  [1]{\csname bibitem#1\endcsname}%
\let\auto@bib@innerbib\@empty
\bibitem [{\citenamefont {{Van Renterghem}}\ and\ \citenamefont
  {Uytdenhouwen}(2016)}]{VanRenterghem2016}%
  \BibitemOpen
  \bibfield  {author} {\bibinfo {author} {\bibfnamefont {W.}~\bibnamefont {{Van
  Renterghem}}}\ and\ \bibinfo {author} {\bibfnamefont {I.}~\bibnamefont
  {Uytdenhouwen}},\ }\href@noop {} {\bibfield  {journal} {\bibinfo  {journal}
  {Journal of Nuclear Materials}\ }\textbf {\bibinfo {volume} {477}},\ \bibinfo
  {pages} {77} (\bibinfo {year} {2016})}\BibitemShut {NoStop}%
\bibitem [{\citenamefont {Ferroni}\ \emph {et~al.}(2015)\citenamefont
  {Ferroni}, \citenamefont {Yi}, \citenamefont {Arakawa}, \citenamefont
  {Fitzgerald}, \citenamefont {Edmondson},\ and\ \citenamefont
  {Roberts}}]{Ferroni2015}%
  \BibitemOpen
  \bibfield  {author} {\bibinfo {author} {\bibfnamefont {F.}~\bibnamefont
  {Ferroni}}, \bibinfo {author} {\bibfnamefont {X.}~\bibnamefont {Yi}},
  \bibinfo {author} {\bibfnamefont {K.}~\bibnamefont {Arakawa}}, \bibinfo
  {author} {\bibfnamefont {S.~P.}\ \bibnamefont {Fitzgerald}}, \bibinfo
  {author} {\bibfnamefont {P.~D.}\ \bibnamefont {Edmondson}}, \ and\ \bibinfo
  {author} {\bibfnamefont {S.~G.}\ \bibnamefont {Roberts}},\ }\href@noop {}
  {\bibfield  {journal} {\bibinfo  {journal} {Acta Materialia}\ }\textbf
  {\bibinfo {volume} {90}},\ \bibinfo {pages} {380} (\bibinfo {year}
  {2015})}\BibitemShut {NoStop}%
\bibitem [{\citenamefont {Fukumoto}\ \emph {et~al.}(2013)\citenamefont
  {Fukumoto}, \citenamefont {Iwasaki},\ and\ \citenamefont
  {Xu}}]{Fukumoto2013}%
  \BibitemOpen
  \bibfield  {author} {\bibinfo {author} {\bibfnamefont {K.}~\bibnamefont
  {Fukumoto}}, \bibinfo {author} {\bibfnamefont {M.}~\bibnamefont {Iwasaki}}, \
  and\ \bibinfo {author} {\bibfnamefont {Q.}~\bibnamefont {Xu}},\ }\href@noop
  {} {\bibfield  {journal} {\bibinfo  {journal} {Journal of Nuclear Materials}\
  }\textbf {\bibinfo {volume} {442}},\ \bibinfo {pages} {360} (\bibinfo {year}
  {2013})}\BibitemShut {NoStop}%
\bibitem [{\citenamefont {Nagasaka}\ \emph {et~al.}(2005)\citenamefont
  {Nagasaka}, \citenamefont {Muroga}, \citenamefont {Watanabe}, \citenamefont
  {Yamasaki}, \citenamefont {Heo}, \citenamefont {Shinozaki},\ and\
  \citenamefont {Narui}}]{Nagasaka2005}%
  \BibitemOpen
  \bibfield  {author} {\bibinfo {author} {\bibfnamefont {T.}~\bibnamefont
  {Nagasaka}}, \bibinfo {author} {\bibfnamefont {T.}~\bibnamefont {Muroga}},
  \bibinfo {author} {\bibfnamefont {H.}~\bibnamefont {Watanabe}}, \bibinfo
  {author} {\bibfnamefont {K.}~\bibnamefont {Yamasaki}}, \bibinfo {author}
  {\bibfnamefont {N.-J.}\ \bibnamefont {Heo}}, \bibinfo {author} {\bibfnamefont
  {K.}~\bibnamefont {Shinozaki}}, \ and\ \bibinfo {author} {\bibfnamefont
  {M.}~\bibnamefont {Narui}},\ }\href@noop {} {\bibfield  {journal} {\bibinfo
  {journal} {Materials Transactions}\ }\textbf {\bibinfo {volume} {46}},\
  \bibinfo {pages} {498} (\bibinfo {year} {2005})}\BibitemShut {NoStop}%
\bibitem [{\citenamefont {Byun}\ \emph {et~al.}(2014)\citenamefont {Byun},
  \citenamefont {Baek}, \citenamefont {Anderoglu}, \citenamefont {Maloy},\ and\
  \citenamefont {Toloczko}}]{Byun2014}%
  \BibitemOpen
  \bibfield  {author} {\bibinfo {author} {\bibfnamefont {T.~S.}\ \bibnamefont
  {Byun}}, \bibinfo {author} {\bibfnamefont {J.~H.}\ \bibnamefont {Baek}},
  \bibinfo {author} {\bibfnamefont {O.}~\bibnamefont {Anderoglu}}, \bibinfo
  {author} {\bibfnamefont {S.~A.}\ \bibnamefont {Maloy}}, \ and\ \bibinfo
  {author} {\bibfnamefont {M.~B.}\ \bibnamefont {Toloczko}},\ }\href@noop {}
  {\bibfield  {journal} {\bibinfo  {journal} {Journal of Nuclear Materials}\
  }\textbf {\bibinfo {volume} {449}},\ \bibinfo {pages} {263} (\bibinfo {year}
  {2014})}\BibitemShut {NoStop}%
\bibitem [{\citenamefont {Rovelli}\ \emph {et~al.}(2017)\citenamefont
  {Rovelli}, \citenamefont {Dudarev},\ and\ \citenamefont
  {Sutton}}]{Rovelli2017}%
  \BibitemOpen
  \bibfield  {author} {\bibinfo {author} {\bibfnamefont {I.}~\bibnamefont
  {Rovelli}}, \bibinfo {author} {\bibfnamefont {S.~L.}\ \bibnamefont
  {Dudarev}}, \ and\ \bibinfo {author} {\bibfnamefont {A.~P.}\ \bibnamefont
  {Sutton}},\ }\href@noop {} {\bibfield  {journal} {\bibinfo  {journal}
  {Journal of the Mechanics and Physics of Solids}\ }\textbf {\bibinfo {volume}
  {103}},\ \bibinfo {pages} {121} (\bibinfo {year} {2017})}\BibitemShut
  {NoStop}%
\bibitem [{\citenamefont {Gu}\ \emph {et~al.}(2015)\citenamefont {Gu},
  \citenamefont {Xiang}, \citenamefont {Quek},\ and\ \citenamefont
  {Srolovitz}}]{Gu2015}%
  \BibitemOpen
  \bibfield  {author} {\bibinfo {author} {\bibfnamefont {Y.}~\bibnamefont
  {Gu}}, \bibinfo {author} {\bibfnamefont {Y.}~\bibnamefont {Xiang}}, \bibinfo
  {author} {\bibfnamefont {S.~S.}\ \bibnamefont {Quek}}, \ and\ \bibinfo
  {author} {\bibfnamefont {D.~J.}\ \bibnamefont {Srolovitz}},\ }\href@noop {}
  {\bibfield  {journal} {\bibinfo  {journal} {Journal of the Mechanics and
  Physics of Solids}\ }\textbf {\bibinfo {volume} {83}},\ \bibinfo {pages}
  {319} (\bibinfo {year} {2015})}\BibitemShut {NoStop}%
\bibitem [{\citenamefont {Sand}\ \emph {et~al.}(2013)\citenamefont {Sand},
  \citenamefont {Dudarev},\ and\ \citenamefont {Nordlund}}]{Sand2013}%
  \BibitemOpen
  \bibfield  {author} {\bibinfo {author} {\bibfnamefont {A.~E.}\ \bibnamefont
  {Sand}}, \bibinfo {author} {\bibfnamefont {S.~L.}\ \bibnamefont {Dudarev}}, \
  and\ \bibinfo {author} {\bibfnamefont {K.}~\bibnamefont {Nordlund}},\
  }\href@noop {} {\bibfield  {journal} {\bibinfo  {journal} {Europhysics
  Letters}\ }\textbf {\bibinfo {volume} {103}},\ \bibinfo {pages} {46003}
  (\bibinfo {year} {2013})}\BibitemShut {NoStop}%
\bibitem [{\citenamefont {Yi}\ \emph {et~al.}(2015)\citenamefont {Yi},
  \citenamefont {Sand}, \citenamefont {Mason}, \citenamefont {Kirk},
  \citenamefont {Roberts}, \citenamefont {Nordlund},\ and\ \citenamefont
  {Dudarev}}]{Yi2015}%
  \BibitemOpen
  \bibfield  {author} {\bibinfo {author} {\bibfnamefont {X.}~\bibnamefont
  {Yi}}, \bibinfo {author} {\bibfnamefont {A.~E.}\ \bibnamefont {Sand}},
  \bibinfo {author} {\bibfnamefont {D.~R.}\ \bibnamefont {Mason}}, \bibinfo
  {author} {\bibfnamefont {M.~A.}\ \bibnamefont {Kirk}}, \bibinfo {author}
  {\bibfnamefont {S.~G.}\ \bibnamefont {Roberts}}, \bibinfo {author}
  {\bibfnamefont {K.}~\bibnamefont {Nordlund}}, \ and\ \bibinfo {author}
  {\bibfnamefont {S.~L.}\ \bibnamefont {Dudarev}},\ }\href@noop {} {\bibfield
  {journal} {\bibinfo  {journal} {Europhysics Letters}\ }\textbf {\bibinfo
  {volume} {110}},\ \bibinfo {pages} {36001} (\bibinfo {year}
  {2015})}\BibitemShut {NoStop}%
\bibitem [{\citenamefont {Beavan}\ \emph {et~al.}(1971)\citenamefont {Beavan},
  \citenamefont {Scanlan},\ and\ \citenamefont {Seidman}}]{Beavan1971}%
  \BibitemOpen
  \bibfield  {author} {\bibinfo {author} {\bibfnamefont {L.~A.}\ \bibnamefont
  {Beavan}}, \bibinfo {author} {\bibfnamefont {R.~M.}\ \bibnamefont {Scanlan}},
  \ and\ \bibinfo {author} {\bibfnamefont {D.~N.}\ \bibnamefont {Seidman}},\
  }\href@noop {} {\bibfield  {journal} {\bibinfo  {journal} {Acta
  Metallurgica}\ }\textbf {\bibinfo {volume} {19}},\ \bibinfo {pages} {1339}
  (\bibinfo {year} {1971})}\BibitemShut {NoStop}%
\bibitem [{\citenamefont {Liu}\ \emph {et~al.}(2017)\citenamefont {Liu},
  \citenamefont {He}, \citenamefont {Zhai}, \citenamefont {Tyburska-Püschel},
  \citenamefont {Voyles}, \citenamefont {Sridharan}, \citenamefont {Morgan},\
  and\ \citenamefont {Szlufarska}}]{Liu2017}%
  \BibitemOpen
  \bibfield  {author} {\bibinfo {author} {\bibfnamefont {C.}~\bibnamefont
  {Liu}}, \bibinfo {author} {\bibfnamefont {L.}~\bibnamefont {He}}, \bibinfo
  {author} {\bibfnamefont {Y.}~\bibnamefont {Zhai}}, \bibinfo {author}
  {\bibfnamefont {B.}~\bibnamefont {Tyburska-Püschel}}, \bibinfo {author}
  {\bibfnamefont {P.~M.}\ \bibnamefont {Voyles}}, \bibinfo {author}
  {\bibfnamefont {K.}~\bibnamefont {Sridharan}}, \bibinfo {author}
  {\bibfnamefont {D.}~\bibnamefont {Morgan}}, \ and\ \bibinfo {author}
  {\bibfnamefont {I.}~\bibnamefont {Szlufarska}},\ }\href@noop {} {\bibfield
  {journal} {\bibinfo  {journal} {Acta Materialia}\ }\textbf {\bibinfo {volume}
  {125}},\ \bibinfo {pages} {377} (\bibinfo {year} {2017})}\BibitemShut
  {NoStop}%
\bibitem [{\citenamefont {Edwards}(1958)}]{Edwards1958}%
  \BibitemOpen
  \bibfield  {author} {\bibinfo {author} {\bibfnamefont {S.~F.}\ \bibnamefont
  {Edwards}},\ }\href@noop {} {\bibfield  {journal} {\bibinfo  {journal}
  {Philosophical Magazine}\ }\textbf {\bibinfo {volume} {3}},\ \bibinfo {pages}
  {1020} (\bibinfo {year} {1958})}\BibitemShut {NoStop}%
\bibitem [{\citenamefont {Salin}\ and\ \citenamefont
  {Caillol}(2000)}]{Salin2000}%
  \BibitemOpen
  \bibfield  {author} {\bibinfo {author} {\bibfnamefont {G.}~\bibnamefont
  {Salin}}\ and\ \bibinfo {author} {\bibfnamefont {J.~M.}\ \bibnamefont
  {Caillol}},\ }\href@noop {} {\bibfield  {journal} {\bibinfo  {journal}
  {Journal of Chemical Physics}\ }\textbf {\bibinfo {volume} {113}},\ \bibinfo
  {pages} {10459} (\bibinfo {year} {2000})}\BibitemShut {NoStop}%
\bibitem [{\citenamefont {Vitos}\ \emph {et~al.}(1998)\citenamefont {Vitos},
  \citenamefont {Ruban}, \citenamefont {Skriver},\ and\ \citenamefont
  {Koll{\'{a}}r}}]{Vitos1998}%
  \BibitemOpen
  \bibfield  {author} {\bibinfo {author} {\bibfnamefont {L.}~\bibnamefont
  {Vitos}}, \bibinfo {author} {\bibfnamefont {A.~V.}\ \bibnamefont {Ruban}},
  \bibinfo {author} {\bibfnamefont {H.~L.}\ \bibnamefont {Skriver}}, \ and\
  \bibinfo {author} {\bibfnamefont {J.}~\bibnamefont {Koll{\'{a}}r}},\
  }\href@noop {} {\bibfield  {journal} {\bibinfo  {journal} {Surface Science}\
  }\textbf {\bibinfo {volume} {411}},\ \bibinfo {pages} {186} (\bibinfo {year}
  {1998})}\BibitemShut {NoStop}%
\bibitem [{\citenamefont {Kaye}\ and\ \citenamefont {Laby}(1995)}]{KayeLaby}%
  \BibitemOpen
  \bibfield  {author} {\bibinfo {author} {\bibfnamefont {G.~W.~C.}\
  \bibnamefont {Kaye}}\ and\ \bibinfo {author} {\bibfnamefont {T.~H.}\
  \bibnamefont {Laby}},\ }\href@noop {} {\emph {\bibinfo {title} {Tables of
  Physical and Chemical Constants}}},\ \bibinfo {edition} {16th}\ ed.\
  (\bibinfo  {publisher} {Longman},\ \bibinfo {year} {1995})\BibitemShut
  {NoStop}%
\bibitem [{\citenamefont {Dudarev}(2013)}]{DudarevDFT2013}%
  \BibitemOpen
  \bibfield  {author} {\bibinfo {author} {\bibfnamefont {S.~L.}\ \bibnamefont
  {Dudarev}},\ }\href@noop {} {\bibfield  {journal} {\bibinfo  {journal}
  {Annual Review of Materials Research}\ }\textbf {\bibinfo {volume} {43}},\
  \bibinfo {pages} {35} (\bibinfo {year} {2013})}\BibitemShut {NoStop}%
\bibitem [{\citenamefont {Mundy}\ \emph {et~al.}(1978)\citenamefont {Mundy},
  \citenamefont {Rothman}, \citenamefont {Lam}, \citenamefont {Hoff},\ and\
  \citenamefont {Nowicki}}]{Mundy1978}%
  \BibitemOpen
  \bibfield  {author} {\bibinfo {author} {\bibfnamefont {J.~N.}\ \bibnamefont
  {Mundy}}, \bibinfo {author} {\bibfnamefont {S.~J.}\ \bibnamefont {Rothman}},
  \bibinfo {author} {\bibfnamefont {N.~Q.}\ \bibnamefont {Lam}}, \bibinfo
  {author} {\bibfnamefont {H.~A.}\ \bibnamefont {Hoff}}, \ and\ \bibinfo
  {author} {\bibfnamefont {L.~J.}\ \bibnamefont {Nowicki}},\ }\href@noop {}
  {\bibfield  {journal} {\bibinfo  {journal} {Physical Review B}\ }\textbf
  {\bibinfo {volume} {18}},\ \bibinfo {pages} {6566} (\bibinfo {year}
  {1978})}\BibitemShut {NoStop}%
\bibitem [{\citenamefont {Iijima}\ \emph {et~al.}(1988)\citenamefont {Iijima},
  \citenamefont {K.},\ and\ \citenamefont {K.}}]{Iijima1988}%
  \BibitemOpen
  \bibfield  {author} {\bibinfo {author} {\bibfnamefont {Y.}~\bibnamefont
  {Iijima}}, \bibinfo {author} {\bibfnamefont {K.}~\bibnamefont {K.}}, \ and\
  \bibinfo {author} {\bibfnamefont {H.}~\bibnamefont {K.}},\ }\href@noop {}
  {\bibfield  {journal} {\bibinfo  {journal} {Acta Metallurgica}\ }\textbf
  {\bibinfo {volume} {36}},\ \bibinfo {pages} {2811} (\bibinfo {year}
  {1988})}\BibitemShut {NoStop}%
\bibitem [{\citenamefont {Dupouy}\ \emph {et~al.}(1966)\citenamefont {Dupouy},
  \citenamefont {Mathie},\ and\ \citenamefont {Adda}}]{Dupouy1966}%
  \BibitemOpen
  \bibfield  {author} {\bibinfo {author} {\bibfnamefont {J.~M.}\ \bibnamefont
  {Dupouy}}, \bibinfo {author} {\bibfnamefont {J.}~\bibnamefont {Mathie}}, \
  and\ \bibinfo {author} {\bibfnamefont {Y.}~\bibnamefont {Adda}},\ }\href@noop
  {} {\bibfield  {journal} {\bibinfo  {journal} {Memoires Scientifiques de la
  Revue de Metallurgie}\ }\textbf {\bibinfo {volume} {63}},\ \bibinfo {pages}
  {481} (\bibinfo {year} {1966})}\BibitemShut {NoStop}%
\bibitem [{\citenamefont {Glensk}\ \emph {et~al.}(2014)\citenamefont {Glensk},
  \citenamefont {Grabowski}, \citenamefont {Hickel},\ and\ \citenamefont
  {Neugebauer}}]{Glensk2014}%
  \BibitemOpen
  \bibfield  {author} {\bibinfo {author} {\bibfnamefont {A.}~\bibnamefont
  {Glensk}}, \bibinfo {author} {\bibfnamefont {B.}~\bibnamefont {Grabowski}},
  \bibinfo {author} {\bibfnamefont {T.}~\bibnamefont {Hickel}}, \ and\ \bibinfo
  {author} {\bibfnamefont {J.}~\bibnamefont {Neugebauer}},\ }\href@noop {}
  {\bibfield  {journal} {\bibinfo  {journal} {Physical Review X}\ }\textbf
  {\bibinfo {volume} {4}},\ \bibinfo {pages} {011018} (\bibinfo {year}
  {2014})}\BibitemShut {NoStop}%
\bibitem [{\citenamefont {Hirth}\ and\ \citenamefont
  {Lothe}(1982)}]{Hirth1982}%
  \BibitemOpen
  \bibfield  {author} {\bibinfo {author} {\bibfnamefont {J.~P.}\ \bibnamefont
  {Hirth}}\ and\ \bibinfo {author} {\bibfnamefont {J.}~\bibnamefont {Lothe}},\
  }\href@noop {} {\emph {\bibinfo {title} {Theory of dislocations}}},\ \bibinfo
  {edition} {2nd}\ ed.\ (\bibinfo  {publisher} {Wiley, New York},\ \bibinfo
  {year} {1982})\BibitemShut {NoStop}%
\bibitem [{\citenamefont {Landau}\ and\ \citenamefont
  {Lifshitz}(1970)}]{Landau1970}%
  \BibitemOpen
  \bibfield  {author} {\bibinfo {author} {\bibfnamefont {L.~D.}\ \bibnamefont
  {Landau}}\ and\ \bibinfo {author} {\bibfnamefont {E.~M.}\ \bibnamefont
  {Lifshitz}},\ }\href@noop {} {\emph {\bibinfo {title} {Theory of
  elasticity}}},\ \bibinfo {edition} {2nd}\ ed.\ (\bibinfo  {publisher}
  {Pargamon Press, Oxford, England, U.K.},\ \bibinfo {year} {1970})\BibitemShut
  {NoStop}%
\bibitem [{\citenamefont {Trinkaus}(1972)}]{Trinkaus1972}%
  \BibitemOpen
  \bibfield  {author} {\bibinfo {author} {\bibfnamefont {H.}~\bibnamefont
  {Trinkaus}},\ }\href@noop {} {\bibfield  {journal} {\bibinfo  {journal}
  {Physica Status Solidi (b)}\ }\textbf {\bibinfo {volume} {54}},\ \bibinfo
  {pages} {209} (\bibinfo {year} {1972})}\BibitemShut {NoStop}%
\bibitem [{\citenamefont {Dudarev}(2000)}]{Dudarev2000}%
  \BibitemOpen
  \bibfield  {author} {\bibinfo {author} {\bibfnamefont {S.~L.}\ \bibnamefont
  {Dudarev}},\ }\href@noop {} {\bibfield  {journal} {\bibinfo  {journal}
  {Physical Review B}\ }\textbf {\bibinfo {volume} {62}},\ \bibinfo {pages}
  {9325} (\bibinfo {year} {2000})}\BibitemShut {NoStop}%
\bibitem [{\citenamefont {Gilbert}\ \emph {et~al.}(2008)\citenamefont
  {Gilbert}, \citenamefont {Dudarev}, \citenamefont {Derlet},\ and\
  \citenamefont {Pettifor}}]{Gilbert2008}%
  \BibitemOpen
  \bibfield  {author} {\bibinfo {author} {\bibfnamefont {M.~R.}\ \bibnamefont
  {Gilbert}}, \bibinfo {author} {\bibfnamefont {S.~L.}\ \bibnamefont
  {Dudarev}}, \bibinfo {author} {\bibfnamefont {P.~M.}\ \bibnamefont {Derlet}},
  \ and\ \bibinfo {author} {\bibfnamefont {D.~G.}\ \bibnamefont {Pettifor}},\
  }\href@noop {} {\bibfield  {journal} {\bibinfo  {journal} {Journal of
  Physics: Condensed Matter}\ }\textbf {\bibinfo {volume} {20}},\ \bibinfo
  {pages} {345214} (\bibinfo {year} {2008})}\BibitemShut {NoStop}%
\bibitem [{\citenamefont {Alexander}\ \emph {et~al.}(2016)\citenamefont
  {Alexander}, \citenamefont {Proville}, \citenamefont {Willaime},
  \citenamefont {Arakawa}, \citenamefont {Gilbert},\ and\ \citenamefont
  {Dudarev}}]{Alexander2016}%
  \BibitemOpen
  \bibfield  {author} {\bibinfo {author} {\bibfnamefont {R.}~\bibnamefont
  {Alexander}}, \bibinfo {author} {\bibfnamefont {L.}~\bibnamefont {Proville}},
  \bibinfo {author} {\bibfnamefont {F.}~\bibnamefont {Willaime}}, \bibinfo
  {author} {\bibfnamefont {K.}~\bibnamefont {Arakawa}}, \bibinfo {author}
  {\bibfnamefont {M.~R.}\ \bibnamefont {Gilbert}}, \ and\ \bibinfo {author}
  {\bibfnamefont {S.~L.}\ \bibnamefont {Dudarev}},\ }\href@noop {} {\bibfield
  {journal} {\bibinfo  {journal} {Physical Review B}\ }\textbf {\bibinfo
  {volume} {94}},\ \bibinfo {pages} {024103} (\bibinfo {year}
  {2016})}\BibitemShut {NoStop}%
\bibitem [{\citenamefont {Dudarev}\ \emph {et~al.}(2004)\citenamefont
  {Dudarev}, \citenamefont {Semenov},\ and\ \citenamefont {Woo}}]{Dudarev2004}%
  \BibitemOpen
  \bibfield  {author} {\bibinfo {author} {\bibfnamefont {S.~L.}\ \bibnamefont
  {Dudarev}}, \bibinfo {author} {\bibfnamefont {A.~A.}\ \bibnamefont
  {Semenov}}, \ and\ \bibinfo {author} {\bibfnamefont {C.~H.}\ \bibnamefont
  {Woo}},\ }\href@noop {} {\bibfield  {journal} {\bibinfo  {journal} {Physical
  Review B}\ }\textbf {\bibinfo {volume} {70}},\ \bibinfo {pages} {094115}
  (\bibinfo {year} {2004})}\BibitemShut {NoStop}%
\bibitem [{\citenamefont {Marqusee}(1984)}]{Marqusee1984}%
  \BibitemOpen
  \bibfield  {author} {\bibinfo {author} {\bibfnamefont {J.~A.}\ \bibnamefont
  {Marqusee}},\ }\href@noop {} {\bibfield  {journal} {\bibinfo  {journal}
  {Journal of Chemical Physics}\ }\textbf {\bibinfo {volume} {81}},\ \bibinfo
  {pages} {976} (\bibinfo {year} {1984})}\BibitemShut {NoStop}%
\bibitem [{\citenamefont {Fradkov}\ \emph {et~al.}(1996)\citenamefont
  {Fradkov}, \citenamefont {Glicksman},\ and\ \citenamefont
  {Marsh}}]{Fradkov1996}%
  \BibitemOpen
  \bibfield  {author} {\bibinfo {author} {\bibfnamefont {V.~E.}\ \bibnamefont
  {Fradkov}}, \bibinfo {author} {\bibfnamefont {M.~E.}\ \bibnamefont
  {Glicksman}}, \ and\ \bibinfo {author} {\bibfnamefont {S.~P.}\ \bibnamefont
  {Marsh}},\ }\href@noop {} {\bibfield  {journal} {\bibinfo  {journal}
  {Physical review E}\ }\textbf {\bibinfo {volume} {53}},\ \bibinfo {pages}
  {3925} (\bibinfo {year} {1996})}\BibitemShut {NoStop}%
\bibitem [{\citenamefont {Wang}\ \emph {et~al.}(2004)\citenamefont {Wang},
  \citenamefont {Glicksman},\ and\ \citenamefont {Rajan}}]{Wang2004}%
  \BibitemOpen
  \bibfield  {author} {\bibinfo {author} {\bibfnamefont {K.~G.}\ \bibnamefont
  {Wang}}, \bibinfo {author} {\bibfnamefont {M.~E.}\ \bibnamefont {Glicksman}},
  \ and\ \bibinfo {author} {\bibfnamefont {K.}~\bibnamefont {Rajan}},\
  }\href@noop {} {\bibfield  {journal} {\bibinfo  {journal} {Physical review
  E}\ }\textbf {\bibinfo {volume} {69}},\ \bibinfo {pages} {1} (\bibinfo {year}
  {2004})}\BibitemShut {NoStop}%
\bibitem [{\citenamefont {Wang}(2008)}]{Wang2008}%
  \BibitemOpen
  \bibfield  {author} {\bibinfo {author} {\bibfnamefont {K.~G.}\ \bibnamefont
  {Wang}},\ }\href@noop {} {\bibfield  {journal} {\bibinfo  {journal} {Physica
  A}\ }\textbf {\bibinfo {volume} {387}},\ \bibinfo {pages} {3084} (\bibinfo
  {year} {2008})}\BibitemShut {NoStop}%
\bibitem [{\citenamefont {Yi}\ \emph {et~al.}(2013)\citenamefont {Yi},
  \citenamefont {Jenkins}, \citenamefont {Briceno}, \citenamefont {Roberts},
  \citenamefont {Zhou},\ and\ \citenamefont {Kirk}}]{Yi2013}%
  \BibitemOpen
  \bibfield  {author} {\bibinfo {author} {\bibfnamefont {X.}~\bibnamefont
  {Yi}}, \bibinfo {author} {\bibfnamefont {M.~L.}\ \bibnamefont {Jenkins}},
  \bibinfo {author} {\bibfnamefont {M.}~\bibnamefont {Briceno}}, \bibinfo
  {author} {\bibfnamefont {S.~G.}\ \bibnamefont {Roberts}}, \bibinfo {author}
  {\bibfnamefont {Z.}~\bibnamefont {Zhou}}, \ and\ \bibinfo {author}
  {\bibfnamefont {M.~A.}\ \bibnamefont {Kirk}},\ }\href@noop {} {\bibfield
  {journal} {\bibinfo  {journal} {Philosophical Magazine}\ }\textbf {\bibinfo
  {volume} {93}},\ \bibinfo {pages} {1715} (\bibinfo {year}
  {2013})}\BibitemShut {NoStop}%
\end{thebibliography}%
\end{document}